\documentclass[12pt]{article} 
\usepackage{amsmath}
\usepackage{amssymb,amsfonts,euscript}
\usepackage{hyperref}
\renewcommand{\baselinestretch}{1.2}
\newcommand\nn{\nonumber}

\newcommand{\Ord}{{\cal{O}}}

\newcommand{\cL}{\cal{L}}

\newcommand{\tg}{\tilde g}
\newcommand{\Zint}{\mathbb{Z}}

\newcommand{\bJ}{\bar{J}}
\newcommand{\bz}{\bar{z}}

\def\a{\alpha}
\def\be{\beta}

\def\m{\mu}
\def\n{\nu}
\def\r{\rho}

\def\de{\delta}

\def\hB{\hat{B}}

\def\hg{\hat{g}}

\def\hh{\hat{H}}
\def\ha{\hat{A}}

\def\part{\partial}
\def\half{{ \srac{1}{2} }}

\def\om{\omega}
\def\thickone{{\rm 1\mskip-4.5mu l}}

\newcommand{\sa}{\mathop{\vtop{\ialign{##\crcr
$\hfil\displaystyle{\longrightarrow}\hfil$\crcr\noalign{\kern-1pt\nointerlineski
p}

\hspace{.12in}$^\sigma$\hskip6pt\crcr\noalign{\kern3pt}}}}}
\newcommand{\slra}{\mathop{\vtop{\ialign{##\crcr
$\hfil\displaystyle{\longleftrightarrow}\hfil$\crcr\noalign{\kern-1pt\nointerlin
eskip}

\hspace{.12in}$^\sigma$\hskip6pt\crcr\noalign{\kern3pt}}}}}
\newcommand{\sat}{\mathop{\vtop{\ialign{##\crcr
$\hfil\displaystyle{\longrightarrow}\hfil$\crcr\noalign{\kern-1pt\nointerlineski
p}

\hspace{.12in}$^\sigma$\hskip6pt\crcr\noalign{\kern3pt}}}}}
\newcommand{\pa}{\mathop{\vtop{\ialign{##\crcr
$\hfil\displaystyle{\oplus}\hfil$\crcr\noalign{\kern+1pt\nointerlineskip}
\hspace{.08in}$^{\alpha=0}$\hskip6pt\crcr\noalign{\kern3pt}}}}}
\newcommand{\pan}{\mathop{\vtop{ialign{##\crcr
$\hfil\displaystyle{\oplus}\hfil$\crcr\noaligan{\kern+2pt\nointerlinkeskip}

\hspace{.03in} $^{\alpha}$\hskip6pt\crcr\noalign{\kern3pt}}}}}
\newcommand{\ka}{\mathop{\vtop{\ialign{##\crcr
$\hfil\displaystyle{\longleftrightarrow}\hfil$\crcr\noalign{\kern-1pt\nointerlin
eskip}

\hspace{.12in}$^K$\hskip6pt\crcr\noalign{\kern3pt}}}}}
\newcommand{\bp}{\mathop{\vtop{ialign{##\crcr
$\hfil\displaystyle{}\hfil$\crcr\noalign{\kern-13pt\nointerlineskip}
\big{(}\hskip0pt\crcr\noalign{\kern3pt}}}}}
\newcommand{\cbp}{\mathop{\vtop{ialign{##\crcr
$\hfil\displaystyle{}\hfil$\crcr\noalign{\kern-13pt\nointerlineskip}
\big{)}\hskip0pt\crcr\noalign{\kern3pt}}}}}

\newcommand{\s}{\sigma}

\newcommand{\ws}{\omega (h_\s)}

\newcommand{\hc}{$\hat{J}_{\gst}$}

\newcommand{\tp}{{2\pi i}}

\newcommand{\sgb}{{\mbox{\scriptsize{\gb}}}}

\def\gb            {\mbox{$\hat{\mathfrak g}$}}

\def\sm#1      {\mbox{\scriptsize $#1$}}

\def\srac#1#2{\smal{\frac{#1}{#2}}}

\def\foot#1{\mbox{\footnotesize $#1$}}
\def\scrs#1{\mbox{\scriptsize $#1$}}
\def\tyny#1{\mbox{\tiny $#1$}}
\def\smal#1{\mbox{\small $#1$}}
\def\big#1{\mbox{\large $#1$}}
\def\Big#1{\mbox{\Large $#1$}}

\def\hjb{\hat{\bar{J}}}

\def\gfrakh{\hat{\mathfrak g}}
\def\hfrakh{\hat{\mathfrak h}}
\def\hfrak{\mathfrak h}

\def\bfrak{\mathfrak b}

\def\dual{\underset{\s}{\longrightarrow}}
\def\duals{\!{\underset{\s}{\rightarrow}}}

\def\sg{\smal{\EuScript{G}}}

\def\hc{^\dagger}

\def\one{{\mathchoice {\rm 1\mskip-4mu l} {\rm 1\mskip-4mu} {\rm
1\mskip-4.5mu l}
{\rm 1\mskip-5mu l}}}

\def\nrm{{n(r)\m}}

\def\mnrn{{-n(r),\n}}
\def\nsn{{n(s)\n}}
\def\ntd{{n(t)\d}}

\def\nrrs{{\srac{n(r)}{\r(\s)}}}
\def\nsrs{{\srac{n(s)}{\r(\s)}}}
\def\scf{{\cal F}}
\def\sG{{\cal G}}

\def\gfrak{\mbox{$\mathfrak g$}}

\def\hj{\hat{J}}

\def\nsn{{n(s)\n}}
\def\schi{{\foot{\chi}}}
\def\schisig{{\foot{\chi(\s)}}}

\def\ntd{{n(t)\delta}}
\def\hc{^\dagger}
\def\st{{\cal T}}
\def\0b{\ }
\def\pl{\partial}

\def\Nrm{{N(r)\m}}

\def\Nsn{{N(s)\n}}

\def\sm{{\cal M}}

\def\ho{{\hat{\Omega}}}
\def\sx{\smal{\EuScript{X}}}
\def\sxfoot{{\foot{\EuScript{X}}}}
\def\sxh{{\hat{\smal{\EuScript{X}}}}}
\def\sxtiny{\tyny{\EuScript{X}}}
\def\sgtiny{\scrs{\EuScript{G}}}

\def\he{{\hat{e}}}

\def\heb{{\hat{\bar{e}\hspace{.02in}}\hspace{-.02in}}}

\def\hx{{\hat{x}}}
\def\hh{\hat{H}}

\def\sh{{\cal H}}
\def\hG{\hat{G}}
\def\hpl{\hat{\pl}}

\def\bh{{ \hat{\beta}}}

\def\su{{ \mathfrak{su} }}
\def\so{{ \mathfrak{so} }}

\def\bigspc{{ \quad \quad \quad \quad}}

\def\ep{{ \epsilon}}

\def\bT{{\bar{T}}}

\def\nrrsf{{ \frac{n(r)}{\r(\s)}}}

\def\Id{{ \dot{I}}}
\def\Jd{{ \dot{J}}}
\def\Kd{{ \dot{K}}}
\def\nrmu{{ \nrm u}}
\def\nsnv{{ \nsn v}}
\def\ntdw{{ \ntd w}}
\def\tg{{ \tilde{g}}}
\def\se{{\cal E}}
\def\sB{{\cal B}}
\def\sH{{\cal H}}
\def\sY{{\cal Y}}
\def\sW{{\cal W}}
\def\hO{{\hat{\Omega}}}
\def\hY{{\hat{Y}}}
\def\hnrhm{{\hat{n}(r)\hat{\m}}}
\def\hnrhmu{{\hat{n}(r)\hat{\m}u}}
\def\hnshnv{{\hat{n}(s)\hat{\n}v}}
\def\hnrmu{{\hat{n}(r)\m u}}
\def\hnsnv{{\hat{n}(s)\n v}}
\def\sA{\EuScript{A}}
\def\om{{\omega}}

\makeatletter
\renewcommand{\@makefnmark}{\mbox{$^{\ddagger\@thefnmark}$}}
\renewcommand{\subsection}{\@startsection
{subsection}{2}{0pt
}{-\baselineskip}{0.5\baselineskip}
{\normalfont\normalsize\bf}}
\renewcommand{\section}{\@startsection
{section}{2}{0pt
}{-\baselineskip}{0.5\baselineskip}
{\bf\large}}

\makeatother
\numberwithin{equation}{section}
\numberwithin{table}{section}

\newcounter{myfigctr}
\def\myfig#1{\refstepcounter{myfigctr}%
 \label{#1}%
}

\newcommand{\publititle}[8]
{ 
  \vspace*{-3cm}
  \begin{flushright} #1 \\ {\tt #2} \end{flushright}
  \vfill
  \begin{center}{\Large
    \bfseries #3}\end{center}
  \vskip 8mm
  \begin{center}{\large #4}\end{center}
  \begin{center}{\normalsize #5}\end{center}
  \vskip 8mm
  \nopagebreak
  \noindent #6
  \vfill
  \begin{flushleft} #7
  \end{flushleft}
  \hrule width 6.7cm \vskip.1mm
  {\small #8}
  \thispagestyle{empty}
  \clearpage
}

\begin{document}

\publititle{ ${}$ \\ UCB-PTH-03/22 \\ LBNL-53759 \\ hep-th/0309101}{}{On
the Target-Space Geometry of\\ Open-String Orientation-Orbifold Sectors
}{M.B.Halpern$^{a}$ and C. Helfgott$^{b}$}
{{\em Department of Physics, University of California and \\
Theoretical Physics Group,  Lawrence Berkeley National
Laboratory \\
University of California, Berkeley, California 94720, USA}
\\[2mm]} {Including world-sheet orientation-reversing automorphisms in the
orbifold program, we recently reported the twisted operator algebra and
twisted KZ equations in each open-string sector of the general WZW
orientation orbifold. In this paper we work out the corresponding
classical description
of these sectors, including the {\it WZW orientation-orbifold action} --
which is naturally defined on the solid half cylinder -- and its associated
WZW orientation-orbifold branes. As a generalization, we also obtain the
{\it sigma-model orientation-orbifold action}, which describes a much larger
class
of open-string orientation-orbifold sectors. As special cases, this class
includes twisted open-string {\it free boson} examples, the open-string WZW
sectors above and the open-string sectors of the
{\it general coset orientation orbifold}. Finally, we derive the {\it
orientation-orbifold Einstein equations}, in terms of twisted Einstein
tensors -- which hold when the twisted open-string sigma model sectors are
1-loop conformal.

} {$^a${\tt halpern@physics.berkeley.edu} \\ $^b${\tt
helfgott@socrates.berkeley.edu}
}

\clearpage

\renewcommand{\baselinestretch}{.4}\rm
{\footnotesize
\tableofcontents
}
\renewcommand{\baselinestretch}{1.2}\rm

\section{Introduction}

In the last few years there has been a quiet revolution in the local
theory of {\it current-algebraic orbifolds}.
Building on the discovery of {\it orbifold affine algebras}
\cite{Chr,AffVir} in the cyclic
permutation orbifolds, Refs.~[3-5] gave the twisted
currents and stress tensor in each twisted sector of any
current-algebraic orbifold $A(H)/H$ - where $A(H)$ is any
current-algebraic conformal field theory [6-13] with a
discrete
symmetry group $H$. The construction treats all current-algebraic
orbifolds at the same time, using the method of {\it eigenfields}
and the {\it principle of local isomorphisms} to map OPEs in the symmetric
theory to OPEs in the orbifold. The orbifold results
are expressed in terms of a set of twisted tensors or {\it duality
transformations}, which are discrete Fourier transforms
constructed from the eigendata of the $H$-{\it eigenvalue problem}.

More recently, the special case of the WZW orbifolds
\begin{gather}
\frac{A_g(H)}{H} ,\quad H\subset Aut(g) \label{Eq1.1}
\end{gather}
was worked out in further detail [14-17], introducing the
{\it extended $H$-eigenvalue problem} and the {\it linkage relation} to
include the {\it twisted affine primary fields}, the twisted vertex
operator equations and the {\it twisted KZ equations} of the WZW
orbifolds.
Ref.~\cite{so2n} includes a review of the general left- and right-mover
twisted KZ systems. For detailed
information on particular classes of WZW orbifolds, we direct the reader
to the following references:

$\bullet$ the WZW permutation orbifolds [14-16]

$\bullet$ the inner-automorphic WZW orbifolds \cite{Big, Perm}

$\bullet$ the (outer-automorphic) charge conjugation orbifold on $\su
(n\geq 3)$ \cite{Big'}

$\bullet$ the outer-automorphic WZW orbifolds on $\so (2n)$, including the
triality orbifolds \linebreak
\indent $\quad$ on $\so(8)$ \cite{so2n}.

\noindent Ref.~\cite{Big'} also solved the twisted vertex operator
equations and the twisted KZ systems in an abelian
limit\footnote{An abelian twisted KZ equation for the inversion orbifold
$x \rightarrow -x$ was given earlier in Ref.~\cite{Froh}.}
to obtain the {\it twisted vertex operators} for each sector of a large
class of orbifolds on abelian $g$. Moreover, Ref.~\cite{Perm}
found the {\it general orbifold Virasoro algebra} (twisted Virasoro
operators \cite{Chr,DV2}) of the WZW permutation
orbifolds and used the general twisted KZ system to study {\it
reducibility} of the general twisted affine primary field. Recent progress
at
the level of characters has been also reported in
Refs.~\cite{KacTod,Chr,Ban,Birke}.

In addition to the operator formulation, there have been a number of
discussions of {\it orbifold geometry} at the action level. In particular,
Ref.~\cite{Big} also gave the {\it general WZW orbifold action}, special
cases of which are further discussed in
Refs.~\cite{Big',so2n}. The general WZW orbifold action provides the
classical description of each sector of every WZW orbifold in terms of the
so-called {\it group orbifold elements} with
diagonal monodromy, which are the classical (high-level) limit of the
twisted affine
primary fields. Moreover, Ref.~\cite{Fab} gauged the general WZW orbifold
action by general twisted gauge groups to obtain the {\it
general coset orbifold action}, which describes each sector of the general
coset orbifold $A_{g/h} (H)/H$. Finally, the geometric description was
extended in Ref.~\cite{Geom} to include a large class of {\it sigma-model
orbifolds} and their corresponding {\it twisted Einstein equations}.

In Ref.~\cite{Orient1}, the orbifold program was extended beyond the space-time
orbifolds above to construct a  new class of orbifolds called the {\it WZW
orientation orbifolds}
\begin{gather}
\frac{A_g (H_-)}{H_-} ,\quad H_- \subset Aut (g\oplus g) \label{Eq1.2}
\end{gather}
where $H_-$ is any automorphism group which contains {\ world-sheet
orientation-
reversing automorphisms}. The orientation-orbifold sectors which arise by
twisting the orientation-reversing automorphisms are twisted open WZW
strings, each of which comes with an {\it extended Virasoro algebra} (which
is in
fact an orbifold Virasoro algebra) at {\it twice} the original closed-string
central charge. The twisted open-string KZ equations of the WZW orientation
orbifolds combine
salient features of the untwisted open-string KZ equations \cite{Giusto}
and the closed-string twisted KZ equations [14-17] of space-time
orbifold theory.

The present paper supplements the operator construction of
Ref.~\cite{Orient1} by discussing the {\it target-space geometry} of
open-string
 orientation orbifold
sectors. In particular, we begin by constructing the {\it WZW orientation
orbifold action} for each open-string sector of $A_g (H_-)/H_-$.
This action, which is a functional of appropriate group orbifold elements,
is naturally
defined on the solid half-cylinder. In a generalization, we also obtain
the {\it sigma-model orientation orbifold action}, which describes a much
larger set of open-string orientation orbifold sectors -- including the
open-string WZW orientation orbifold sectors as well as the open-string sectors
 of the {\it general coset
 orientation orbifold}. The corresponding {\it orientation-orbifold
 Einstein equations} are also worked out  -- which hold when the twisted
open-string
 sigma model sectors are one-loop conformal.

 The less technically-inclined reader may prefer to look first at
 Subsec.~$5.6$, where we discuss the simple case of twisted open-string
 {\it free boson} actions -- which result from the twisting of closed strings
 on abelian $g$.

\section{Group Elements and Group Orbifold Elements}

\subsection{Orientation-Reversing Automorphisms}

To fix notation, we begin with the affine Lie algebra \cite{Kac, Moody,
BH, ICFT} associated to
Lie$(g \oplus g)$
\begin{subequations}
\label{Eq2.1}
\begin{gather}
[ J_a (m) ,J_b (n) ]= if_{ab}{}^c J_c (m+n) + mG_{ab} \de_{m+n,0} \\
[ \bJ_a (m) ,\bJ_b (n) ] =if_{ab}{}^c \bJ_c (m+n) +mG_{ab} \de_{m+n,0} \\
[ J_a (m) ,\bJ_b (n) ] =0 ,\quad m,n \in \Zint ,\quad a,b,c =1,\ldots
,\text{dim }g \\
g=\oplus_I \gfrak^I ,\quad f_{ab}{}^c =\oplus_I f_{a(I),b(I)}^I {}^{c(I)}
,\quad G_{ab} =\oplus_I k_I \eta^I_{a(I),b(I)}
\end{gather}
\end{subequations}
where $\gfrak^I$ is any simple compact Lie algebra, $f_{ab}{}^c$ are the
structure constants of $g$ and $G_{ab}$ is the generalized
metric of affine $g$. The local currents on the cylinder $0\leq \xi \leq
2\pi$ corresponding to these modes are:
\begin{gather}
J_a (\xi,t) =\sum_m J_a (m) e^{-im (t+\xi)} ,\quad \bJ_a (\xi,t) =\sum_m
\bJ_a (m) e^{-im(t-\xi)} \,. \label{Eq2.2}
\end{gather}
For the classical dynamics of this paper, we will need the
standard\footnote{Our brackets are rescaled by an $i$ so that $[\,,]
\rightarrow \{ ,\}$.}
equal-time bracket form of affine $(g \oplus g)$, including the brackets
with the WZW group elements $g(T)$
\begin{subequations}
\label{Eq2.3}
\begin{gather}
\{ J_a (\xi,t) ,J_b (\eta,t) \} =\tp \left( f_{ab}{}^c J_c (\eta,t) +
G_{ab} \pl_\xi \right) \de (\xi-\eta) \\
\{ \bJ_a (\xi,t) ,\bJ_b (\eta,t) \} =\tp \left( f_{ab}{}^c \bJ_c (\eta,t)
-G_{ab} \pl_\xi \right) \de (\xi-\eta) \\
\{ J_a (\xi,t) ,\bJ_b (\eta,t) \} =0
\end{gather}
\begin{gather}
\{ J_a (\xi,t) ,g(T,\eta,t) \} =2\pi g(T,\eta,t) T_a \de (\xi-\eta) \\
\{ \bJ_a (\xi,t) ,g(T,\eta,t) \} =-2\pi T_a g(T,\eta,t) \de (\xi-\eta) \\
\{ J_a (\xi,t) ,g^{-1} (T,\eta,t) \} =-2\pi T_a g^{-1} (T,\eta,t) \de
(\xi-\eta) \\
\{ \bJ_a (\xi,t) ,g^{-1} (T,\eta,t) \} =2\pi g^{-1} (T,\eta,t) T_a \de
(\xi-\eta) \\
[T_a ,T_b ]=if_{ab}{}^c T_c ,\quad Tr (MT_a T_b )=G_{ab} ,\quad [M,T_a ]=0
,\quad [M,g(T,\xi,t)]=0  \label{Eq 2.3h} \\
g(T) =\{ g(T)_\a{}^\be \} ,\quad T_a = \{ (T_a )_\a{}^\be \} ,\quad \a,\be
=1,\ldots ,\text{dim }T
\end{gather}
\end{subequations}
where $T$ is any matrix rep of Lie $g$. As shown in Eq.~\eqref{Eq 2.3h},
the representation matrices $T$ are normalized with the
invertible data matrix M \cite{Big}
\begin{gather}
M \equiv M(k,T) =\oplus_I \frac{k_I}{y(T^I)} ,\quad T= \oplus_I T^I ,\quad
g(T,\xi,t) =\oplus_I g^I (T^I,\xi,t) \label{Eq2.4}
\end{gather}
which records the affine levels $\{k\}$ of each simple component of $g$
and the Dynkin indices $\{y\}$ of rep $T$. The group elements are
$2\pi$-periodic
\begin{gather}
g(T,\xi+2\pi ,t) = g(T,\xi,t) \label{Eq2.5}
\end{gather}
and we will also introduce the following tangent-space form of the group
elements
\begin{gather}
g(T,\xi,t) = e^{i\be^a (\xi,t) T_a} ,\quad \be^a (\xi+2\pi,t) = \be^a
(\xi,t) \label{Eq2.6}
\end{gather}
where $\be^a ,\,a=1\ldots \text{dim }g$ are the tangent-space coordinates.

Then the classical (high-level) limit of the affine primary fields
$g(T,\bz,z) ,\,g(\bT,z,\bz)^T$ of Ref.~\cite{Orient1} on the sphere translate
into the
following forms of the group elements on the cylinder
\begin{subequations}
\label{Eq2.7}
\begin{gather}
g(T,\bz,z) \longrightarrow g(T,\xi,t) = e^{i\be^a (\xi,t) T_a} \\
g(\bT,z,\bz)^t \longrightarrow g(\bT,-\xi,t)^t = e^{i\be^a (-\xi,t) \bT_a^t} 
   = e^{-i\be^a (-\xi,t) T_a} = g^{-1} (T,-\xi,t) \\
\bT \equiv -T^t ,\quad (T_a^t )_\a{}^\be \equiv (T_a )_\be{}^\a
\end{gather}
\end{subequations}
where superscript $t$ is matrix transpose.

With the correspondence \eqref{Eq2.7}, we
may then read off from Ref.~\cite{Orient1} the cylinder form of the general
{\it world-sheet orientation-reversing automorphism} $\hat{h}_\s =\tau_1
\times h_\s \in H_-$ of affine $(g\oplus g)$
\begin{subequations}
\label{Eq2.8}
\begin{gather}
g(T,\xi,t)' = W(h_\s ;T) g^{-1} (T,-\xi,t) W\hc(h_\s;T) \label{Eq 2.8a} \\
g^{-1}(T,-\xi,t)' =W(h_\s;T) g(T,\xi,t) W\hc(h_\s;T) \\
\be^a (\xi,t)' = -\be^b (-\xi,t) \omega\hc (h_\s)_b{}^a ,\quad \be^a (-\xi,t)'
=-\be^b (\xi,t) \omega\hc (h_\s)_b{}^a \label{Eq 2.8c} \\
J_a (\xi,t)' =\ws_a{}^b \bJ_b (-\xi,t) ,\quad \bJ_a (\xi,t)' =\ws_a{}^b
J_b (-\xi,t) \label{Eq 2.8d} \\
\ws_a{}^c \ws_b{}^d G_{cd} =G_{ab} ,\quad \ws_a{}^d \ws_b{}^e f_{de}{}^c
=f_{ab}{}^d \ws_d{}^c
\end{gather}
\begin{gather}
W\hc (h_\s;T) T_a W(h_\s;T) =\ws_a{}^b T_b \label{Eq 2.8f} \\
\ws \ws\hc =\one ,\quad W(h_\s;T) W\hc (h_\s;T) =\one \\
[W(h_\s ;T),M(k,T)]=0
\end{gather}
\end{subequations}
where $\ws, W(h_\s;T)$ are the actions of $h_\s \in Aut(g)$ in the adjoint
rep and in matrix rep $T$ respectively. The linkage relation \cite{Big}
in Eq.~\eqref{Eq 2.8f} guarantees the consistency of Eqs.~($2.8$a,b) with
\eqref{Eq 2.8c} and \eqref{Eq 2.8d}. In what follows, we deal exclusively
with the orientation-reversing automorphisms \eqref{Eq2.8}, which lead to
open-string
sectors. We remind the reader however that $H_-$ always contains an equal
number of orientation-preserving automorphisms, which leads to an equal number
of closed-string sectors in each orientation orbifold.

Following the two-component notation of Ref.~\cite{Orient1}, we can further
define a {\it matrix group element} and its tangent-space coordinates:
\begin{subequations}
\label{Eq2.9}
\begin{gather}
\tg (T,\xi,t) \equiv \left( \begin{array}{cc} g(T,\xi,t) & 0 \\ 0&
g^{-1}(T,-\xi,t) \end{array} \right) =e^{i\be^{a\Id}(\xi,t) T_a \r_\Id}
\label{Eq 2.9a} \\
\r_0 =\left( \begin{array}{cc} 1&0\\0&0 \end{array} \right) ,\quad \r_1
=\left( \begin{array}{cc} 0&0\\0&1 \end{array} \right) ,\quad Tr (\r_\Id
   \r_\Jd )=\de_{\Id \Jd} ,\quad \r_\Id \r_\Jd =\de_{\Id \Jd} \r_\Id
\label{Eq 2.9b}  \\
\be^{a,0} (\xi,t) \equiv \be^a (\xi,t) ,\quad \be^{a,1} (\xi,t) \equiv
-\be^a (-\xi,t) ,\quad \be^{a\Id} (\xi+2\pi ,t) =\be^{a\Id}(\xi,t)
\end{gather}
\begin{gather}
\tg(T,-\xi,t) = \tau_1 \tg^{-1} (T,\xi,t) \tau_1 \label{Eq 2.9d} \\
\be^{a\Id} (-\xi,t) =-\be^{a\Jd} (\xi,t) (\tau_1 )_{\Jd}{}^{\Id} \label{Eq
2.9e} \\
[\tau_3 ,\tg(T,\xi,t)]=[M\otimes \one_2 ,\tg(T,\xi,t)]= 0
\end{gather}
\begin{gather}
\tg(T,\xi,t)' = \tau_1 W(h_\s;T) \tg (T,\xi,t) W\hc(h_\s;T) \tau_1
\label{Eq 2.9g} \\
\tg^{-1} (T,\xi,t)' = \tau_1 W(h_\s;T) \tg^{-1} (T,\xi,t) W\hc(h_\s;T)
\tau_1 \\
\be^{a\Id} (\xi,t)' = \be^{b\Jd} (\xi,t) \omega\hc (h_\s)_b{}^a (\tau_1
)_{\Jd}{}^{\Id} \label{Eq 2.9i}
\end{gather}
\end{subequations}
\begin{gather}
W\hc (h_\s;T) T_a W(h_\s;T) =\ws_a{}^b T_b ,\quad \tau_1 \r_\Id \tau_1 =
(\tau_1 )_{\Id \Jd} \r_\Jd ,\quad \Id ,\Jd \in \{0,1\} \,. \label{Eq2.10}
\end{gather}
Here $\vec{\tau}$ are the Pauli matrices, and the automorphic response
\eqref{Eq 2.9g} of the matrix group element is the same as that of the
matrix affine primary field in Ref.~\cite{Orient1}. The linkage relations in
\eqref{Eq2.10} guarantee the consistency of Eqs.~($2.9$g,h) with
\eqref{Eq 2.9i}. The corresponding two-component form $J_{a\Id}(\xi,t)
=(J_a (\xi,t) ,\bJ_a (-\xi,t))$ of the currents is discussed in
Ref.~\cite{Orient1}.

We call attention to the behavior under {\it world-sheet parity} $\xi
\rightarrow -\xi$ of the matrix group element in \eqref{Eq 2.9d} and
the tangent-space coordinates in \eqref{Eq 2.9e}. These relations tell us
that the strip $0\leq \xi \leq \pi$ is the natural fundamental range for
the
two-component variables. In what follows we often suppress the time label
$t$.

\subsection{Eigengroup Elements}

The orientation-reversing automorphisms in Eq.~\eqref{Eq2.9} are now in
 the proper form required for
the orbifold program, which follows the dotted line in the commuting
diagram \cite{Dual, Geom} of Fig.~\ref{fig:goe}.

\begin{picture}(370,178)(0,0)
\put(148,165){$\tg$}
\put(157,158){\line(1,0){5}}
\put(178,158){\line(1,0){5}}
\put(198,158){\line(1,0){5}}
\put(218,158){\line(1,0){5}}
\put(238,158){\line(1,0){5}}
\put(167,158){\line(1,0){5}}
\put(187,158){\line(1,0){5}}
\put(208,158){\line(1,0){5}}
\put(228,158){\line(1,0){5}}
\put(248,158){\line(1,0){5}}
\put(259,150){\oval(15,15)[tr]}
\put(266,147){\line(0,-1){5}}
\put(266,139){\line(0,-1){5}}
\put(266,131){\line(0,-1){5}}
\put(266,123){\line(0,-1){5}}
\put(266,114){\line(0,-1){5}}
\put(266,106){\vector(0,-1){10}}
\put(184,165){$U\tg U\hc = \sg$}
\thicklines
\put(151,158){\vector(0,-1){60}}
\put(151,98){\vector(0,1){60}}
\put(148,86){$\hat{\sg}$}
\put(266,165){$\sg$}
\put(184,86){$ U\hat{\sg}U\hc=\hg$}
\put(270,158){\vector(0,-1){60}}
\put(270,98){\vector(0,1){60}}
\put(266,86){$\hg$}

\put(98,70) {{\footnotesize Each vertical double arrow is a local
isomorphism }}
\put(87,59) {$\foot{\tg}$}
\put(98,59) {{\footnotesize = Lie group elements: mixed under
automorphisms}}
\put(87,47) {$\sgtiny$}
\put(98,47) {{\footnotesize = eigengroup elements: diagonal under
automorphisms}}
\put(87,35) {$\foot{\hg}$}
\put(98,35) {{\footnotesize = group orbifold elements}}
\put(87,23) {${\hat{\sgtiny}}$}
\put(98,23) {{\footnotesize = group orbifold elements with twisted
boundary conditions}}
\put(98,4) {Fig.\,\ref{fig:goe}: Group and group orbifold elements}
\end{picture}
\myfig{fig:goe}
\vspace{.1in}

\noindent In particular, our next task is the construction of the
eigenfields $\sg$, for which we will need the analogues of the
$H$-eigenvalue problem of
Ref.~\cite{More} and the extended $H$-eigenvalue problem of Ref.~\cite{Big}:
\begin{subequations}
\label{Eq2.11}
\begin{gather}
\ws U\hc (\s) = U\hc (\s) E(\s),\quad E (\s)_\nrm{}^\nsn = \de_\nrm{}^\nsn
E_{n(r)}(\s) \label{Eq 2.11a} \\
W(h_\s;T) U\hc (T,\s) = U\hc (T,\s) E(T,\s) ,\quad
E(T,\s)_\Nrm{}^{\!\!\Nsn} \equiv \de_\Nrm{}^\Nsn E_{N(r)}(T,\s)
   \label{Eq 2.11b} \\
\de_\nrm{}^\nsn \equiv \de_\m{}^\n \de_{n(r)-n(s),0\, \text{mod }\r(\s)}
,\quad \de_\Nrm{}^\Nsn \equiv \de_\m{}^\n \de_{N(r)-N(s),0\,
   \text{mod }R(\s)} \\
U\hc (\s) =\{ U\hc (\s)_a{}^\nrm \} ,\quad U\hc (T,\s) =\{ U\hc
(T,\s)_\a{}^\Nrm \} \\
E_{n(r)}(\s) \equiv e^{-\tp \nrrs} ,\quad E_{N(r)}(T,\s) \equiv e^{-\tp
\srac{N(r)}{R(\s)}} \,.
\end{gather}
\end{subequations}
Here $\r(\s)$ and $R(\s)$ are the orders of $h_\s$ in the adjoint rep and
rep $T$ respectively. The unitary eigenvector matrices $U\hc (\s), U\hc
(T,\s)$ are periodic
\begin{gather}
n(r) \,\rightarrow \,n(r) \pm \r(\s) ,\quad N(r) \,\rightarrow \,N(r)\pm
R(\s) \label{Eq2.12}
\end{gather}
in their respective spectral indices $n(r)$ and $N(r)$, and the same is
true below for any object with these indices. In what follows, the barred
quantities
$\bar{n}(r)$ and $\bar{N}(r)$ are the pullbacks of the spectral indices to
their fundamental ranges
\begin{subequations}
\label{Eq2.13}
\begin{gather}
\srac{\bar{n}(r)}{\r(\s)} =\nrrs -\lfloor \nrrs \rfloor ,\quad
\srac{\bar{N}(r)}{R(\s)} = \srac{N(r)}{R(\s)} -\lfloor \srac{N(r)}{R(\s)}
\rfloor \\
\bar{n}(r) \in \{0,\ldots ,\r(\s)-1\} ,\quad \bar{N}(r) \in \{0,\ldots
,R(\s)-1\}
\end{gather}
\end{subequations}
where $\lfloor x \rfloor$ is the floor of $x$. Explicit solutions of these
eigenvalue problems are given in Refs.~[3,5,14-17].

For this discussion, we will also need the standard {\it duality
transformations} \cite{Dual,More,Big} or twisted tensors of sector $\s$
\begin{subequations}
\label{Eq2.14}
\begin{align}
&\sG_{\nrm;\nsn} (\s) \equiv \schisig_\nrm \schisig_\nsn U(\s)_\nrm{}^a
U(\s)_\nsn{}^b G_{ab} \nn \\
&\bigspc \bigspc = \sG_{\nsn;\nrm} (\s)= \de_{n(r)+n(s) ,0\, \text{mod }
\r(\s)} \sG_{\nrm;\mnrn} (\s) \label{Eq 2.14a} \\
&\scf_{\nrm;\nsn}{}^{\!\!\!\ntd} (\s) \!\equiv \schisig_\nrm \schisig_\nsn
\schisig^{-1}_\ntd U(\s)_\nrm{}^a U(\s)_\nsn{}^b f_{ab}{}^c U\hc
(\s)_c{}^\ntd \quad \quad \nn \\
&\bigspc = \!-\scf_{\nsn;\nrm}{}^{\!\!\!\ntd}(\s)
\!=\!\de_{n(r)+n(s)-n(t),0\,\text{mod } \r(\s)}
   \scf_{\nrm;\nsn}{}^{\!\!\!\!n(r)+n(s),\de} (\s) \quad \label{Eq 2.14b}
\end{align}
\begin{gather}
\st_\nrm (T,\s) \equiv \schisig_\nrm U(\s)_\nrm{}^a U(T,\s) T_a U\hc
(T,\s) \label{Eq 2.14c} \\
E_{n(r)} (\s)^\ast \st_\nrm (T,\s) = E(T,\s) \st_\nrm (T,\s) E^\ast(T,\s)
\label{Eq 2.14d} \\
[\st_\nrm (T,\s) ,\st_\nsn (T,\s)]
=i\scf_{\nrm;\nsn}{}^{\!\!n(r)+n(s),\de} (\s) \st_{n(r)+n(s),\de} (T,\s)
\label{Eq 2.14e} \\
\sm(\st(T,\s),\s) \equiv U(T,\s) M(k,T) U\hc (T,\s) \label{Eq 2.14f} \\
[\sm (\st(T,\s),\s) ,\st_\nrm (T,\s)]= [\sm(\st(T,\s),\s) ,E(T,\s)]=0
\label{Eq 2.14g} \\
\widehat{Tr} \left( \sm(\st(T,\s),\s) \st_\nrm (T,\s) \st_\nsn (T,\s)
\right) =\sG_{\nrm;\nsn}(\s)
\end{gather}
\end{subequations}
which are well-known in ordinary (space-time) current-algebraic orbifold
theory. Listed here in particular are the twisted structure constants
$\scf(\s)$,
the twisted representation matrices $\st$ (which satisfy the orbifold Lie
algebra $\hg(\s)$ in \eqref{Eq 2.14e}), and the (invertible) twisted data
matrix
$\sm$ -- where $\schisig_\nrm$ are arbitrary normalization constants.
These duality transformations have been evaluated [3,5,14-17] for
all basic orbifold types, except for a few outer-automorphic orbifolds of
$\Zint_2$ type. In the case of permutation-invariant systems
\begin{gather}
g=\oplus \gfrak^I ,\quad \gfrak^I \simeq \gfrak ,\quad k_I =k ,\quad T^I
\simeq T \label{Eq2.15}
\end{gather}
one has $M(k,T) \propto \thickone$ and the twisted data matrix satisfies
$\sm (\st,\s) =M(k,T)$.

We are now in a position to define the {\it eigengroup elements}
\begin{subequations}
\label{Eq2.16}
\begin{gather}
\sg(\st,\xi,\s) \equiv UU(T,\s) \tg(T,\xi) U\hc(T,\s)U\hc =\left(
\begin{array}{cc} \sg_{(0)}(\st,\xi,\s) &\sg_{(1)}(\st,\xi,\s) \\
  \sg_{(1)}(\st,\xi,\s)& \sg_{(0)}(\st,\xi,\s) \end{array} \right)
\label{Eq 2.16a} \\
\sg_{(u)} (\st,\xi,\s) \equiv \srac{1}{2} U(T,\s) \left( g(T,\xi,t)
+(-1)^u g^{-1} (T,-\xi,t) \right) U\hc(T,\s) ,\quad \bar{u}=0,1
\end{gather}
\begin{gather}
\sg^{-1} (\st,\xi,\s) = \tau_3 \sg(\st,-\xi,\s) \tau_3= \left(
\begin{array}{cc} \sg_{(0)} (\st,-\xi,\s) & -\sg_{(1)} (\st,-\xi,\s) \\
    -\sg_{(1)} (\st,-\xi,\s) & \sg_{(0)} (\st,-\xi,\s) \end{array} \right)
\label{Eq 2.16c} \\
[\tau_1 ,\sg (\st,\xi,\s)]=[\sm(\st,\s) \otimes \one_2 ,\sg(\st,\xi,\s)]
=0 \\
U=U\hc \equiv \frac{1}{\sqrt{2}} \left( \begin{array}{cc} 1&1\\1&-1
\end{array} \right) ,\quad \tau_1 U\hc =U\hc \tau_3  \label{Eq 2.16e}
\end{gather}
\end{subequations}
which (like the matrix group elements) are also two-component fields. The
eigengroup elements are the high-level limits of the eigenprimary fields
of Ref.~\cite{Orient1}. Here $U\hc$ plays the same role in the two-dimensional
space that $U\hc(T,\s)$ plays in the space of rep $T$, and the product
$U\hc (T,\s)U\hc$
plays the role of the total $U\hc$ shown in Fig.~\ref{fig:goe}. The form
of the inverse in \eqref{Eq 2.16c} follows directly from
Eq.~\eqref{Eq 2.9d}, but further non-local quadratic relations can be
obtained by comparing this form with \eqref{Eq 2.16a}.

The eigengroup elements are defined to diagonalize the action of the
automorphism $\hat{h}_\s$:
\begin{subequations}
\label{Eq2.17}
\begin{gather}
\sg (\st,\xi,\s)' = \tau_3 E(T,\s) \sg(\st,\xi,\s) E^\ast (T,\s) \tau_3
\label{Eq 2.17a} \\
\sg^{-1} (\st,\xi,\s)' = \tau_3 E(T,\s) \sg^{-1}(\st,\xi,\s) E^\ast (T,\s)
\tau_3 \\
\sg_{(u)} (\st,\xi,\s)' = (-1)^u E(T,\s) \sg_{(u)} (\st,\xi,\s) E^\ast
(T,\s) \,.
\end{gather}
\end{subequations}
At the tangent-space level, one finds the corresponding structure
\begin{subequations}
\label{Eq2.18}
\begin{gather}
\sg(\st(T,\s),\xi,\s) \!=\!e^{i \bfrak^\nrmu (\xi,\s) \st_\nrmu (T,\s)}
,\quad \st_\nrmu (T,\s) \!\equiv \!\st_\nrm (T,\s) \tau_u ,\quad u\!=\!0,1
\label{Eq 2.18a} \\
\tau_0 =\one ,\quad \vec{\tau} =\text{ Pauli matrices} \\
\bfrak^\nrmu (\xi,\s) \equiv \be^{a\Id} (\xi) \schisig_\nrm^{-1} U\hc
(\s)_a{}^\nrm (\srac{1}{\sqrt{2}} (U\hc )_\Id{}^u ) \bigspc \nn \\
\bigspc \bigspc \,\, = \srac{1}{2} \left( \be^a (\xi) +(-1)^{u+1} \be^a
(-\xi) \right) \schisig_\nrm^{-1} U\hc (\s)_a{}^\nrm \\
\quad \quad \,\, = \srac{1}{2} \left( \bfrak^\nrm (\xi,\s) +(-1)^{u+1}
\bfrak^\nrm (-\xi,\s) \right) \\
\bfrak^\nrm (\xi,\s) \equiv \be^a (\xi) \schisig_\nrm^{-1} U\hc
(\s)_a{}^\nrm \\
\bfrak^\nrmu (-\xi,\s) = (-1)^{u+1} \bfrak^\nrmu (\xi,\s) \label{Eq 2.18e}
\\
\bfrak^\nrmu (\xi,\s)' = \bfrak^\nrmu (\xi,\s) e^{\tp (\nrrs
+\srac{u}{2})}  \label{Eq 2.18f}
\end{gather}
\end{subequations}
in terms of the tangent-space eigencoordinates $\bfrak$. The world-sheet
parity \eqref{Eq 2.18e} of the tangent-space eigencoordinates implies
the world-sheet parity \eqref{Eq 2.16c} of the eigengroup elements. With
the $\st$-selection rule \eqref{Eq 2.14d}, the (diagonal) automorphic
response \eqref{Eq 2.18f} similarly implies the automorphic response
\eqref{Eq 2.17a} of the eigengroup elements. We also note that the {\it
total} twisted
representation matrices in \eqref{Eq 2.18a} satisfy the total {\it
orbifold Lie algebra} $\hg_O (\s)$ of sector $\hat{h}_\s$
\begin{subequations}
\label{Eq2.19}
\begin{gather}
[\st_\nrmu (T,\s) ,\st_\nsnv (T,\s)]
=i\scf_{\nrm;\nsn}{}^{n(r)+n(s),\de}(\s) \st_{n(r)+n(s),\de,u+v}(T,\s)
\label{Eq 2.19a} \\
\widehat{Tr} \left( (\sm \otimes \one_2 )\st_\nrmu (T,\s) \st_\nsnv (T,\s)
\right) =\sG_{\nrmu;\nsnv} (\s) \label{Eq 2.19b}  \\
\bigspc \bigspc \bigspc \equiv 2\de_{u+v ,0\,\text{mod }2}
\sG_{\nrm;\nsn}(\s)  \label{Eq 2.19c}
\end{gather}
\end{subequations}
where $\scf(\s)$ are the twisted structure constants defined in \eqref{Eq
2.14b} and $\sG(\s)$ in \eqref{Eq 2.19b} is
the total twisted metric of sector $\hat{h}_\s$.

\subsection{Group Orbifold Elements and Definite Monodromies}

We now invoke the principle of local isomorphisms \cite {Dual,More,Big}
to go from the eigenfield formulation to twisted
open-string sector $\hat{h}_\s$ of the orientation orbifold
\begin{gather}
\sg (\st,\xi,t,\s) \dual \hg(\st,\xi,t,\s) ,\quad \bfrak (\xi,t,\s) \dual
\bh (\xi,t,\s) \label{Eq2.20}
\end{gather}
where $\hg$ and $\bh$ are respectively the {\it group orbifold elements}
and the {\it twisted tangent-space coordinates} of sector $\hat{h}_\s$.
Both of these twisted fields inherit the index structure and the
constraints of the corresponding eigenfields
\begin{subequations}
\label{Eq2.21}
\begin{gather}
\hg(\st,\xi,\s) =\{ \hg(\st,\xi,\s)_{\Nrm u}{}^{\!\!\!\Nsn v} \} \\
[\sm \otimes \one_2 ,\hg(\st,\xi,\s)]= 0 \\
[\tau_1 ,\hg(\st,\xi,\s)]= 0 \label{Eq 2.21c} \\
\hg (\st,-\xi,\s) = \tau_3 \hg^{-1} (\st,\xi,\s) \tau_3 \label{Eq 2.21d} \\
\hg(\st,-\xi,\s)_{\Nrm u}{}^{\Nsn v} =(-1)^{u-v} \hg^{-1}
(\st,\xi,\s)_{\Nrm u}{}^{\Nsn v}
\end{gather}
\end{subequations}
\begin{subequations}
\label{Eq2.22}
\begin{gather}
\hg(\st,\xi,\s) = e^{i\bh^\nrmu (\xi,\s) \st_\nrmu (T,\s)} \label{Eq
2.22a} \\
\bh^\nrmu (-\xi,\s) =(-1)^{u+1} \bh^\nrmu (\xi,\s) ,\quad \bar{u}=0,1
\label{Eq 2.22b}
\end{gather}
\end{subequations}
where Eq.~\eqref{Eq 2.22a} is the twisted tangent-space form of the group
orbifold element. The tangent-space form guarantees that the group
orbifold
elements form a group because the total matrices $\st$ satisfy the total
orbifold Lie algebra $\hg_O (\s)$. Moreover, Eqs.~\eqref{Eq 2.22a} and
\eqref{Eq 2.22b} together satisfy all the conditions on $\hg$ in
Eq.~\eqref{Eq2.21}. Finally, the {\it world-sheet parities}
\eqref{Eq 2.21d} and \eqref{Eq 2.22b} tell us that the natural range of
$\xi$ is the strip $0\leq \xi \leq \pi$, and restriction to the
strip is equivalent to the conventional orbifold identification $-\xi \sim
\xi$.

The relations \eqref{Eq2.21} also give the more explicit two-component
structure of the group orbifold elements
\begin{subequations}
\label{Eq2.23}
\begin{gather}
\hg(\st,\xi,\s) = \left( \begin{array}{cc} \hg_{(0)} (\st,\xi,\s)
&\hg_{(1)} (\st,\xi,\s) \\ \hg_{(1)} (\st,\xi,\s) &\hg_{(0)}
   (\st,\xi,\s) \end{array} \right) ,\quad [\sm ,\hg_{(u)}(\st,\xi,\s) ]=0
\label{Eq 2.23a} \\
\hg^{-1} (\st,\xi,\s) = \left( \begin{array}{cc} \hg_{(0)} (\st,-\xi,\s)
&-\hg_{(1)} (\st,-\xi,\s) \\ -\hg_{(1)} (\st,-\xi,\s) &\hg_{(0)}
   (\st,-\xi,\s) \end{array} \right) \label{Eq 2.23b}
\end{gather}
\begin{gather}
\hg(\st,\xi,\s)_{\Nrm u}{}^{\Nsn v} =\hg_{(u-v)} (\st,\xi,\s)_\Nrm{}^\Nsn
\\
\hg^{-1}(\st,\xi,\s)_{\Nrm u}{}^{\Nsn v} =(-1)^{u-v}
\hg_{(u-v)}(\st,-\xi,\s)
\end{gather}
\end{subequations}
and the form  \eqref{Eq 2.23b} of the inverse implies the further
non-local quadratic relations
\begin{subequations}
\label{Eq2.24}
\begin{gather}
\hg_{(0)} (\st,\xi,\s) \hg_{(0)} (\st,-\xi,\s) -\hg_{(1)}(\st,\xi,\s)
\hg_{(1)}(\st,-\xi,\s) =\one \\
\hg_{(0)} (\st,\xi,\s) \hg_{(1)} (\st,-\xi,\s) -\hg_{(1)}(\st,\xi,\s)
\hg_{(0)}(\st,-\xi,\s) =0
\end{gather}
\end{subequations}
among the reduced components $\hg_{(u)}$ of the group orbifold elements.

In addition to these properties, one may attempt to impose definite
monodromies on $\hg$ and $\bh$, as they would follow from Eqs.~\eqref{Eq
2.17a},
\eqref{Eq 2.18f} by a standard application of local isomorphisms
\begin{subequations}
\label{Eq2.25}
\begin{gather}
\text{automorphic responses of } \sg ,\bfrak \,\dual \,\text{monodromies
of } \hg,\bh \quad \label{Eq 2.25a} \\
\hg(\st(T,\s),\xi+2\pi,\s) = \tau_3 E(T,\s) \hg(\st(T,\s),\xi,\s) E^\ast
(T,\s) \tau_3 \quad \label{Eq 2.25b} \\
\hg^{-1} (\st(T,\s),\xi +2\pi,\s) =\tau_3 E(T,\s) \hg^{-1}
(\st(T,\s),\xi,\s) E^\ast (T,\s) \tau_3 \\
\hg(\st,\xi+2\pi,\s)_{\Nrm u}{}^{\Nsn v} =e^{-\tp (\frac{N(r)-N(s)}{R(\s)}
+\frac{u-v}{2})} \hg(\st,\xi,\s)_{\Nrm u}{}^{\Nsn v} \label{Eq 2.25d} \\
\hg_{(u)} (\st(T,\s),\xi+2\pi,\s) =(-1)^u E(T,\s) \hg_{(u)}
(\st(T,\s),\xi,\s) E^\ast (T,\s) \\
\bh^\nrmu (\xi+2\pi,\s) = \bh^\nrmu (\xi,\s) e^{\tp (\nrrs +\srac{u}{2})}
\label{Eq 2.25f}
\end{gather}
\end{subequations}
where the eigenvalue matrix $E(T,\s)$ is defined in Eq.~\eqref{Eq2.11}. As
above, Eqs.~\eqref{Eq 2.25f} and the $\st$-selection rule \eqref{Eq 2.14d}
imply Eq.~\eqref{Eq 2.25b}, and moreover, all these monodromies are
consistent with each other.

In Ref.~\cite{Orient1} however, we argued from the twisted currents (see
Subsec.~$3.3$) that the group orbifold elements in the open-string sectors
of
the WZW orientation orbifolds have the definite monodromy \eqref{Eq 2.25b}
{\it only} in the case
\begin{gather}
h_\s^2 =1, \quad E^2 (T,\s) =1 ,\quad \srac{\bar{n}(r)}{\r(\s)} \in
\{0,\half \}   \label{Eq2.26}
\end{gather}
and will have {\it mixed} monodromy when $h_\s^2 \neq 1$.

We present here another version of this argument which shows that the
definite monodromies in Eq.~\eqref{Eq2.25} are not consistent with the
world-sheet parities
\eqref{Eq 2.21d}, \eqref{Eq 2.22b} beyond the case $h_\s^2 =1$.

To find this inconsistency, consider {\it generic orbits} of the
world-sheet parity and monodromy, following the steps:
\begin{subequations}
\label{Eq2.27}
\begin{align}
\hg(\st(T,\s),\xi,\s) &=\tau_3 E(T,\s) \hg(\st(T,\s),\xi -2\pi ,\s) E^\ast
(T,\s) \tau_3 \\
&=E(T,\s) \hg^{-1} (\st(T,\s),2\pi -\xi,\s) E^\ast (T,\s) \\
&=\tau_3 E^2 (T,\s) \hg^{-1} (\st(T,\s) ,-\xi,\s) E^{\ast 2}(T,\s) \tau_3
\\
&=E^2 (T,\s) \hg(\st(T,\s) ,\xi,\s) E^{\ast 2}(T,\s)
\end{align}
\begin{align}
\bh^\nrmu (\xi) &= \bh^\nrmu (\xi -2\pi) e^{\tp (\nrrsf +\frac{u}{2})} \\
&= -\bh^\nrmu (2\pi -\xi) e^{\tp \nrrsf}  \\
&= -\bh^\nrmu (-\xi) e^{4\pi i \nrrsf} (-1)^u  \\
&= \bh^\nrmu (\xi) e^{4\pi i\nrrsf}  \,.
\end{align}
\end{subequations}
These relations are easily expressed as the selection rules
\begin{subequations}
\label{Eq2.28}
\begin{gather}
\hg(\st(T,\s),\xi,\s)_{\Nrm u}{}^{\Nsn v} (1- e^{-4\pi i
(\frac{N(r)-N(s)}{R(\s)})}) =0 \bigspc \quad \quad \nn \\
\bigspc \quad \quad \Rightarrow \, \hg(\st(T,\s),\xi,\s)_{\Nrm u}{}^{\Nsn
v}=0 \text{ unless } \srac{N(r)-N(s)}{R(\s)} \in \half \Zint \\
\bh^\nrmu(\xi,\s ) (1-e^{4\pi i\nrrsf}) =0 \,\Rightarrow \, \bh^\nrmu
(\xi,\s )=0 \text{ unless } \srac{n(r)}{\r(\s)} \in \half \Zint
\end{gather}
\end{subequations}
which are again consistent with each other by the $\st$-selection rule
\eqref{Eq 2.14d}. The selection rules tell us however that the twisted
fields
with world-sheet parity and definite monodromy can have no support beyond
the case $h_\s^2 =1$ in Eq.~\eqref{Eq2.26}.

This conclusion is a restriction \cite{Orient1} on that part of the principle
of local isomorphisms which deals with the monodromies of open-string
group
orbifold elements, so that the correct form of Eq.~\eqref{Eq 2.25a} for
the open-string sectors of the orientation orbifolds is:
\begin{gather}
h_\s^2 =1 \!: \quad \text{automorphic responses of } \sg,\bfrak \,\dual \,
\text{monodromies of } \hg ,\bh \,. \bigspc \label{Eq2.29}
\end{gather}
We emphasize with Ref.~\cite{Orient1} that all the other results obtained by
eigenfields and local isomorphisms hold for all $\hat{h}_\s$ -- except for
the
definite $2\pi$-monodromy \eqref{Eq2.25}, which holds only for $h_\s^2 =1$.

In spite of the absence of definite $2\pi$-monodromy for $\hg$ when
$h_\s^2 \!\neq \!1$, each sector $\hat{h}_\s \!=\tau_1 \!\times \!h_\s$ is
a {\it twisted}
open string because each sector contains at least some fractionally-moded
twisted currents (see Subsec.~$3.4$).

For completeness, we finally note that the class of $h_\s^2 =1$ open-string
sectors
can be subdivided into two cases \cite{Orient1}
\begin{subequations}
\label{Eq2.30}
\begin{gather}
h_\s =1 \!:\quad \r(\s)=1 ,\quad \ws =1 ,\quad \bar{n}(r)=\bar{N}(r)=0
\label{Eq 2.30a} \\
\,\,\, h_\s^2=1 ,\,\,h_\s \neq 1\!: \quad \r(\s)=2,\quad \ws^2 =1,\quad
\bar{n}(r), \bar{N}(r) \in \{ 0,1\}  \label{Eq 2.30b}
\end{gather}
\end{subequations}
where \eqref{Eq 2.30a} is the {\it basic} open-string sector, and
\eqref{Eq 2.30b} collects the {\it generic} open-string sectors with
$h_\s^2 =1$. We
will return to further discuss the special properties of the class
$h_\s^2 =1$ in Subsec.~$3.5$.

\section{Actions and Boundary Conditions on the Solid Half Cylinder}

Our next task is to find the action which describes the open-string WZW
orientation-orbifold sectors above. In order to apply the method of
eigenfields
and the principle of local isomorphisms to this problem, we will first
rewrite the standard WZW action in an unconventional but equivalent form
on
the {\it solid half cylinder} $\Gamma_{1/2}$ (see Fig.~\ref{fig:halfcyl}).

\begin{picture}(300,175)
\put(0,0){\begin{picture}(200,175)
\put(212,118){\begin{picture}(42,28)
\qbezier(17,-8)(23,0)(17,8)
\qbezier(17,8)(16,9)(15,10)
\qbezier(15,10)(12,12)(8,14)
\qbezier(8,14)(4,15)(0,15)
\qbezier(0,15)(-4,15)(-8,14)
\qbezier(-8,14)(-12,12)(-15,10)
\qbezier(-15,10)(-16,9)(-17,8)
\end{picture}} %

\put(212,48){\begin{picture}(42,28)
\qbezier(17,-8)(23,0)(17,8)
\qbezier(17,8)(16,9)(15,10)
\qbezier(8,14)(4,15)(0,15)
\qbezier(-8,14)(-12,12)(-15,10)
\qbezier(-15,10)(-16,9)(-17,8)
\end{picture}} %

\put(212,118){\vector(0,1){42}}\put(211,164){$t$}
\put(212,118){\vector(2,1){42}}\put(258,140){$y$}

\put(212,118){\vector(2,-1){42}}\put(258,94){$x$}

\put(212,118){\vector(2,-1){42}}\put(244,108){$\xi=0$}
\put(157,117){$\xi =\pi$}

\put(170,139){\line(2,-1){84}} %
\put(195,56){\line(2,-1){33}} %

\put(232,48){\line(0,1){70}} %
\put(195,56){\line(0,1){70}} %
\put(229,40){\line(0,1){70}} %

\put(130,10){Fig.\,\ref{fig:halfcyl}: The Solid Half Cylinder}

\end{picture}}
\end{picture}

\myfig{fig:halfcyl}
\vspace{.05in}

\subsection{WZW Action on the Solid Half Cylinder}

The standard untwisted WZW action \cite{Nov,W} on the solid cylinder $\Gamma$
is easily generalized to semisimple $g$ \cite{Big}
\begin{align}
&S_{WZW}[M,g;\Gamma] =-\frac{1}{8\pi} \int \!\!dt \!\int_0^{2\pi}
\!\!\!d\xi \,Tr \left( M\!\left( g^{-1}(T,\xi)\pl_+ g(T,\xi) g^{-1}(T,\xi)
   \pl_- g(T,\xi) \right) \right) \nn \\
& \bigspc \bigspc -\frac{1}{12\pi} \int_{\Gamma} Tr \left( M\!\left(
g^{-1}(T,\xi) dg(T,\xi) \right)^3 \right) ,\quad \pl_{\pm} =\pl_t
   \pm \pl_\xi  \label{Eq3.1}
\end{align}
where the data matrix $M$ is defined in Eq.~\eqref{Eq2.4}. As usual, the
integration over $\Gamma$ in the WZW term involves $\int \!dt
\int_0^{2\pi}
\!d\xi \int \!d\r$, where $\r$ is the radial coordinate. Using the
properties of $M$ in Eqs.~\eqref{Eq2.5} and \eqref{Eq2.11}, cyclicity of
$Tr$, and $2\pi$-periodicity in the form
\begin{gather}
f(\xi +2\pi) =f(\xi) \,\Rightarrow \, \int_0^{2\pi} \!\!d\xi f(-\xi)
=\int_0^{2\pi} \!\!d\xi f(\xi) \label{Eq3.2}
\end{gather}
it is then straightforward to check that the WZW action is invariant
\begin{gather}
S_{WZW} [M,g' ;\Gamma] =S_{WZW} [M,g;\Gamma] \label{Eq3.3}
\end{gather}
under the general orientation-reversing automorphism in Eq.~\eqref{Eq2.8}.

The WZW action can in fact be reexpressed in terms of the matrix group
element $\tilde{g}$ on the solid half cylinder $\Gamma_{1/2}$
\begin{align}
&S_{WZW}[M \otimes \one_2 ,\tg ;\Gamma_{\half}] \nn \\
&\bigspc \equiv -\frac{1}{8\pi} \int \!\!dt \int_0^{\pi} \!\!\!d\xi
\,\widehat{Tr} \left( (M\otimes
  \one_2 ) \left( \tg^{-1}(T,\xi)\pl_+ \tg(T,\xi) \tg^{-1}(T,\xi) \pl_-
\tg(T,\xi) \right) \right) \quad \quad \nn \\
& \bigspc \bigspc \!\!-\frac{1}{12\pi} \int_{\Gamma_{\half}}
\!\!\!\widehat{Tr} \left( (M\otimes \one_2 ) \left( \tg^{-1}(T,\xi)
d\tg(T,\xi)
   \right)^3 \right) =S_{WZW}[M,g;\Gamma] \label{Eq3.4}
\end{align}
where integration over $\Gamma_{1/2}$ involves $\int \!dt \int_0^{\pi}
\!d\xi \int \!d\r$. To see this equality for the WZW term in particular,
follow the steps
\begin{subequations}
\label{Eq3.5}
\begin{gather}
\gamma(t,\xi,\r) \equiv \ep^{ABC} Tr \left( Mg^{-1}(T,\xi) \pl_A g(T,\xi)
g^{-1}(T,\xi) \pl_B g(T,\xi) g^{-1}(T,\xi) \pl_C g(T,\xi) \right) \\
\{A,B,C\} = \{ t,\xi,\r \} ,\quad \ep^{t\xi \r} =1  \label{Eq 3.5b}
\end{gather}
\begin{align}
\int_{\Gamma} Tr \left( M \left( g^{-1} (T,\xi) dg(T,\xi) \right)^3
\right) & =\int \!\!dt \int_0^{2\pi} \!\!\!d\xi \int \!\!d\r
   \,\gamma (t,\xi,\r) \nn \\
& =\int \!\!dt \int_0^{\pi} \!\!d\xi \int \!\!d\r \left( \gamma (t,\xi,\r)
+\gamma (t,-\xi,\r) \right) \label{Eq 3.5c} \\
& =\int_{\Gamma_{\half}} \widehat{Tr} \left( (M \otimes \one_2) \left(
\tg^{-1} (T,\xi) d\tg (T,\xi) \right)^3 \right) \label{Eq 3.5d}
\end{align}
\end{subequations}
where $2\pi$-periodicity of the quantity $\gamma (t,\xi,\r)$ was used to
obtain \eqref{Eq 3.5c}. For the last equality, the reader may find it
helpful to use the form
\eqref{Eq 2.9a} of the matrix group element to write out \eqref{Eq 3.5d}
as a sum of two traces $Tr$ and use cyclicity. A similar argument suffices
to rewrite the kinetic term of the WZW action as
\begin{gather}
-\frac{1}{8\pi} \!\int \!\!dt \!\int_0^\pi \!\!\!d\xi \,\Big{[} Tr \left(
M \left( g^{-1}(T,\xi) \pl_+ g(T,\xi) g^{-1}(T,\xi) \pl_- g(T,\xi) \right)
  \right) +(\xi \rightarrow -\xi) \Big{]}  \label{Eq3.6}
\end{gather}
which sums to the $\tg$ form of the kinetic term in Eq.~\eqref{Eq3.4}.

The $\Gamma_{1/2}$ form \eqref{Eq3.4} of the WZW action is transparently
invariant under the two-component form \eqref{Eq2.9} of the general
orientation-reversing automorphism, this time without use of periodicity.
For reference below, we also note that both integrands in the action
\eqref{Eq3.4} are even under world-sheet parity
$\xi \leftrightarrow -\xi$.

The next step in the orbifold program is to express the $\Gamma_{1/2}$
form \eqref{Eq3.4} of the action in terms of the eigengroup elements
\begin{align}
&S_{WZW}[\sm \otimes \one_2 ,\sg ;\Gamma_{\half}] \nn \\
&\quad \,\, \equiv -\frac{1}{8\pi} \int \!\!dt \int_0^{\pi} \!\!d\xi
\,\widehat{Tr} \left( (\sm \otimes \one_2 ) \left( \sg^{-1}(\st,\xi)\pl_+
  \sg(\st,\xi) \sg^{-1}(\st,\xi) \pl_- \sg(\st,\xi) \right) \right) \quad
\quad \nn \\
&\quad \quad \quad -\frac{1}{12\pi} \int_{\Gamma_{\half}} \!\!\widehat{Tr}
\left( (\sm \otimes \one_2 ) \left( \sg^{-1}(\st,\xi) d\sg(\st,\xi)
   \right)^3 \right) = S_{WZW} [M \otimes \one_2 ,\tg(\sg);\Gamma_{\half}]
\label{Eq3.7}
\end{align}
where the eigengroup element $\sg$ and the twisted data matrix $\sm$ are
defined in Eqs.~\eqref{Eq 2.16a} and \eqref{Eq 2.14f} respectively. Using
the properties of $\sm$ in Eq.~\eqref{Eq 2.14g}, this form of the WZW
action is invariant under the diagonal action \eqref{Eq2.17} of the
orientation-reversing automorphism.

\subsection{The WZW Orientation-Orbifold Action}

We are now prepared to apply the principle of local isomorphisms to the
eigengroup elements $\sg$ in the context of the  half-cylinder
WZW action \eqref{Eq3.7}
\begin{subequations}
\label{Eq3.8}
\begin{gather}
\sg (\st,\xi,t,\s) \,\dual \,\hg (\st,\xi,t,\s) \\
S_{WZW} [\sm \otimes \one_2 ,\sg ;\Gamma_{\half}] \, \dual
\,\hat{S}_{\hg_O (\s)}[\sm \otimes \one_2 , \hg ;\Gamma_{\half}]
\end{gather}
\end{subequations}
where $\hg (\st,\xi,t,\s)$ are the group orbifold elements of Sec.~$2$.
This gives the {\it WZW orientation-orbifold action} of open-string sector
$\hat{h}_\s$ on the solid half cylinder $\Gamma_{1/2}$
\begin{subequations}
\label{Eq3.9}
\begin{align}
&\!\hat{S}_{\hg_O (\s)}[\sm \otimes \one_2 , \hg ;\Gamma_{\half}] \!\equiv
\nn \\
& \bigspc -\!\frac{1}{8\pi} \int \!\!dt \!\int_0^{\pi} \!\!\!\!\!d\xi
\,\widehat{Tr}\left( (\sm \otimes \one_2 ) \left( \hg^{-1}(\st,\xi)
  \pl_+ \hg(\st,\xi) \hg^{-1}(\st,\xi) \pl_- \hg(\st,\xi) \right) \right)
\quad \nn \\
&\bigspc \bigspc -\frac{1}{12\pi} \int_{\Gamma_{\half}}
\!\!\!\widehat{Tr}\left( (\sm \otimes \one_2 ) \left(
\hg^{-1}(\st,\xi)d\hg(\st,\xi) \right)^3 \right) \label{Eq 3.9a}
\end{align}
\begin{gather}
\hg (\st,-\xi,\s) =\tau_3 \hg^{-1} (\st,\xi,\s) \tau_3 \label{Eq 3.9b} \\
[\tau_1 ,\hg(\st(T,\s),\xi,\s)]=[\sm \otimes \one_2
,\hg(\st(T,\s),\xi,\s)]=0
\end{gather}
\end{subequations}
where the second term in Eq.~\eqref{Eq 3.9a} is the {\it
orientation-orbifold WZW term}.
Here we have followed the convention of the orbifold program
\cite{Big,Big',Fab,Geom} in labelling the orientation-orbifold action of
sector
$\hat{h}_\s$ by its
associated orbifold Lie algebra $\hg_O (\s)$ in \eqref{Eq 2.19a}. It is
not difficult to check with Eq.~\eqref{Eq 3.9b} that each integrand in
this action is
symmetric under world-sheet parity $\xi \leftrightarrow -\xi$.

Although the WZW action \eqref{Eq3.4} and the WZW orientation-orbifold
action \eqref{Eq3.9} are both defined on the solid half cylinder
$\Gamma_{1/2}$,
only the latter truly describes an open string. The fundamental reason for
this is that the orientation-orbifold sector has only a {\it single}
untwisted
Virasoro algebra \cite{Orient1}, which we will review at the classical
level in
Subsec.~$3.4$.

The WZW orientation-orbifold action \eqref{Eq3.9}, which describes all
open-string WZW orientation-orbifold sectors, is a
central result of this paper. Closed-string sectors of the WZW orientation
orbifolds are described by the standard WZW orbifold action \cite{Big}.

\subsection{Variation of the Half-Cylinder Actions}

In this subsection, we study the variation (with special attention to
boundaries) of the half-cylinder form of the WZW action \eqref{Eq3.4}
and the WZW orientation-orbifold action \eqref{Eq3.9}. For brevity, we
will study only the WZW orientation-orbifold action explicitly, but
the same steps can be followed for the simpler case of the WZW action by
the substitution $\sm \rightarrow M ,\, \hg \rightarrow \tg$. For this
discussion, we will need
the general variations
\begin{subequations}
\label{Eq3.10}
\begin{align}
&\de \widehat{Tr} \left( (\sm \otimes \one_2) \hg^{-1} \pl_+ \hg \hg^{-1}
\pl_- \hg \right) \nn \\
& \quad = -\widehat{Tr} \left[ (\sm \otimes \one_2 ) \left( \pl_- \left(
\hg^{-1} \pl_+ \hg \right) +\pl_+ \left(
   \hg^{-1} \pl_- \hg \right) \right) \hg^{-1} \de \hg \right] \nn \\
& \quad \quad +2\widehat{Tr} \left[ (\sm \otimes \one_2 ) \left( \pl_t
\left( \hg^{-1} \pl_t \hg \hg^{-1} \de
   \hg \right) -\pl_\xi \left( \hg^{-1} \pl_\xi \hg \hg^{-1} \de \hg
\right) \right) \right] \label{Eq 3.10a} \\
&\ep^{ABC} \de \widehat{Tr} \left( (\sm \otimes \one_2 ) \hg^{-1} \pl_A
\hg \hg^{-1} \pl_B \hg \hg^{-1} \pl_C \hg \right) =\pl_A \hat{R}^A \\
& \hat{R}^A \equiv 3\ep^{ABC} \widehat{Tr} \left( (\sm \otimes \one_2 )
\hg^{-1} \pl_B \hg \hg^{-1} \pl_C \hg \hg^{-1} \de \hg \right) \label{Eq
3.10c} \\
&\hat{R}^\r = \frac{3}{2} \widehat{Tr} \left[ (\sm \otimes \one_2 ) \left(
\pl_- (\hg^{-1} \pl_+ \hg ) -\pl_+ (\hg^{-1} \pl_- \hg) \right) \right]
\end{align}
\end{subequations}
which are standard in WZW theory, except that we have kept the total
derivative terms in Eq.~\eqref{Eq 3.10a}.

Using the world-sheet parity \eqref{Eq 3.9b} and cyclicity of
$\widehat{Tr}$, we can now verify the following behavior under world-sheet
parity
\begin{subequations}
\label{Eq3.11}
\begin{align}
\hat{K}(\xi) &\equiv \widehat{Tr} \left( (\sm \otimes \one_2)
\hg^{-1}(\xi) \pl_\xi \hg(\xi) \hg^{-1}(\xi) \de \hg(\xi) \right) \\
&=\widehat{Tr} \left( (\sm \otimes \one_2 ) \hg(-\xi) \pl_\xi
\hg^{-1}(-\xi) \hg(-\xi) \de \hg^{-1}(-\xi) \right) \nn \\
&=\widehat{Tr} \left( (\sm \otimes \one_2 ) \hg^{-1} (-\xi) \pl_\xi
\hg(-\xi) \hg^{-1}(-\xi) \de\hg(-\xi) \right) \nn \\
&= -\hat{K}(-\xi) \label{Eq 3.11b}
\end{align}
\begin{gather}
\hat{R}^{t,\r}(\xi) = \hat{R}^{t,\r}(-\xi) ,\quad \hat{R}^\xi (\xi)
=-\hat{R}^\xi (-\xi) \label{Eq 3.11c}
\end{gather}
\end{subequations}
where $\hat{K}(\xi)$ appears in the variation \eqref{Eq 3.10a} of the
kinetic term, and $\hat{R}(\xi)$ is defined in Eq.~\eqref{Eq 3.10c}.

The variation of the WZW orientation-orbifold action now proceeds as
usual, requiring that the bulk and boundary terms
cancel independently.

The boundary contribution of the total derivative term in Eq.~\eqref{Eq
3.10a} to the variation of the {\it kinetic term} in the action
\eqref{Eq3.9} is
\begin{subequations}
\label{Eq3.12}
\begin{gather}
\frac{1}{4\pi} \int \!\!dt \int_0^{\pi} \!\!d\xi\, \widehat{Tr} \left(
\pl_\xi (\hg^{-1} \pl_\xi \hg \hg^{-1} \de \hg) \right) =
  \frac{1}{4\pi} \!\int \!\!dt \left( \hat{K} (\pi) -\hat{K} (0) \right)
\equiv 0 \\
\rightarrow \hat{K}(\pi) =\hat{K} (0) =0 \label{Eq 3.12b}
\end{gather}
\end{subequations}
where Eq.~\eqref{Eq 3.12b} is the local form of the variational boundary
condition. The boundary condition at $\xi =0$ can also be obtained
immediately from the world-sheet parity
\eqref{Eq 3.11b}.

For the variation of the {\it orientation-orbifold WZW term} in
\eqref{Eq3.9}, it is convenient to introduce the standard Cartesian
coordinates $(x,y,t)$
shown in Fig.~2. In these coordinates, the flat side of $\Gamma_{1/2}$ is
the plane $y=0$, which is composed of the two half-planes at $\xi=0$ and
at
$\xi=\pi$. Then the boundary contribution from the flat side of
$\Gamma_{1/2}$ is
\begin{subequations}
\label{Eq3.13}
\begin{gather}
\vec{e}_y \Big{|}_{\xi =0} =\vec{e}_\xi ,\quad \vec{e}_y \Big{|}_{\xi=\pi}
=-\vec{e}_\xi \\
\int_{\Gamma_{\frac{1}{2}}} \!\!\pl_A \hat{R}^A =
\int_{\pl\Gamma_{\frac{1}{2}}} \!\!d\vec{S} \cdot \vec{\hat{R}} =\int
\!\!dt \int_0^{\pi}
   \!\!d\xi \,\hat{R}^\r +\int_{y=0} \!\!\!dt dx (-\vec{e}_y \cdot
\vec{\hat{R}}) \\
\int_{y=0} \!\!\!dt dx \left( -\vec{e}_y \cdot \vec{\hat{R}} \right)
=\int_{y=0} \!\!\!dt dx \left( \hat{R}^\xi (\pi) -\hat{R}^\xi (0) \right)
\equiv 0 \\
\rightarrow \hat{R}^\xi (\pi) =\hat{R}^\xi (0) =0 \label{Eq 3.13d}
\end{gather}
\end{subequations}
where $\vec{e}_i$ denotes the unit vector in the $i$ direction. At $\xi
=0$, the local form of this variational boundary condition also follows
immediately from the
world-sheet parity \eqref{Eq 3.11c}.

The local forms of the boundary conditions in Eqs.~\eqref{Eq 3.12b} and
\eqref{Eq 3.13d} are important parts of the description of the {\it WZW
orientation-orbifold branes}, on which the twisted open WZW strings end (see
also Subsecs.~$3.4$ and $3.5$).

Having dealt with the boundary terms, this leaves only the total bulk
contribution
\begin{gather}
\de \hat{S}_{\hg_O (\s)}[\sm \otimes \one_2 ,\hg ;\Gamma_{\half}]
=\frac{1}{4\pi} \int \!\!dt \int_0^{\pi} \!\!d\xi \,\widehat{Tr} \left(
(\sm \otimes \one_2 )
   \pl_- (\hg^{-1} \pl_+ \hg ) \hg^{-1} \de\hg \right)  \label{Eq3.14}
\end{gather}
to the variation of the WZW orientation-orbifold action. Then, because the
invertible matrix $\sm$ commutes with the group orbifold elements, we find
the
open-string equations of motion
\begin{subequations}
\label{Eq3.15}
\begin{gather}
\pl_- \hj (\st,\xi,t) =0 ,\quad \hj(\st,\xi,t) \equiv -\frac{i}{2}
\hg^{-1}(\st,\xi,t) \pl_+ \hg(\st,\xi,t) \\
\,\longrightarrow \, \pl_+ \hjb (\st,\xi,t) =0 \label{Eq 3.15b} \\
\hjb (\st,\xi,t) \equiv -\frac{i}{2} \hg(\st,\xi,t) \pl_- \hg^{-1}
(\st,\xi,t) =\tau_3 \hj(\st,-\xi,t) \tau_3 \label{Eq 3.15c} \\
[\sm \otimes \one_2 ,\hj(\st) ]=[\sm \otimes \one_2 ,\hjb(\st)] =[\tau_1
,\hj(\st)]=[\tau_1 ,\hjb(\st)]=0
\end{gather}
\end{subequations}
where $\hj(\st) ,\hjb(\st)$ are the {\it twisted open-string matrix
currents}. As indicated in \eqref{Eq 3.15b}, $\hjb(\st)$ conservation
follows from the
conservation of $\hj(\st)$. The world-sheet parity relation \eqref{Eq
3.15c} between $\hj$ and $\hjb$ follows from Eq.~\eqref{Eq 2.21d}, and
this relation tells
us that the twisted currents share the same set of twisted current modes
(see the following subsection), as is appropriate for open strings. The
twisted matrix
currents are also two-component fields, and we find in particular
\begin{subequations}
\label{Eq3.16}
\begin{gather}
\hj(\st,\xi,t) =\left( \begin{array}{cc} \hj_{(0)}(\st,\xi,t) &
\hj_{(1)}(\st,\xi,t) \\ \hj_{(1)} (\st,\xi,t) & \hj_{(0)} (\st,\xi,t)
\end{array} \right) \\
\hjb (\st,\xi,t) =\left( \begin{array}{cc} \hj_{(0)}(\st,-\xi,t) &
-\hj_{(1)}(\st,-\xi,t) \\ -\hj_{(1)} (\st,-\xi,t) & \hj_{(0)} (\st,-\xi,t)
   \end{array} \right) \\
\hj_{(u)} (\st,\xi,t) \equiv -\frac{i}{2} \sum_{v=0}^1 (-1)^v \hg_{(v)}
(\st,-\xi,t) \pl_+ \hg_{(u-v)} (\st,\xi,t) ,\quad \bar{u}=0,1 \,.
\end{gather}
\end{subequations}
where the reduced components $\hg_{(u)}$ of the group orbifold elements were
defined in Eq.~\eqref{Eq2.23}.

Returning to the simpler case of the $\Gamma_{1/2}$ form \eqref{Eq3.4} of
the untwisted WZW action, we find that the bulk contributions to the
variation give the classical
equation of motion
\begin{gather}
\pl_- (\tilde{g}^{-1} (T,\xi,t) \pl_+ \tilde{g} (T,\xi,t))=0
\label{Eq3.17}
\end{gather}
which implies the conservation of the usual untwisted left- and
right-mover WZW currents. We can also explicitly verify the required local
boundary conditions
\begin{gather}
K(\pi) =K(0) =0 ,\quad R^\xi (\pi) =R^\xi (0)=0 \label{Eq3.18}
\end{gather}
from the world-sheet parities
\begin{gather}
K(\xi) =-K(-\xi) ,\quad R^{t,\r} (\xi) =R^{t,\r} (-\xi) ,\quad R^\xi (\xi)
=-R^\xi (-\xi) \label{Eq3.19}
\end{gather}
and the $2\pi$-periodicity of the untwisted fields. Here the quantities
$K,R$ are obtained from those in Eqs.~\eqref{Eq3.10}, \eqref{Eq3.11} by
the substitution
$\sm \rightarrow M$ and $\hg \rightarrow \tg$.

The variational boundary conditions \eqref{Eq 3.12b}, \eqref{Eq 3.13d}
can also be explicitly verified from the monodromies in the case $h_\s^2 =1$
(see Subsec.~$3.5$).

\subsection{More about the Twisted Open-String Currents}

In this subsection, we further discuss the classical forms of the twisted
open-string currents and their corresponding twisted open-string stress
tensors.

We begin this discussion by noting that the functional form of the twisted
currents in ($3.15$a,c) can be rewritten as the classical equations of
motion
of the group orbifold elements:
\begin{subequations}
\label{Eq3.20}
\begin{gather}
\pl_+ \hg(\st,\xi,t) =2i \hg(\st,\xi,t) \hj(\st,\xi,t) ,\quad \pl_-
\hg(\st,\xi,t) =-2i \hjb(\st,\xi,t) \hg(\st,\xi,t) \\
\hj(\st,\xi,t) =\hj_\nrmu(\xi,t) \sG^{\nrmu;\nsnv} (\s) \st_\nsn \tau_v \\
\hjb(\st,\xi,t) =\hj_\nrmu(-\xi,t) \sG^{\nrmu;\nsnv}(\s) \st_\nsn (-1)^v
\tau_v \label{Eq 3.20c} \\
\sG^{\nrmu;\nsnv}(\s) =\srac{1}{2} \de_{u+v,0\,\text{mod }2}
\sG^{\nrm;\nsn}(\s) \label{Eq 3.20d} \\
\sG_{\nrm;\ntd}(\s) \sG^{\ntd;\nsn}(\s) = \de_\m{}^\n
\de_{n(r)-n(s),0\,\text{mod }\r(\s)} \,.
\end{gather}
\end{subequations}
In this system, the index-current expansions ($3.20$b,c) of each matrix
current are taken from the high-level limit of the corresponding twisted
vertex operator equations in Eq.~$(6.1)$ of Ref.~\cite{Orient1}. As expected,
the index-current expansions of $\hj$ and $\hjb$ satisfy the world-sheet
parity relation in Eq.~\eqref{Eq 3.15c}.

For completeness, we also read off from Ref.~\cite{Orient1} the bracket form of
some important algebraic relations
\begin{subequations}
\label{Eq3.21}
\begin{gather}
\hj_\nrmu (\xi,t,\s) =\sum_m \hj_\nrmu (m\!+\!\nrrs \!+\!\srac{u}{2})
e^{-i (m+\nrrsf +\frac{u}{2})(t+\xi)} ,\quad \bar{u}=0,1 \label{Eq 3.21a}
\\
\hj_\nrmu (\xi+2\pi ,t,\s) =e^{-\tp (\frac{n(r)}{\r(\s)} +\frac{u}{2})}
\hj_\nrmu (\xi,t,\s) \label{Eq 3.21b}
\end{gather}
\begin{align}
& \{ \hj_\nrmu (m\!+\!\nrrs \!+\!\srac{u}{2}) ,\hj_\nsnv (n\!+\!\nsrs
\!+\!\srac{v}{2}) \} \nn \\
&\bigspc \bigspc = i\scf_{\nrm;\nsn}{}^{\!\!\!n(r)+n(s),\de}(\s)
\hj_{n(r)+n(s),\de,u+v} (m\!+\!n\!+\!\srac{n(r)+n(s)}{\r(\s)}
\!+\!\srac{u+v}{2}) \nn \\
& \bigspc \bigspc +(m\!+\!\nrrs \!+\!\srac{u}{2})
\de_{m+n+\frac{n(r)+n(s)}{\r(\s)} +\frac{u+v}{2},0} (2\de_{u+v
,0\,\text{mod }2} \sG_{\nrm;\nsn}(\s)) \label{Eq 3.21c} \\
&\{ \hj_\nrmu (m\!+\!\nrrs \!+\!\srac{u}{2}), \hg(\st(T),\xi,t) \} =
\hg(\st(T),\xi,t) (\st_\nrm(T) \tau_u) e^{i(m+\nrrsf +\frac{u}{2})(t+\xi)}
\nn \\
& \bigspc \bigspc \bigspc -(\st_\nrm(T) (-1)^u \tau_u) \hg(\st(T),\xi,t)
e^{i(m+\nrrsf +\frac{u}{2})(t-\xi)}  \label{Eq 3.21d}
\end{align}
\end{subequations}
where the mode expansion \eqref{Eq 3.21a} and the monodromy \eqref{Eq
3.21b} of the currents are {\it valid for all} $\hat{h}_\s$, without
restriction
to $h_\s^2 =1$.
Note that the open-string twisted current algebra $\gfrakh_O (\s)$ in
\eqref{Eq 3.21c} shares the same twisted structure constants $\scf (\s)$
with the
orbifold Lie algebra $\hg_O (\s)$ in \eqref{Eq 2.19a}, and that as
expected for open strings, the twisted currents act simultaneously as
left- and right-movers
in Eq.~\eqref{Eq 3.21d}. Moreover, because $\bar{u}$ takes the values $0$
and $1$, each open-string sector $\hat{h}_\s$ contains fractionally-moded
currents.

We also note the monodromy of the twisted open-string matrix currents \cite{Orient1}
\begin{subequations}
\label{Eq3.22}
\begin{gather}
\hj(\st(T,\s),\xi+2\pi,t) =\tau_3 E(T,\s) \hj(\st(T,\s),\xi,t) E^\ast
(T,\s) \tau_3 \\
\hjb(\st(T,\s),\xi+2\pi,t) =\tau_3 E^\ast (T,\s) \hjb(\st(T,\s),\xi,t)
E(T,\s) \tau_3
\end{gather}
\end{subequations}
which follow  (by the $\st$-selection rule \eqref{Eq 2.14d}) from the
 monodromy \eqref{Eq 3.21b} of the index currents --
 and hence are also true for all $\hat{h}_\s$.

With Ref.~\cite{Orient1}, we emphasize that the definite but {\it clashing}
monodromies \eqref{Eq3.22} of the twisted open-string matrix currents are
precisely the
reason that the group orbifold elements $\hg$ cannot have definite
monodromy beyond $h_\s^2 =1$.

Using the general monodromies \eqref{Eq3.22} and the world-sheet parity
\eqref{Eq 3.15c}, we may now give the boundary conditions of the twisted
currents
on the strip:
\vspace{-0.10in}
\begin{subequations}
\label{Eq3.23}
\begin{gather}
\hjb(\st(T),0) \!=\!\tau_3 \hj(\st(T),0) \tau_3  \\
\hjb(\st(T),\pi) \!=\!E^\ast (T,\s) \hj(\st(T),\pi) E(T,\s) \,.
\end{gather}
\end{subequations}
In addition to the variational boundary conditions of the previous
subsection, these twisted current boundary conditions should be included
as another
important part of the description of the {\it orientation-orbifold branes}
in open-string sector $\hat{h}_\s$.

As promised, we finally mention the classical form of the twisted
open-string stress tensors
\begin{subequations}
\label{Eq3.24}
\begin{gather}
\hat{\Theta}_u (\xi,t,\s) =\srac{1}{8\pi} \sG^{\nrm ;\nsn}(\s) \sum_{v=0}^1
\hj_{\nrm v} (\xi,t,\s) \hj_{\nsn ,u-v}(\xi,t,\s) ,\quad \bar{u} =0,1
\label{Eq 3.24a} \\
\pl_- \hat{\Theta}_u (\xi,t,\s) =0 \\
\hat{\Theta}_u (\xi+2\pi,t,\s) =(-1)^u \hat{\Theta}_u (\xi,t,\s) \\
\hat{\Theta}_u (\xi,t,\s) =\srac{1}{2\pi} \sum_m \hat{L}_u (m\!+\!\srac{u}{2})
e^{-i(m+\frac{u}{2}) (t+\xi)} \\
\{ \hat{L}_u (m\!+\!\srac{u}{2}) ,\hat{L}_v (n\!+\!\srac{v}{2}) \}
=(m\!-\!n\! +\!\srac{u-v}{2}) \hat{L}_{u+v} (m\!+\!n\!+\!\srac{u+v}{2})
\label{Eq 3.24e}
\end{gather}
\end{subequations}
where Eq.~\eqref{Eq 3.24a} is the classical (high-level) limit of the twisted
operator stress tensors of Ref.~\cite{Orient1}. The unique physical stress
tensor of
sector $\hat{h}_\s$ is the component $\hat{\Theta}_0 (\xi,t,\s)$ with
trivial monodromy. Following Ref.~\cite{Orient1}, we emphasize that the
classical form
\eqref{Eq 3.24e} of the {\it orbifold Virasoro algebra}\footnote{The
operator form of the orbifold Virasoro algebra is given in
Ref.~\cite{Orient1},
where we
discuss the quantum doubling $\hat{c}=2c$ of the closed-string central
charge $c$.}
correspondingly contains only the single classical untwisted Virasoro
subalgebra generated
by $\{ \hat{L}_0 (m) \}$ -- which tells us that sector $\hat{h}_\s$ is a
twisted open string.

\subsection{More about the Branes when $h_\s^2 =1$}

In our discussion above, we have given two important parts of the
description of the WZW orientation-orbifold branes for all $\hat{h}_\s$,
namely the variational
boundary conditions in Subsec.~$3.3$ and the twisted current boundary
conditions in Subsec.~$3.4$. In what follows, we consider further
structure of the branes,
concentrating primarily but not exclusively on the case $h_\s^2 =1$.

For the case $h_\s^2 =1$, the monodromies in \eqref{Eq2.25} give us a
simple way to determine the boundary conditions of the twisted fields $\hg
,\bh$ on the
strip $0\!\leq \!\xi \!\leq \!\pi$. As in Subsec.~$2.3$, we consider the
combined action of the world-sheet parity \eqref{Eq 2.21d} and the
monodromy
\eqref{Eq 2.25b} of the group orbifold elements, this time for the {\it
special orbits} associated to $\xi= 0$ and $\pi$. This gives the boundary
conditions
on the group orbifold elements
\begin{subequations}
\label{Eq3.25}
\begin{gather}
 \hg^{-1} (\st(T,\s),0,\s) =\tau_3 \hg(\st(T,\s),0,\s) \tau_3  \label{Eq
3.25a} \\
h_\s^2 =1\!:\bigspc \hg^{-1} (\st(T,\s),\pi,\s) =E(T,\s)
\hg(\st(T,\s),\pi,\s) E^\ast(T,\s) \bigspc \label{Eq 3.25b}
\end{gather}
\end{subequations}
which describe further structure of the branes at the ends of each twisted
open string. The boundary condition \eqref{Eq 3.25a} at $\xi =0$ follows
immediately from the world-sheet parity, and hence holds for all
$\hat{h}_\s$, while the following steps for $h_\s^2 =1$
\begin{gather}
\hg^{-1}(\st(T),\pi) =\tau_3 E(T) \hg^{-1}(\st(T),-\pi) E^\ast(T) \tau_3 =
E(T) \hg(\st(T),\pi) E^\ast(T) \label{Eq3.26}
\end{gather}
were used to obtain the boundary condition at $\xi =\pi$.

Moreover, the $\tau_1$-constraint \eqref{Eq 2.21c} holds for all
$\hat{h}_\s$ in the bulk and at the boundary
\begin{gather}
[\tau_1 ,\hg (\st(T,\s),0,\s)] =0 ,\quad [\tau_1 ,\hg (\st(T,\s),\pi
,\s)]=0 \label{Eq3.27}
\end{gather}
which allows us to rewrite the boundary conditions in Eq.~\eqref{Eq3.25}
in terms of the reduced components $\hg_{(u)} ,\,\bar{u}=0,1$ of the group
orbifold elements:
\begin{subequations}
\label{Eq3.28}
\begin{gather}
[\hg_{(0)}(\st),\hg_{(1)}(\st)]=0 ,\quad \hg_{(0)}(\st)^2
-\hg_{(1)}(\st)^2 =\one \text{ at } \xi=0 \\
\left. \begin{array}{ccc}
\hg_{(0)}(\st(T)) E(T) \hg_{(0)}(\st(T)) +\hg_{(1)}(\st(T)) E(T)
\hg_{(1)}(\st(T)) =&E(T) \\
\hg_{(0)}(\st(T)) E(T) \hg_{(1)}(\st(T)) +\hg_{(1)}(\st(T)) E(T)
\hg_{(0)}(\st(T)) =&0 \end{array} \right\} \text{ at } \xi=\pi \,.
\label{Eq 3.28b}
\end{gather}
\end{subequations}
To see Eq.~\eqref{Eq3.28}, begin by multiplying both sides of \eqref{Eq
3.25a} and \eqref{Eq 3.25b} by $\hg$; this set of boundary
conditions can also be understood as the residuals at the boundaries of
the non-local quadratic relations in Eq.~\eqref{Eq2.24}. As above, the
relations
in \eqref{Eq 3.28b} apply only to the case $h_\s^2 =1$.

For the twisted tangent-space coordinates $\bh$, we find the corresponding
boundary conditions:
\vspace{-0.10in}
\begin{subequations}
\label{Eq3.29}
\begin{gather}
\bh^{\nrm 0} (0,\s)=0 \label{Eq 3.29a} \\
h_\s^2 =1\!: \quad \bh^\nrmu (\pi,\s)=0 \text{ unless }
\srac{n(r)}{\r(\s)} \in \Zint +\half \,. \bigspc \label{Eq 3.29b}
\end{gather}
\end{subequations}
The result in Eq.~\eqref{Eq 3.29a} holds for all $\hat{h}_\s$ because it
is obtained directly from the world-sheet parity \eqref{Eq 2.22b} at
$\xi=0$,
while the following steps
\begin{gather}
\bh^\nrmu (\pi,\s) = \bh^\nrmu (-\pi ,\s) e^{\tp (\nrrsf +\frac{u}{2})} =
-\bh^\nrmu (\pi,\s) e^{\tp \nrrsf} \label{Eq3.30}
\end{gather}
give the result in Eq.~\eqref{Eq 3.29b} for $h_\s^2 =1$. We remark that
the result in Eq.~\eqref{Eq 3.29b} can be solved simply as $\bh^{0\m
u}(\pi,\s)=0$,
but the more abstract language of \eqref{Eq 3.29b} will be useful below.
With the
tangent-space form of $\hg$ in \eqref{Eq 2.22a} and the $\st$-selection
rule
\eqref{Eq 2.14d}, it is not difficult to check that the $\bh$ boundary
conditions \eqref{Eq3.29} imply those of $\hg(\st)$ in Eq.~\eqref{Eq3.25}.

The special orbits also allow us to derive boundary conditions for {\it
derivatives} of the twisted fields, for example:
\begin{subequations}
\label{Eq3.31}
\begin{gather}
(\pl_t \bh^{\nrm 0}) (0,\s) = 0\\
h_\s^2 =1\!: \quad (\pl_t \bh^\nrmu) (\pi ,\s) =0 \text{ unless }
\srac{n(r)}{\r(\s)} \in \Zint +\half \bigspc
\end{gather}
\begin{gather}
(\pl_\xi \bh^{\nrm 1}) (0,\s) =0 \\
h_\s^2 =1\!: \quad (\pl_\xi \bh^\nrmu) (\pi,\s)=0 \text{ unless }
\srac{n(r)}{\r(\s)} \in \Zint \,. \bigspc
\end{gather}
\end{subequations}
We note in particular that the $\pl_t \bh$ boundary conditions are the
same as those of $\bh$. The boundary conditions obtained in this way for
$\pl_t \hg$ and $\pl_\xi \hg$ are more complicated however, except when
taken in the combinations which correspond to the {\it twisted currents}
(see Subsec.~$3.4$) and the {\it variational boundary conditions} (see
Subsec.~$3.3$).

We may also use these $h_\s^2 =1$ monodromies to explicitly verify, for
this specific case, various properties given above for all $\hat{h}_\s$.
For example, the monodromies of the twisted currents
\begin{subequations}
\label{Eq3.32}
\begin{gather}
\hj (\st(T,\s),\xi +2\pi,t) =\tau_3 E(T,\s) \hj (\st(T,\s) ,\xi,t) E^\ast
(T,\s) \tau_3 \\
\hjb (\st(T,\s),\xi +2\pi,t) =\tau_3 E(T,\s) \hjb (\st(T,\s) ,\xi,t)
E^\ast (T,\s) \tau_3
\end{gather}
\end{subequations}
follow from the $\hg$ monodromies \eqref{Eq2.25} and the $\hg$ forms of
the twisted currents in Eq.~\eqref{Eq3.15}. The monodromies \eqref{Eq3.32}
agree
with the general monodromies \eqref{Eq3.22} of the twisted currents when
$h_\s^2 =1 ,\, E(T,\s)^2 =1$.

Finally, we explicitly check the variational boundary conditions \eqref{Eq
3.12b} and \eqref{Eq 3.13d} at $\xi =\pi$:
\begin{gather}
\hat{K} (\pi) =\hat{R}^\xi (\pi) =0 \,.  \label{Eq3.33}
\end{gather}
In fact, these boundary conditions follow immediately when $h_\s^2 =1$
from the world-sheet parities in Eq.~\eqref{Eq3.11} and the trivial
monodromies
\begin{gather}
\hat{K} (\xi +2\pi) = \hat{K}(\xi) ,\quad \hat{R}^A (\xi +2\pi) =\hat{R}^A
(\xi)  \label{Eq3.34}
\end{gather}
which themselves are obtained from the $\hg$ monodromy.\footnote{Similarly,
 one easily checks from the $\hg$ monodromies and Eq.~\eqref{Eq
2.14g} that both integrands of the WZW orientation-orbifold action
\eqref{Eq3.9} are $2\pi$-periodic; this conclusion holds of course only for
$h_\s^2 =1$.}

In what follows we will find, as illustrated above, that {\it all}
boundary conditions at $\xi =0$ follow from world-sheet parity alone, and
hence are true for all $\hat{h}_\s$.

\section{Einstein Geometry on the Solid Half Cylinder}

Our task in this section is to study the Einstein geometry, including the
twisted Einstein tensors \cite{Geom}, of the open-string sectors of the WZW
orientation orbifolds above.

\subsection{Two-Component WZW Geometry}

We begin this discussion in the untwisted theory, with the familiar\footnote{See for
example Refs.~\cite{Yam,Club,Geom}.} geometric quantities of ordinary WZW
models on the solid cylinder $\Gamma$:
\begin{subequations}
\label{Eq4.1}
\begin{gather}
g(T,x(\xi)) =e^{ix^i (\xi)e_i{}^a (0) T_a} ,\quad e_i{}^a(0) =\de_i{}^a
,\quad \pl_i =\frac{\pl}{\pl x^i} \\
e_i (T,x(\xi)) =-i g^{-1} (T,x(\xi)) \pl_i g(T,x(\xi)) = e (x(\xi))_i{}^a
T_a
\end{gather}
\begin{gather}
\bar{e}_i (T,x(\xi)) =-i g(T,x(\xi)) \pl_i g^{-1}(T,x(\xi)) =\bar{e}
(x(\xi))_i{}^a T_a \\
\bar{e} (x(\xi))_i{}^a =-e (x(\xi))_i{}^b \Omega (x(\xi))_b{}^a ,\quad
g(T,x(\xi)) T_a g^{-1}(T,x(\xi)) =\Omega(x(\xi))_a{}^b T_b \\
\Omega (x(\xi)) = g^{-1} (T^{adj},x(\xi))
\end{gather}
\begin{align}
&G_{ij} (x(\xi)) =e (x(\xi))_i{}^a G_{ab} e (x(\xi))_j{}^b =\bar{e}
(x(\xi))_i{}^a G_{ab} \bar{e} (x(\xi))_j{}^b ,\quad Tr (MT_a T_b)=G_{ab} \\
&H_{ijk}(x(\xi)) =\pl_i B_{jk} (x(\xi)) +\text{ cyclic} =-iTr \left( Me_i
(T,x(\xi)) [e_j(T,x(\xi)) ,e_k(T,x(\xi))] \right) \nn \\
& \bigspc = iTr \left( M\bar{e}_i (T,x(\xi)) [\bar{e}_j(T,x(\xi))
,\bar{e}_k(T,x(\xi))] \right) =e_i{}^a e_j{}^b e_k{}^c f_{ab}{}^d G_{dc}
\end{align}
\begin{gather}
\gamma (\xi) = \ep^{ABC} Tr \!\left( \!Mg^{-1}\pl_A gg^{-1} \pl_B gg^{-1}
\pl_C g \right) =\pl_A j^A (\xi) \label{Eq 4.1h} \\
j^A(\xi) \equiv \srac{3}{2} \ep^{ABC} \pl_B x^i (\xi) \pl_C x^j (\xi)
B_{ij}(x(\xi)) \label{Eq 4.1i} \\
A = \{ t, \xi ,\r \} ,\quad \ep^{t \xi \r} =1 \,.  \label{Eq 4.1j}
\end{gather}
\end{subequations}
Here $e,\bar{e},\Omega ,G$ and $H$ are respectively the left- and
right-invariant vielbeins, the
adjoint action, the Einstein metric and the torsion field. Note that we
have chosen
the Einstein coordinates
\begin{gather}
\{ x^i =\be^a e(0)_a{}^i ,i=1 \ldots \text{ dim } g \}  \label{Eq4.2}
\end{gather}
which correspond to $e(0)=\thickone$ at the origin. The identities in
Eqs.~\eqref{Eq 4.1h}, \eqref{Eq 4.1i} are the standard Gauss' law which
relates the
winding-number current $j^a (\xi)$ to the density $\gamma (\xi)$ defined
earlier
 in Eq.~\eqref{Eq 3.5b}.

Following Ref.~\cite{Orient1} and our development above, we next define the
corresponding two-component fields:
\begin{subequations}
\label{Eq4.3}
\begin{gather}
x^{i\Id} ,\, \Id =0,1\!: \quad x^{i0}(\xi) \equiv x^i (\xi) ,\quad
x^{i1}(\xi) \equiv -x^i (-\xi) \label{Eq 4.3a} \\
\pl_{i\Id} (\xi) \equiv \frac{\pl}{\pl x^{i\Id}(\xi)} ,\quad \pl_{i\Id}
(\xi) x^{j\Jd}(\xi) =\de_i^j \de_{\Id}{}^\Jd \\
\tg(T,x) =\left( \begin{array}{cc} g(T,x^0 (\xi)) &0\\0& g(T,x^1 (\xi))
\end{array} \right) ,\quad \pl_{\pm} \tg (T,x)
   =\pl_{i\Id} \tg(T,x) \pl_{\pm} x^{i\Id} (\xi) \\
e_{i\Id} (T,x) \equiv -i \tg^{-1} (T,x) \pl_{i\Id} \tg(T,x) \equiv
e(x)_{i\Id}{}^{a\Jd} T_a \r_\Jd \\
e(x)_{i\Id}{}^{a\Jd} =-i \widehat{Tr} \left( (M\otimes \one_2 ) \tg^{-1}
\pl_{i\Id} \tg  G^{ab} \de^{\Jd \Kd} T_b \r_\Kd \right) =\de_\Id{}^\Jd
e(x^\Id)_i{}^a
\end{gather}
\begin{gather}
\tg(T,x) T_a \r_\Id \tg^{-1}(T,x) =\tilde{\Omega}(x)_{a\Id}{}^{b\Jd} T_b
\r_\Jd ,\quad \tilde{\Omega}(x) =\tg^{-1} (T^{adj},x) \\
\tilde{\Omega}(x)_{a\Id}{}^{b\Jd} =\de_\Id{}^\Jd \Omega(x^\Id)_a{}^b
\end{gather}
\begin{align}
G_{i\Id ;j\Jd} (x) &\equiv -\widehat{Tr} \left( (M \otimes \one_2 )
\tg^{-1}(T,x) \pl_{i\Id} \tg(T,x) \tg^{-1}(T,x)
  \pl_{j\Jd} \tg(T,x(\xi)) \right) \nn \\
&=e(x)_{i\Id}{}^{a\Kd} e(x)_{j\Jd}{}^{b \dot{L}} G_{a\Kd ,b\dot{L}} ,\quad
\quad G_{a\Kd ;b\dot{L}} =\de_{\Kd \dot{L}} G_{ab} \nn \\
&=\de_{\Id \Jd} G_{ij}(x^\Id) \label{Eq 4.3h}
\end{align}
\begin{align}
B_{i\Id ;j\Jd} (x) &=\de_{\Id \Jd} B_{ij} (x^\Id ) \label{Eq 4.3i} \\
H_{i\Id ;j\Jd ;k\Kd} (x) &\equiv \pl_{i\Id} B_{j\Jd ;k\Kd} (x) +\text{
cyclic} \nn \\
&= -i\widehat{Tr} \left( (M\otimes \one_2 ) e_{i\Id} (T,x) [e_{j\Jd} (T,x)
,e_{k\Kd} (T,x)] \right)\nn \\
& =e_{i\Id}{}^{a\dot{L}} e_{j\Jd}{}^{b\dot{M}} e_{k\Kd}{}^{c\dot{N}}
f_{a\dot{L};b\dot{M};c\dot{N}} ,\quad \quad f_{a\Id ;b\Jd ;c\Kd}
   =\de_{\Id \Jd} \de_{\Jd \Kd} f_{ab}{}^d G_{dc} \quad \nn \\
&= e(x^\Id )_i{}^a e(x^\Id )_j{}^b e(x^\Id)_k{}^c f_{ab}{}^d G_{dc}
=\de_{\Id \Jd} \de_{\Jd \Kd} H_{ijk} (x^\Id )
\end{align}
\begin{gather}
\ep^{ABC} \widehat{Tr} \left( (M \otimes \one_2 ) \tg^{-1} (T,x) \pl_A
\tg(T,x) \tg^{-1}(T,x) \pl_B \tg(T,x) \tg^{-1}(T,x) \pl_C \tg(T,x) \right)
\nn \\
 =\pl_A \tilde{j}^A (\xi) \label{Eq 4.3k} \\
\tilde{j}^A (\xi) \equiv \srac{3}{2} \ep^{ABC} \pl_B x^{i\Id}(\xi) \pl_C
x^{j\Jd}(\xi) B_{i\Id ;j\Jd} (x(\xi)) \,. \label{Eq 4.3l}
\end{gather}
\end{subequations}
The matrices $\r_\Id$ are defined in Eq.~\eqref{Eq 2.9b}. We
emphasize that, although many of the fields carry more than one
two-component index
$\Id ,\Jd$, each field is {\it diagonal} in these indices and therefore
has only {\it two} independent components in the two-dimensional space.

From the definitions of these two-component fields, one finds the
following behavior under world-sheet parity $\xi \rightarrow -\xi$:
\begin{subequations}
\label{Eq4.4}
\begin{gather}
x^{i\Id} (-\xi) =-x^{i\Jd} (\xi) (\tau_1 )_\Jd{}^\Id ,\quad \pl_{i\Id}
(-\xi) = -(\tau_1 )_\Id{}^\Jd \pl_{i\Jd} (\xi) \label{Eq 4.4a}
\end{gather}
\begin{gather}
\tilde{\Omega} (x(-\xi))_{a\Id}{}^{b\Jd} =(\tau_1 )_{\Id}{}^{\Kd}
\tilde{\Omega}^{-1} (x(\xi))_{a\Kd}{}^{b\dot{L}} (\tau_1 )_{\dot{L}}{}^\Jd
\\
e(x(-\xi))_{i\Id}{}^{a\Jd} =(\tau_1 )_\Id{}^\Kd \left( e(x(\xi))
\tilde{\Omega} (x(\xi)) \right)_{\!i\Kd}^{\,\,a\dot{L}} (\tau_1
)_{\dot{L}}{}^\Jd
\end{gather}
\begin{gather}
G_{i\Id ;j\Jd} (x(-\xi)) =(\tau_1 )_\Id{}^{\Kd} (\tau_1 )_\Jd{}^{\dot{L}}
G_{i\Kd ;j\dot{L}} (x(\xi)) \\
H_{i\Id ;j\Jd ;k\Kd} (x(-\xi)) =(\tau_1 )_\Id{}^{\dot{L}} (\tau_1
)_\Jd{}^{\dot{M}} (\tau_1 )_\Kd{}^{\dot{N}}
H_{i\dot{L};j\dot{M};k\dot{N}}(x(\xi)) \\
B_{i\Id ;j\Jd} (x(-\xi)) =-(\tau_1 )_\Id{}^{\Kd} (\tau_1 )_\Jd{}^{\dot{L}}
B_{i\Kd;j\dot{L}} (x(\xi))
\end{gather}
\begin{gather}
\tilde{j}^\xi (-\xi) =-\tilde{j}^\xi (\xi) ,\quad \tilde{j}^{t,\r} (-\xi)
=\tilde{j}^{t,\r} (\xi) \,. \label{Eq 4.4g}
\end{gather}
\end{subequations}
In particular, the results in Eqs.~($4.4$c-g) follow by using the
world-sheet parity \eqref{Eq 2.9d} in the relevant trace formulae in
Eq.~\eqref{Eq4.3}.

Taken together, the diagonal forms in Eq.~\eqref{Eq4.3} and the
world-sheet parities \eqref{Eq4.4} imply {\it consistency relations} such
as
\begin{gather}
G_{ij} (x^\Id (-\xi)) =(\tau_1 )_\Id{}^\Jd G_{ij} (x^\Jd (\xi)) ,\quad
B_{ij} (x^\Id (-\xi)) =-(\tau_1 )_\Id{}^\Jd B_{ij} (x^\Jd (\xi))
\label{Eq4.5}
\end{gather}
which record in our notation (see Eq.~\eqref{Eq 4.3a}) the following
behavior of the original fields under (target-space) {\it space-time
parity}
$x \leftrightarrow -x$:
\begin{subequations}
\label{Eq4.6}
\begin{gather}
g(T,-x) =g^{-1} (T,x) ,\quad \Omega(-x) =\Omega^{-1} (x) ,\quad
e(-x)_i{}^a =-\bar{e}(x)_i{}^a \\
G_{ij} (-x) =G_{ij} (x) ,\quad B_{ij}(-x) =-B_{ij}(x) ,\quad H_{ijk} (-x)
=H_{ijk}(x) \,. \label{Eq 4.6b}
\end{gather}
\end{subequations}
These space-time parities hold in the general WZW model, as is easily
checked from the following explicit forms \cite{Club,Geom} of the geometric
quantities:
\begin{subequations}
\label{Eq4.7}
\begin{gather}
\Omega (x) =e^{-iY(x)} \\
e(x)_i{}^a =e(0)_i{}^b \left( \srac{e^{iY(x)}-1}{iY(x)}
\right)_{\!b}^{\,a} ,\quad \bar{e}(x)_i{}^a =e(0)_i{}^b \left(
\srac{e^{-iY(x)}-1}
{iY(x)} \right)_{\!b}^{\,a} \\
G_{ij} (x) =e(0)_i{}^a \left( \srac{e^{iY(x)} +e^{-iY(x)} -2}{(iY(x))^2}
\right)_{\!a}^{\,b} G_{bc} e(0)_j{}^c  \\
B_{ij} (x) =e(0)_i{}^a \left( \srac{e^{iY(x)} -e^{-iY(x)}
-2iY(x)}{(iY(x))^2} \right)_{\!a}^{\,b} G_{bc} e(0)_j{}^c  \\
Y(x) \equiv x^i e(0)_i{}^a T_a^{adj} ,\quad (T_a^{adj} )_b{}^c
=-if_{ab}{}^c \,.
\end{gather}
\end{subequations}
Although the space-time parities \eqref{Eq4.6} appear here as a byproduct
of our formulation, we will find below and in Sec.~5 that space-time
parity in
fact plays a fundamental role in the construction of orientation orbifolds.

\subsection{From the Solid Half Cylinder to the Strip}

Using the two component fields of the previous subsection, we turn next to
find the two-dimensional or sigma-model form of the half-cylinder action in Eq.~\eqref{Eq3.4}.

For this discussion, we note first the boundary conditions on the two-component winding-number current
$\tilde{j}(\xi)$ in Eq.~\eqref{Eq 4.3l}
\begin{gather}
\tilde{j}^\xi (0) =\tilde{j}^\xi (\pi) =0 \label{Eq4.8}
\end{gather}
which are obtained from $2\pi$-periodicity of the untwisted fields and the
world-sheet parity \eqref{Eq 4.4g} of $\tilde{j}(\xi)$. It follows that
\begin{gather}
 \int_{y=0} \!\!dt dx ( -\vec{e}_y \cdot \tilde{j} ) = \int_{y=0} \!\!dt
dx (\tilde{j}^\xi (\pi) -\tilde{j}^\xi (0)) =0 \label{Eq4.9}
\end{gather}
so, with the Gauss' law \eqref{Eq 4.3k}, one sees for the WZW term that
the contribution from the flat side of the solid half cylinder vanishes.
Then
one finds the two-dimensional form of the WZW term on $\Gamma_{1/2}$:
\begin{gather}
\int_{\Gamma_{\frac{1}{2}}} \!\!\widehat{Tr} \left( (M\otimes \one_2 )
(\tg^{-1}(T,\xi) d\tg (T,\xi))^3 \right) =-\srac{3}{2} \int \!\!dt
   \int_0^\pi \!\!d\xi \, B_{i\Id ;j\Jd} (x) \pl_+ x^{i\Id} \pl_- x^{j\Jd}
\,. \label{Eq4.10}
\end{gather}
With this result and Eq.~\eqref{Eq 4.3h} for the kinetic term, we finally
obtain the {\it
sigma-model form} of the WZW action on the {\it strip}
\begin{subequations}
\label{Eq4.11}
\begin{align}
S_{WZW}^{\text{ strip}} &= \frac{1}{8\pi} \int \!\!dt \int_0^\pi \!\!d\xi
\left( G_{i\Id ;j\Jd} (x) +B_{i\Id ;j\Jd} (x) \right) \pl_+ x^{i\Id} \pl_-
x^{j\Jd} \label{Eq 4.11a} \\
&= \frac{1}{8\pi} \int \!\!dt \int_0^\pi \!\!d\xi \sum_{\Id =0}^1 \left(
G_{ij} (x^\Id ) +B_{ij} (x^\Id ) \right) \pl_+ x^{i\Id} \pl_- x^{j\Jd}
\label{Eq 4.11b} \\
& =S_{WZW} [M\!\otimes \!\one_2 ,\tg;\Gamma_\half ]
\end{align}
\end{subequations}
where $G_{ij}(x)$ and $B_{ij}(x)$ are the original WZW fields.

It is instructive to check that the two-component action \eqref{Eq4.11} on
the strip is
equivalent to the standard sigma-model form of the WZW action on the
cylinder:
\begin{gather}
S_{WZW} =\frac{1}{8\pi} \int \!\!dt \int_0^{2\pi} \!\!d\xi \left( G_{ij}
(x(\xi)) +B_{ij} (x(\xi)) \right) \pl_+ x^i (\xi) \pl_- x^j (\xi) \,.
\label{Eq4.12}
\end{gather}
To see this equivalence, begin with the form of the strip action in
Eq.~\eqref{Eq 4.11b} and use the definition of $x^{\Id}$ in Eq.~\eqref{Eq
4.3a} to follow the steps
\begin{subequations}
\label{Eq4.13}
\begin{align}
S^{\text{ strip}}_{WZW}&=\frac{1}{8\pi} \int \!\!dt \int_0^\pi \!\!d\xi
\Big{[} \left( G_{ij} (x(\xi)) +B_{ij}(x(\xi)) \right) \pl_+ x^i(\xi)
\pl_- x^j(\xi) \nn \\
& \bigspc \bigspc +\left( G_{ij} (-x(-\xi)) +B_{ij} (-x(-\xi)) \right)
\pl_+ x^i (-\xi) \pl_- x^j (-\xi) \Big{]} \label{Eq 4.13a} \\
&=\frac{1}{8\pi} \int \!\!dt \int_0^\pi \!\!d\xi \Big{[} \left( G_{ij}
(x(\xi)) +B_{ij} (x(\xi)) \right) \pl_+ x^i (\xi) \pl_- x^j (\xi) \nn \\
& \bigspc \bigspc +\left( G_{ij} (x(-\xi)) -B_{ij} (x(-\xi)) \right) \pl_+
x^i (-\xi) \pl_- x^j (-\xi) \Big{]} \label{Eq 4.13b}
\end{align}
\begin{align}
&=\frac{1}{8\pi} \int \!\!dt \int_0^\pi \!\!d\xi \Big{[} \left( G_{ij}
(x(\xi)) +B_{ij} (x(\xi)) \right) \pl_+ x^i (\xi) \pl_- x^j (\xi) \nn \\
& \bigspc \bigspc +\left( G_{ij} (x(-\xi)) +B_{ij} (x(-\xi)) \right) \pl_-
x^i (-\xi) \pl_+ x^j (-\xi) \Big{]} \ \label{Eq 4.13c} \\
&=\frac{1}{8\pi} \int \!\!dt \left( \int_0^\pi +\int_{-\pi}^0 \right) d\xi
\left( G_{ij} (x(\xi)) +B_{ij} (x(\xi)) \right) \pl_+ x^i (\xi) \pl_- x^j
(\xi) =S_{WZW}
  \label{Eq 4.13d}
\end{align}
\end{subequations}
where \eqref{Eq 4.13b} is obtained from \eqref{Eq 4.13a} by the space-time
parities \eqref{Eq 4.6b}. The last form in Eq.~\eqref{Eq 4.13d} is indeed
equal to the standard form
\eqref{Eq4.12} of the sigma model action on the cylinder -- because the
integrands of both are $2\pi$-periodic.

\subsection{Automorphic Responses of the Geometric Quantities}

We turn next to the action of the general orientation-reversing
automorphism $\hat{h}_\s$ on the two-component geometric quantities:
\begin{subequations}
\label{Eq4.14}
\begin{gather}
\ws_i{}^j \equiv e(0)_i{}^a \ws_a{}^b e(0)_b{}^j ,\quad \ws \in Aut(g) \\
x^{i\Id}(\xi)' = x^{j\Jd}(\xi) (\tau_1 )_\Jd{}^\Id \om\hc (h_\s)_j{}^i
,\quad \pl_{i\Id}(\xi)' = \ws_i{}^j (\tau_1 )_\Id{}^\Jd \pl_{j\Jd} ,\quad
\Id =0,1
\end{gather}
\begin{gather}
\tilde{\Omega}(x)_{a\Id}{}^{b\Jd}{}' =\tilde{\Omega}(x')_{a\Id}{}^{b\Jd}
=\ws_a{}^c (\tau_1 )_\Id{}^{\Kd} \tilde{\Omega}(x)_{c\Kd}
   {}^{d\dot{L}} (\tau_1 )_{\dot{L}}{}^\Jd \om\hc (h_\s)_d{}^b \\
e_{i\Id}(T,x)' =e_{i\Id} (T,x') =\ws_i{}^j (\tau_1 )_\Id{}^\Jd \tau_1
W(h_\s;T) e_{j\Jd} (T,x) W\hc (h_\s;T) \tau_1 \\
e(x)_{i\Id}{}^{a\Jd}{}' =e(x')_{i\Id}{}^{a\Jd} = \ws_i{}^j (\tau_1
)_\Id{}^{\Kd} e(x)_{j\Kd}{}^{b\dot{L}} (\tau_1 )_{\dot{L}}{}^\Jd
   \om\hc (h_\s)_b{}^a
\end{gather}
\begin{gather}
G_{i\Id ;j\Jd}(x)' =G_{i\Id ;j\Jd}(x') =\ws_i{}^{k} \ws_j{}^{l} (\tau_1
)_\Id{}^{\Kd} (\tau_1 )_\Jd{}^{\dot{L}} G_{k\Kd ;l\dot{L}}(x) \\
B_{i\Id ;j\Jd}(x)' =B_{i\Id ;j\Jd}(x') =\ws_i{}^{k} \ws_j{}^{l} (\tau_1
)_\Id{}^{\Kd} (\tau_1 )_\Jd{}^{\dot{L}} B_{k\Kd ;l\dot{L}}(x) \\
H_{i\Id ;j\Jd ;k\Kd}(x)' \!=H_{i\Id ;j\Jd ;k\Kd}(x') \bigspc \bigspc
\bigspc \bigspc \bigspc \nn \\
\!=\!\ws_i{}^l \ws_j{}^m \ws_k{}^n (\tau_1 )_\Id{}^{\!\dot{L}} (\tau_1
)_\Jd{}^{\!\dot{M}} (\tau_1 )_\Kd{}^{\!\dot{N}}
   H_{l\dot{L};m\dot{M};n\dot{N}}(x) \,.
\end{gather}
\end{subequations}
These automorphic responses follow by substitution of Eqs.~($2.9$g-i) into
\eqref{Eq4.2} and the $\widehat{Tr}$ formulae in Eq.~\eqref{Eq4.3}.
As expected, the strip form \eqref{Eq 4.11a} of the WZW
sigma-model action is invariant under the general orientation-reversing
automorphism above.

Finally, it is instructive to consider the action of $\hat{h}_\s$ on the
original fields:
\begin{subequations}
\label{Eq4.15}
\begin{gather}
x^i (\xi)' = -x^j (-\xi) \om\hc(h_\s)_j{}^i \\
g(T,x(\xi))' = g(T,x' (\xi)) =g(T,-x(-\xi)\om\hc ) = W(h_\s ;T)
g(T,-x(-\xi)) W\hc (h_\s ;T) \\
\Omega (x(\xi))_a{}^b {}' =\Omega (-x(-\xi)\om\hc)_a{}^b = \ws_a{}^c \Omega
(-x(-\xi))_c{}^d \om\hc (h_\s )_d{}^b \\
e(x(\xi))_i{}^a{}' =e(-x(-\xi) \om\hc)_i{}^a = \ws_i{}^j e(-x(-\xi))_j{}^b
\om\hc (h_\s)_b{}^a
\end{gather}
\begin{gather}
G_{ij} (x (\xi))' =G_{ij} (-x(-\xi)\om\hc) =\ws_i{}^k \ws_j{}^l
G_{kl}(-x(-\xi)) \\
B_{ij} (x (\xi))' =B_{ij} (-x(-\xi)\om\hc) =\ws_i{}^k \ws_j{}^l
B_{kl}(-x(-\xi)) \\
H_{ijk} (x (\xi))' =H_{ijk} (-x(-\xi)\om\hc) =\ws_i{}^l \ws_j{}^m \ws_k{}^n
H_{lmn}(-x(-\xi)) \,.
\end{gather}
\end{subequations}
We note in particular another set of {\it consistency relations} on the
original fields
\begin{subequations}
\label{Eq4.16}
\begin{gather}
g(T,x\om\hc) = W(T)g(T,x) W\hc (T) \\
\Omega (x\om\hc)_a{}^b = \om_a{}^c \Omega (x)_c{}^d (\om\hc )_d{}^b ,\quad
e(x\om\hc)_i{}^a = \om_i{}^j e(x)_j{}^b (\om\hc)_b{}^a
\end{gather}
\begin{gather}
G_{ij} (x\om\hc) =\om_i{}^k \om_j{}^l G_{kl}(x) ,\quad B_{ij} (x\om\hc) =\om_i{}^k
\om_j{}^l B_{kl}(x) \\
H_{ijk} (x\om\hc) =\om_i{}^l \om_j{}^m \om_k{}^n H_{lmn}(x) \\
\om \equiv \ws ,\quad W(T) \equiv W(h_\s;T) ,\quad h_\s \in Aut(g)
\end{gather}
\end{subequations}
which are implied by comparing the last two entries of each line in
Eq.~\eqref{Eq4.15}. This set of consistency relations  expresses the Lie
symmetry of the WZW models, and these relations can be easily
checked from the explicit forms given in Eq.~\eqref{Eq4.7}. In fact, as we
will emphasize in
 Subsec.~$5.1$, such
consistency relations together with the behavior of the fields under
space-time parity
can be considered as the fundamental symmetries which allow
orientation-reversing automorphisms.

\subsection{Geometric Eigenfields}

Following the familiar procedure \cite{Dual,More,Big,Geom,Orient1} of the
orbifold program, we now define the eigenfields associated to the
two-component
Einstein coordinates and tensors
\begin{subequations}
\label{Eq4.17}
\begin{gather}
U(\s)_\nrm{}^i = U(\s)_\nrm{}^a e(0)_a{}^i ,\quad U\hc(\s)_i{}^\nrm =
e(0)_i{}^a U\hc(\s)_a{}^\nrm \label{Eq 4.17a}
\end{gather} \vspace{-0.1in}
\begin{align}
&\bigspc \quad \sx_\s^\nrmu (\xi) \!\equiv \!x^{i\Id}(\xi)
\schisig_\nrm^{-1} U\hc (\s)_i{}^\nrm (\srac{1}{\sqrt{2}} U\hc {}_\Id{}^u
) ,\quad \bar{u}=0,1 \\
&\bigspc \pl_\nrmu (\xi) \!\equiv \!\frac{\pl}{\pl \sx^\nrmu (\xi)}
=\schisig_\nrm U(\s)_\nrm{}^i (\sqrt{2} U_u{}^\Id ) \pl_{i\Id}(\xi) \\
&\bigspc \quad \pl_\nrmu (\xi) \sx_\s^\nsnv (\xi)= \de_{n(r)-n(s)
,0\,\text{mod }\r(\s)} \de_\m{}^\n \de_{u-v,0\,\text{mod }2}
\end{align}
\begin{align}
&\sW(\sx)_\nrmu{}^{\!\!\!\nsnv} \!\equiv \!\schisig_\nrm U(\s)_\nrm{}^a
(\sqrt{2}U_u{}^\Id) \tilde{\Omega}(x(\sx))_{a\Id}{}^{b\Jd} \nn \\
& \bigspc \bigspc \bigspc \bigspc \times \schisig_\nsn^{-1}
U\hc(\s)_b{}^\nsn (\srac{1}{\sqrt{2}} (U\hc )_\Jd{}^v ) \quad \label{Eq
4.17e} \\
&\se_\nrmu (\st,\sx) \equiv \schisig_\nrm U(\s)_\nrm{}^i (\sqrt{2}
U_u{}^\Id ) U(T,\s) U e_{i\Id} (T,x(\sx)) U\hc U\hc(T,\s) \\
&\se(\sx)_\nrmu{}^{\!\!\!\nsnv} \!\equiv \!\schisig_\nrm U(\s)_\nrm{}^i
(\sqrt{2} U_u{}^\Id ) e(x(\sx))_{i\Id}{}^{a\Jd} \nn \\
& \bigspc \bigspc \bigspc \bigspc \times \schisig_\nsn^{-1} U\hc
(\s)_a{}^\nsn (\srac{1}{\sqrt{2}} (U\hc )_\Jd{}^v ) \quad \label{Eq 4.17g}
\end{align}
\begin{align}
\!\!&\!\!\sG_{\nrmu ;\nsnv}(\sx) \!\equiv \!\schisig_\nrm \schisig_\nsn
U(\s)_\nrm{}^{\!i} U(\s)_\nsn{}^{\!j} (\sqrt{2}U_u{}^{\!\!\Id}
)(\sqrt{2}U_v{}^{\!\!\Jd} ) G_{i\Id ;j\Jd}(x(\sx))  \\
\!\!&\!\!\sB_{\nrmu ;\nsnv}(\sx) \!\equiv \!\schisig_\nrm \schisig_\nsn
U(\s)_\nrm{}^{\!i} U(\s)_\nsn{}^{\!j} (\sqrt{2}U_u{}^{\!\!\Id}
)(\sqrt{2}U_v{}^{\!\!\Jd} ) B_{i\Id ;j\Jd}(x(\sx))  \\
&\sH_{\nrmu ;\nsnv ;\ntd w}(\sx) \equiv \schisig_\nrm \schisig_\nsn
\schisig_\ntd U(\s)_\nrm{}^i U(\s)_\nsn{}^j U(\s)_\ntd{}^k \nn \\
&  \bigspc \bigspc \bigspc \times (\sqrt{2}U_u{}^\Id )(\sqrt{2}U_v{}^\Jd )
(\sqrt{2}U_w{}^\Kd ) H_{i\Id ;j\Jd ;k\Kd}(x(\sx))
\end{align}
\end{subequations}
where the unitary matrices $U(\s), U(T,\s)$ and $U$ are defined in
Eqs.~\eqref{Eq2.11}, \eqref{Eq 2.16e} and $\schisig_\nrm$ are the same
normalization constants which appear in Eq.~\eqref{Eq2.14}.

Relations among these eigenfields include the following
\begin{subequations}
\label{Eq4.18}
\begin{gather}
\sg (\st,\sx) \st_\nrmu \sg^{-1} (\st,\sx) = \sW (\sx)_\nrmu{}^\nsnv
\st_\nsnv \\
\se_\nrmu (\st,\sx) =-i\sg^{-1}(\st,\sx) \pl_\nrmu \sg(\st,\sx) =
\se(\sx)_\nrmu{}^\nsnv \st_\nsnv \\
\!\!\!\!\sG_{\nrmu ;\nsnv}(\sx) \!= \!-\widehat{Tr} \left( (\sm \!\otimes
\!\one_2) \!\left( \sg^{-1} \pl_\nrmu \sg \sg^{-1} \pl_\nsnv \sg \right)
\right) \nn \\
\bigspc =\se(\sx)_\nrmu{}^{n(r')\m 'u'} \se(\sx)_\nsnv{}^{n(s')\n 'v'}
\sG_{n(r')\m 'u';n(s')\n 'v'}(\s) \label{Eq 4.18c} \\
\sH_{\nrmu ;\nsnv ;\ntd w}(\sx) =\pl_\nrmu \sB_{\nsnv ;\ntd w} (\sx) +
\text{ cyclic} \quad \quad \bigspc  \nn \\
\bigspc =\!\se_\nrmu{}^{\!\!\!\!n(r')\m 'u'} \se_\nsnv{}^{\!\!\!\!n(s')\n
'v'} \se_{\ntd w}{}^{\!\!\!\!n(t')\de 'w'}
   \scf_{n(r')\m 'u';n(s')\n 'v';n(t')\de 'w'}(\s) \\
\scf_{\nrmu;\nsnv ;\ntd w}(\s) = 2\de_{u+v+w,0\,\text{mod }2}
\scf_{\nrm;\nsn;\ntd}(\s) \label{Eq 4.18e} \\
\scf_{\nrm;\nsn;\ntd}(\s) = \scf_{\nrm;\nsn}{}^{n(t'),\de '}(\s)
\sG_{n(t'), \de ';\ntd}(\s)
\end{gather}
\begin{gather}
\ep^{ABC} \widehat{Tr} \left( (\sm \otimes \one) \left( \sg^{-1} \pl_A \sg
\sg^{-1} \pl_B \sg \sg^{-1} \pl_C \sg \right) \right) = \pl_A \tilde{j}^A
(\xi) \label{Eq 4.18g} \\
\tilde{j}^A (\xi) = \srac{3}{2} \ep^{ABC} \pl_B \sx^\nrmu (\xi) \pl_C
\sx^\nsnv (\xi) \sB_{\nrmu;\nsnv}(\sx(\xi))
\end{gather}
\end{subequations}
where the eigengroup element, the total twisted metric and the ordinary
twisted metric were defined in Eqs.~\eqref{Eq 2.16a}, \eqref{Eq 2.19b} and
\eqref{Eq2.14} respectively. The $\scf(\s)$ on the right side of
Eq.~\eqref{Eq 4.18e} is the totally antisymmetric form of the ordinary
twisted structure constants defined in Eq.~\eqref{Eq2.14}.

As above, these eigenfields are constructed to have diagonal responses to
the orientation-reversing automorphism:
\begin{subequations}
\label{Eq4.19}
\begin{gather}
\sx_\s^\nrmu (\xi)' =\sx_\s^\nrmu (\xi) e^{\tp (\nrrsf +\frac{u}{2})}
,\quad \pl_\nrmu (\xi)' =e^{-\tp (\nrrsf +\frac{u}{2})} \pl_\nrmu (\xi) \\
\se_\nrmu (\st,\sx)' =\se_\nrmu (\st,\sx ')=e^{-\tp (\nrrsf +\frac{u}{2})}
E(T,\s) \se_\nrmu (\st,\sx) E^\ast (T,\s) \\
\se(\sx)_\nrmu{}^{\!\!\!\nsnv}{}' =\se(\sx ')_\nrmu{}^{\!\!\!\nsnv}
=e^{-\tp (\frac{n(r)-n(s)}{\r(\s)} +\frac{u-v}{2})}
\se(\sx)_\nrmu{}^{\!\!\!\nsnv}
\end{gather}
\begin{gather}
\sG_{\nrmu ;\nsnv} (\sx)' =\sG_{\nrmu ;\nsnv}(\sx ') =e^{-\tp
(\frac{n(r)+n(s)}{\r(\s)} +\frac{u+v}{2})} \sG_{\nrmu ;\nsnv}(\sx) \\
\sB_{\nrmu ;\nsnv} (\sx)' =\sB_{\nrmu ;\nsnv}(\sx ') =e^{-\tp
(\frac{n(r)+n(s)}{\r(\s)} +\frac{u+v}{2})} \sB_{\nrmu ;\nsnv}(\sx) \\
\sH_{\nrmu ;\nsnv ;\ntd w} (\sx)' =\sH_{\nrmu ;\nsnv ;\ntd w}(\sx ')
\bigspc \bigspc \bigspc \quad \quad \nn \\
\bigspc \bigspc \bigspc =e^{-\tp (\frac{n(r)+n(s)+n(t)}{\r(\s)}
+\frac{u+v+w}{2})} \sH_{\nrmu ;\nsnv ;\ntd w}(\sx) \,.
\end{gather}
\end{subequations}
To obtain these responses, we have used the action \eqref{Eq4.14} of
$\hat{h}_\s$, the
definitions in Eq.~\eqref{Eq4.17} and the $H$-eigenvalue problems in
Eq.~\eqref{Eq2.11}.

Using the explicit forms \eqref{Eq4.7} of the original fields, these
eigenfields can also be explicitly evaluated as
\begin{subequations}
\label{Eq4.20}
\begin{gather}
\sW (\sx) = \sg^{-1} (\tilde{\st}^{adj} ,\sx) =e^{-i \sY(\sx)} ,\quad
\sY(\sx) \equiv \sx_\s^\nrmu \tilde{\st}^{adj}_\nrm \tau_u \label{Eq
4.20a} \\
(\tilde{\st}^{adj}_\nrm )_\nsn{}^{\!\ntd} \!\equiv \!
-i\scf_{\nrm;\nsn}{}^{\!\ntd}(\s)\!= \!\schisig_\nsn \schisig_\ntd^{-1}
   \st_\nrm(T^{adj},\s)_\nsn{}^{\!\ntd} \label{Eq 4.20b} \\
[\tilde{\st}^{adj}_\nrm ,\tilde{\st}^{adj}_\nsn ]=
i\scf_{\nrm;\nsn}{}^{n(r)+n(s),\de} (\s) \tilde{\st}^{adj}_{n(r)+n(s),\de}
\end{gather}
\begin{gather}
\se(\sx)_\nrmu{}^{\!\!\nsnv} = \left(
\srac{e^{i\sY(\sxtiny)}-1}{i\sY(\sxtiny)}
\right)_{\!\nrmu}^{\,\,\,\,\,\nsnv} ,\quad \se(0) =\one \\
\sG_{\nrmu;\nsnv}(\sx) =\left( \srac{e^{i\sY(\sxtiny)}
+e^{-i\sY(\sxtiny)}-2}{(i\sY(\sxtiny) )^2}
\right)_{\!\nrmu}^{\,\,\,\,\,\ntd w}
   \sG_{\ntd w;\nsnv}(\s) \\
\sB_{\nrmu;\nsnv}(\sx) =\left(
\srac{e^{i\sY(\sxtiny)}-e^{-i\sY(\sxtiny)}-2i\sY(\sxtiny)}{(i\sY(\sxtiny))^2}
\right)_{\!\nrmu}^{\,\,\,\,\,\ntd w} \sG_{\ntd w;\nsnv}(\s)
\end{gather}
\end{subequations}
where the matrices $\tilde{\st}$ in Eq.~\eqref{Eq 4.20b} are the so-called
rescaled twisted representation matrices \cite{Geom}. Because of the Pauli
matrices
in $\sY$, it is not difficult to check that all these eigenfields have
only two independent components in the two-dimensional space:
\begin{subequations}
\label{Eq4.21}
\begin{gather}
\sW (\sx)_\nrmu{}^\nsnv =\sW^{(u-v)} (\sx)_\nrm{}^\nsn ,\quad
\se(\sx)_\nrmu{}^\nsnv = \se^{(u-v)} (\sx)_\nrm{}^\nsn \\
\sG_{\nrmu ;\nsnv}(\sx) =\sG_{\nrm;\nsn}^{(u+v)}(\sx) ,\quad
\sB_{\nrmu;\nsnv}(\sx) = \sB_{\nrm;\nsn}^{(u+v)}(\sx) \\
\sH_{\nrmu ;\nsnv ;\ntd w}(\sx) = \sH_{\nrm;\nsn;\ntd}^{(u+v+w)}(\sx)
,\quad \bar{u},\bar{v},\bar{w} \in \{0,1\} \,.
\end{gather}
\end{subequations}
This fact is also apparent when the eigenfields are expressed in terms of
the standard eigenfields \cite{Geom} of space-time orbifold theory.
\begin{subequations}
\label{Eq4.22}
\begin{gather}
\sx_\s^\nrmu (\xi) =\srac{1}{2} \left( \sx_\s^\nrm (\xi) -(-1)^u
\sx_\s^\nrm (-\xi) \right) \\
\sW (\sx(x))_\nrmu{}^\nsnv =\srac{1}{2} \left( \sW (x^0)_\nrm{}^\nsn
+(-1)^{u-v} \sW (x^1)_\nrm{}^\nsn \right) \\
\se(\sx(x))_\nrmu{}^\nsnv =\srac{1}{2} \left( \se(x^0)_\nrm{}^\nsn
+(-1)^{u-v} \se(x^1)_\nrm{}^\nsn \right) \\
\sG_{\nrmu ;\nsnv} (\sx(x)) =\sG_{\nrm ;\nsn} (x^0) +(-1)^{u+v}
\sG_{\nrm;\nsn} (x^1) \\
\sB_{\nrmu ;\nsnv} (\sx(x)) =\sB_{\nrm ;\nsn} (x^0) +(-1)^{u+v}
\sB_{\nrm;\nsn} (x^1) \\
\sH_{\nrmu \nsnv ;\ntd w}(\sx(x)) =\sH_{\nrm;\nsn;\ntd}(x^0) +(-1)^{u+v+w}
\sH_{\nrm;\nsn;\ntd}(x^1)
\end{gather}
\begin{gather}
\sx_\s^\nrm (\xi) \equiv x^i(\xi) \schisig_\nrm^{-1} U\hc(\s)_i{}^\nrm \\
\sW (x)_\nrm{}^\nsn \equiv \schisig_\nrm U(\s)_\nrm{}^a \Omega(x)_a{}^b
\schisig_\nsn^{-1} U\hc(\s)_b{}^\nsn \\
\se(x)_\nrm{}^\nsn \equiv \schisig_\nrm U(\s)_\nrm{}^i e(x)_i{}^a
\schisig_\nsn^{-1} U\hc(\s)_a{}^\nsn \\
\sG_{\nrm ;\nsn}(x) \equiv \schisig_\nrm \schisig_\nsn U(\s)_\nrm{}^i
U(\s)_\nsn{}^j G_{ij}(x) \\
\sB_{\nrm ;\nsn}(x) \equiv \schisig_\nrm \schisig_\nsn U(\s)_\nrm{}^i
U(\s)_\nsn{}^j B_{ij}(x)
\end{gather}
\begin{align}
&\!\sH_{\nrm;\nsn;\ntd}(x) \!\equiv \!\schisig_\nrm \schisig_\nsn
\schisig_\ntd U(\s)_\nrm{}^i U(\s)_\nsn{}^j U(\s)_\ntd{}^k H_{ijk}(x)
\end{align}
\end{subequations}
which are obtained by using Eq.~\eqref{Eq A.3c} to do the internal sums on
the two-component indices $\Id$ in Eq.~\eqref{Eq4.17}.

Many of the eigenfields also diagonalize the behavior under world-sheet
parity:
\begin{subequations}
\label{Eq4.23}
\begin{gather}
\sx_\s^\nrmu (-\xi) =-\sx_\s^{\nrm v} (\xi) (\tau_3 )_v{}^u = (-1)^{u+1}
\sx_\s^\nrmu (\xi) \\
\pl_\nrmu (-\xi) =(-1)^{u+1} \pl_\nrmu (\xi) \\
\sY_{\nrmu}{}^\nsnv (\sx(-\xi)) =(-1)^{u+v+1} \sY_\nrmu{}^\nsnv (\sx(\xi))
\label{Eq 4.23c} \\
\sG_{\nrmu ;\nsnv}(\sx(-\xi)) =(-1)^{u+v} \sG_{\nrmu;\nsnv}(\sx(\xi)) \\
\sB_{\nrmu ;\nsnv}(\sx(-\xi)) =(-1)^{u+v+1} \sB_{\nrmu;\nsnv}(\sx(\xi)) \\
\sH_{\nrmu ;\nsnv ;\ntd w}(\sx(-\xi)) =(-1)^{u+v+w} \sH_{\nrmu ;\nsnv
;\ntd w}(\sx(\xi))\,.
\end{gather}
\end{subequations}
These relations follow from the corresponding world-sheet parities in
Eq.~\eqref{Eq4.4}, but the relations in ($4.23$d,e) also follow from
\eqref{Eq 4.23c}
and the explicit forms in Eq.~\eqref{Eq4.20}. More complicated behavior is
obtained for other eigenfields
\begin{subequations}
\label{Eq4.24}
\begin{gather}
\sW(\sx(-\xi))_\nrmu{}^\nsnv =(-1)^{u-v} \sW^{-1} (\sx(\xi))_\nrmu{}^\nsnv
\\
\se(\sx(-\xi))_\nrmu{}^\nsnv =(-1)^{u-v} \left( \se(\sx(\xi))
\sW(\sx(\xi)) \right)_\nrmu{}^\nsnv
\end{gather}
\end{subequations}
where we have used Eqs.~\eqref{Eq4.4}, \eqref{Eq 4.17e} and \eqref{Eq
4.17g}.

Finally, we may reexpress the strip form \eqref{Eq 4.11a} of the WZW
action in terms of the eigenfields:
\begin{subequations}
\label{Eq4.25}
\begin{gather}
S_{WZW}^{\text{ strip}} = \int \!\!dt \int_0^\pi \!\! d\xi \, {\cL} \\
{\cL} =\frac{1}{8\pi} \left( \sG_{\nrmu;\nsnv}(\sx) +\sB_{\nrmu ;\nsnv}
(\sx) \right) \pl_+ \sx^\nrmu \pl_- \sx^\nsnv \bigspc \bigspc \nn \\
 =\frac{1}{8\pi} \left( \sG_{\nrm;\nsn}^{(u+v)} (\sx)
+\sB_{\nrm;\nsn}^{(u+v)} (\sx) \right) \pl_+ \sx_\s^\nrmu \pl_-
\sx_\s^\nsnv \bigspc \bigspc \nn \\
 =\frac{1}{8\pi} \sum_{w=0}^1 \left( \sG_{\nrm;\nsn}^{(w)}(\sx)
+\sB_{\nrm;\nsn}^{(w)}(\sx) \right) \sum_{u=0}^1 \pl_+ \sx^\nrmu \pl_-
\sx^{\nsn ,w-u} \,.
\end{gather}
\end{subequations}
This form can also be obtained from the WZW action \eqref{Eq3.7} on
$\Gamma_{1/2}$ by using the identity \eqref{Eq 4.18c}, the Gauss' law
\eqref{Eq 4.18g} and
Eq.~\eqref{Eq4.9}.

\subsection{Einstein Geometry of the Twisted Open WZW Strings}

We now move to the twisted fields, using the principle of local
isomorphisms \cite{Dual, More, Big, Geom}
\begin{subequations}
\label{Eq4.26}
\begin{gather}
\sg \dual \hg ,\quad \sx \dual \hx ,\quad \sW \dual \hO ,\quad \se \dual
\he \label{Eq 4.26a} \\
\sG \dual \hG ,\quad \sB \dual \hB ,\quad \sH \dual \hh ,\quad \tilde{j}
\dual \hat{j} \label{Eq 4.26b} \\
S_{WZW}^{\text{ strip}} \dual \hat{S}_{\hg_O (\s)}^{\text{ strip}}
\end{gather}
\end{subequations}
to map the eigenfields and the action to the twisted fields and the action
formulation of the twisted open-string sector.

We begin with some useful relations among the twisted fields
\begin{subequations}
\label{Eq4.27}
\begin{gather}
\hx_\s^\nrmu \equiv \hx_\s^\nrmu (\xi,t) ,\quad \hpl_\nrmu (\xi) \equiv
\frac{\pl}{\pl \hx^\nrmu (\xi)} ,\quad \bar{u}=0,1 \label{Eq 4.27a} \\
\hg (\st,\hx) \st_\nrmu \hg^{-1} (\st,\hx) = \hO (\hx)_\nrmu{}^\nsnv
\st_\nsnv \\
\he_\nrmu (\st,\hx) =-i\hg^{-1}(\st,\hx) \hpl_\nrmu \hg(\st,\hx) =
\he(\hx)_\nrmu{}^\nsnv \st_\nsnv \\
\he (\hx)_\nrmu{}^\nsnv = -i\widehat{Tr} \left( (\sm \otimes \one_2)
\hg^{-1} (\st,\hx) \hpl_\nrmu \hg (\st,\hx) \sG^{\nsnv ;\ntd w}(\s)
\st_{\ntd w} \right) \label{Eq 4.27d}
\end{gather}
\begin{gather}
\hG_{\nrmu ;\nsnv}(\hx) =-\widehat{Tr} \left( (\sm \otimes \one_2)
\hg^{-1} \hpl_\nrmu \hg \hg^{-1} \hpl_\nsnv \hg \right) \bigspc \quad
\quad \nn \\
\quad =\he(\hx)_\nrmu{}^{n(r')\m 'u'} \he(\hx)_\nsnv{}^{n(s')\n 'v'}
\sG_{n(r')\m 'u';n(s')\n 'v'}(\s) \label{Eq 4.27e} \\
\bigspc \bigspc =\he(\hx)_\nrmu{}^{n(r')\m 'u'}
\he(\hx)_\nsnv{}^{-n(r'),\n ',-u'} \sG_{n(r')\m 'u';-n(r'),\n ',-u'}(\s)
\quad \label{Eq 4.27f} \\
\hh_{\nrmu ;\nsnv ;\ntd w}(\hx) =\hpl_\nrmu \hB_{\nsnv ;\ntd w}(\hx)
+\text{ cyclic } \bigspc \bigspc \nn \\
\bigspc =\!\he_\nrmu{}^{\!\!\!\!n(r')\m 'u'} \he_\nsnv{}^{\!\!\!\!n(s')\n
'v'} \he_{\ntd w}{}^{\!\!\!\!n(t') \de 'w'}
   \scf_{n(r')\m 'u';n(s')\n 'v';n(t')\de 'w'}(\s)
\end{gather}
\begin{gather}
\ep^{ABC} \widehat{Tr} \left( (\sm \otimes \one) \left( \hg^{-1} \pl_A \hg
\hg^{-1} \pl_B \hg \hg^{-1} \pl_C \hg \right) \right) = \pl_A \hat{j}^A
(\xi) \label{Eq 4.27h} \\
\hat{j}^A (\xi) = \srac{3}{2} \ep^{ABC} \pl_B \hx_\s^\nrmu (\xi) \pl_C
\hx_\s^\nsnv (\xi) \hB_{\nrmu;\nsnv}(\hx(\xi)) \, \label{Eq 4.27i}
\end{gather}
\end{subequations}
which follow by local isomorphisms from the corresponding relations among
the eigenfields. Here $\sG^{\bullet}(\s)$ in \eqref{Eq 4.27d} is the total
inverse twisted
metric defined in Eq.~\eqref{Eq 3.20d}, and we have used the selection
rule \eqref{Eq 2.14a}
for the tangent-space metric $\sG_{\bullet}(\s)$ to obtain the reduced
form \eqref{Eq 4.27f} of the twisted Einstein metric $\hG(\hx)$. A similar
reduced form of the
twisted torsion $\hh(\hx)$ can be obtained by using the selection rules
($2.13$a,b), \eqref{Eq 4.18e}. The quantity $\hat{j}^A (\xi)$ in Eq.~\eqref{Eq 4.27i} 
will be called the twisted winding-number current.

From Eqs.~\eqref{Eq4.23}, \eqref{Eq4.24}, the principle of local
isomorphisms also gives us the world-sheet parities of the twisted fields:
\begin{subequations}
\label{Eq4.28}
\begin{gather}
\hx_\s^\nrmu (-\xi) =-\hx_\s^{\nrm v} (\xi) (\tau_3 )_v{}^u = (-1)^{u+1}
\hx_\s^\nrmu (\xi) \label{Eq 4.28a} \\
\hpl_\nrmu (-\xi) =(-1)^{u+1} \hpl_\nrmu (\xi) \\
\hO(\hx(-\xi))_\nrmu{}^\nsnv = (-1)^{u-v} \hO^{-1}(\hx(\xi))_\nrmu{}^\nsnv
\\
\he(\hx(-\xi))_\nrmu{}^\nsnv =(-1)^{u-v} \left( \he(\hx(\xi))\hO(\hx(\xi))
\right)_\nrmu^{\quad \nsnv} \label{Eq 4.28d} \\
\hG_{\nrmu ;\nsnv}(\hx(-\xi)) =(-1)^{u+v} \hG_{\nrmu;\nsnv}(\hx(\xi)) \\
\hB_{\nrmu ;\nsnv}(\hx(-\xi)) =(-1)^{u+v+1} \hB_{\nrmu;\nsnv}(\hx(\xi)) \\
\hh_{\nrmu ;\nsnv ;\ntd w}(\hx(-\xi)) =(-1)^{u+v+w} \hh_{\nrmu ;\nsnv
;\ntd w}(\hx(\xi)) \\
\hat{j}^\xi (-\xi) = -\hat{j}^\xi (\xi) ,\quad \hat{j}^{t,\r} (-\xi) =
\hat{j}^{t,\r} (\xi) \,.  \label{Eq 4.28h}
\end{gather}
\end{subequations}
We also give the explicit functional forms
\begin{subequations}
\label{Eq4.29}
\begin{gather}
\hO (\hx) =e^{-i\hY (\hx)} ,\quad \hY (\hx) \equiv \hx_\s^\nrmu
\tilde{\st}^{adj}_\nrm \tau_u \\
\he(\hx)_\nrmu{}^{\!\!\!\!\!\nsnv} \!= \!\left(
\srac{e^{i\hY(\hx)}-1}{i\hY(\hx)} \right)_\nrmu^{\quad \nsnv} \!,\,\,
\he(0)_\nrmu{}^{\!\!\!\!\!\nsnv} \!=\!
  \de_\m^\n \de_{n(r)+n(s),0\,\text{mod }\r(\s)} \de_{u+v,0\,\text{mod }2}
\end{gather}
\begin{align}
\hG_{\nrmu;\nsnv}(\hx) &=\left( \srac{e^{i\hY(\hx)}
+e^{-i\hY(\hx)}-2}{(i\hY(\hx))^2} \right)_\nrmu^{\quad \ntd w} \sG_{\ntdw
;\nsnv}(\s) \nn \\
&= \left( \srac{e^{i\hY(\hx)} +e^{-i\hY(\hx)}-2}{(i\hY(\hx))^2}
\right)_\nsnv^{\quad \ntd w} \sG_{\ntdw ;\nrmu}(\s) \nn \\
&= \left( \srac{e^{i\hY(\hx)} +e^{-i\hY(\hx)}-2}{(i\hY(\hx))^2}
\right)_\nrmu^{\quad -n(s),\n ,-v} \sG_{-n(s),\n ,-v;\nsnv}(\s)
\end{align}
\begin{align}
\hB_{\nrmu;\nsnv}(\hx) &=\left( \srac{e^{i\hY(\hx)}
-e^{-i\hY(\hx)}-2i\hY(\hx)}{(i\hY(\hx))^2} \right)_\nrmu^{\quad \ntd w}
   \sG_{\ntdw ;\nsnv}(\s) \nn \\
&= -\left( \srac{e^{i\hY(\hx)} -e^{-i\hY(\hx)}-2i\hY(\hx)}{(i\hY(\hx))^2}
\right)_\nsnv^{\quad \ntd w} \sG_{\ntdw ;\nrmu}(\s) \nn \\
&= \left( \srac{e^{i\hY(\hx)} -e^{-i\hY(\hx)}-2i\hY(\hx)}{(i\hY(\hx))^2}
\right)_\nrmu^{\quad -n(s),\n ,-v} \sG_{-n(s),\n ,-v;\nsnv}(\s)
\end{align}
\end{subequations}
which follow by local isomorphisms from Eq.~\eqref{Eq4.20}. The rescaled
matrices $\tilde{\st}^{adj}$ which appear in these formulae were defined
in Eq.~\eqref{Eq 4.20b}.

Like the group orbifold elements in Eq.~\eqref{Eq2.23} and the twisted
Einstein coordinates \eqref{Eq 4.27a}, the explicit functional forms in
Eq.~\eqref{Eq4.29}
tell us that the other twisted fields are also two-component fields
\begin{subequations}
\label{Eq4.30}
\begin{gather}
\hO(\hx)_\nrmu{}^\nsnv = \hO^{(u-v)} (\hx)_\nrm{}^\nsn ,\quad \ho^{-1}
(\hx)_\nrmu{}^\nsnv =\ho_{(u-v)}^{-1} (\hx)_\nrm{}^\nsn \\
\he(\hx)_\nrmu{}^\nsnv = \he^{(u-v)} (\hx)_\nrm{}^\nsn \\
\hG_{\nrmu;\nsnv}(\hx) =\hG_{\nrm;\nsn}^{(u+v)} (\hx) ,\quad
\hB_{\nrmu;\nsnv}(\hx) =\hB_{\nrm;\nsn}^{(u+v)} (\hx) \\
\hh_{\nrmu;\nsnv ;\ntd w}(\hx) =\hh_{\nrm;\nsn;\ntd}^{(u+v+w)} (\hx)
,\quad \bar{u},\bar{v} ,\bar{w} \in \{0,1\}
\end{gather}
\end{subequations}
and all of the relations above simplify accordingly. For example, the
following world-sheet parities  of the reduced components
\begin{subequations}
\label{Eq4.31}
\begin{gather}
\ho^{(w)} (\hx(-\xi))_\nrm{}^\nsn = (-1)^w \ho^{-1}_{(w)}
(\hx(\xi))_\nrm{}^\nsn \\
\he^{(w)} (\hx(-\xi))_\nrm{}^\nsn = (-1)^w \sum_{v=0}^1 \he^{(v)}
(\hx(\xi))_\nrm{}^\ntd \ho^{(w-v)} (\hx(\xi))_\ntd{}^\nsn \\
\hG_{\nrm;\nsn}^{(w)} (\hx(-\xi)) =(-1)^w \hG_{\nrm;\nsn}^{(w)} (\hx(\xi))
\\
\hB_{\nrm;\nsn}^{(w)} (\hx(-\xi)) =(-1)^{w+1} \hB_{\nrm;\nsn}^{(w)}
(\hx(\xi)) \\
\hh_{\nrm;\nsn;\ntd}^{(w)} (\hx(-\xi)) =(-1)^w \hh_{\nrm;\nsn;\ntd}^{(w)}
(\hx(\xi))
\end{gather}
\end{subequations}
are obtained by substitution of the reduced forms \eqref{Eq4.30}  into
Eq.~\eqref{Eq4.28}.

The open-string sigma-model form of the WZW orientation-orbifold action
can be obtained by derived local isomorphisms from the (eigenfield)
sigma-model form
\eqref{Eq4.25} of the WZW action:
\begin{subequations}
\label{Eq4.32}
\begin{gather}
\hat{S}_{\hg_O (\s)}^{\text{ strip}} \equiv \int \!\!dt \int_0^\pi \!\!
d\xi \, \hat{{\cL}}_\s^{WZW} \\
\hat{{\cL}}_\s^{WZW} =\frac{1}{8\pi} \left( \hG_{\nrmu;\nsnv}(\hx)
+\hB_{\nrmu ;\nsnv} (\hx) \right) \pl_+ \hx_\s^\nrmu \pl_- \hx_\s^\nsnv
\bigspc \bigspc \nn \\
 =\frac{1}{8\pi} \left( \hG_{\nrm;\nsn}^{(u+v)} (\hx)
+\hB_{\nrm;\nsn}^{(u+v)} (\hx) \right) \pl_+ \hx_\s^\nrmu \pl_-
\hx_\s^\nsnv \,\, \bigspc \quad \quad \nn \\
 \quad \,=\frac{1}{8\pi} \sum_{w=0}^1 \left( \hG_{\nrm;\nsn}^{(w)}(\hx)
+\hB_{\nrm;\nsn}^{(w)}(\hx) \right) \sum_{u=0}^1 \pl_+ \hx_\s^\nrmu \pl_-
\hx_\s^{\nsn ,w-u} \\
\hat{S}_{\hg_O (\s)}^{\text{ strip}} = \hat{S}_{\hg_O (\s)} [\sm \!\otimes
\!\one_2 ,\hg ;\Gamma_\half ] \, . \label{Eq 4.32c}
\end{gather}
\end{subequations}
We finally comment on the equivalence \eqref{Eq 4.32c} of the WZW
orientation-orbifold action \eqref{Eq3.9} on $\Gamma_{1/2}$ and this
open-string sigma-model form
on the strip, both of which followed above for all $\hat{h}_\s$ by local
isomorphisms -- without the use of monodromy. Using the twisted Gauss' law
\eqref{Eq 4.27h} and the form of the twisted Einstein metric in
Eq.~\eqref{Eq 4.27e} to compare these two actions, we find the
$(3\!\leftrightarrow \!2)$-dimensional consistency
relation
\begin{subequations}
\label{Eq4.33}
\begin{gather}
\int_{\Gamma_{\frac{1}{2}}} \!\!\!\widehat{Tr} \left( (\sm \otimes \one_2)
(\hg^{-1} (\st,\xi) d\hg (\st,\xi))^3 \right) \bigspc \bigspc \bigspc \nn
\\
   \quad \quad \bigspc = -\srac{3}{2} \!\int \!\!dt \!\! \int_0^\pi
\!\!\!d\xi \, \hB_{\nrmu ;\nsnv} (\hx) \pl_+ \hx^\nrmu \pl_- \hx^\nsnv \\
\Rightarrow \int_{y=0} \!\! dt dx (-\vec{e}_y \cdot \hat{j}) = \int_{y=0}
\!\! dt dx (\hat{j}^\xi (\pi) -\hat{j}^\xi (0)) =0 \label{Eq 4.33b}
\end{gather}
\end{subequations}
which requires (as in Subsec.~$4.2$) a vanishing contribution from the
flat side of the solid half cylinder for all $\hat{h}_\s$. The local form
of this
boundary condition on the twisted winding-number current
\begin{gather}
\hat{j}^\xi (0) =\hat{j}^\xi (\pi) =0 \label{Eq4.34}
\end{gather}
is verified explicitly for the case $h_\s^2 =1$ in the following
subsection. Another set of boundary conditions can be obtained in the
usual fashion by variation
of the open-string sigma-model form \eqref{Eq4.32} of the action, but we
will discuss these variational boundary conditions more generally in
Subsec.~$5.2$.

\subsection{Monodromies and Extra Boundary Conditions}

As above, we can obtain further structure in the case $h_\s^2 =1$, for which
local isomorphisms gives us the {\it monodromies} of all the twisted
open-string
fields:
\begin{subequations}
\label{Eq4.35}
\begin{gather}
h_\s^2 =1\!: \quad \quad \text{automorphic responses } \dual \text{
monodromies} \bigspc \label{Eq 4.35a} \\
\hg(\st,\hx(\xi+2\pi)) =\tau_3 E(T,\s) \hg(\st ,\hx(\xi)) E^\ast (T,\s)
\tau_3 \\
\hx_\s^\nrmu (\xi+2\pi)= \hx_\s^\nrmu (\xi) e^{\tp (\nrrsf +\frac{u}{2})}
,\quad \bar{u}=0,1  \label{Eq 4.35c} \\
\hpl_\nrmu (\xi) \equiv \frac{\pl}{\pl \hx^\nrmu(\xi)} ,\quad \hpl_\nrmu
(\xi+2\pi) = e^{-\tp (\nrrsf +\frac{u}{2})} \hpl_\nrmu (\xi) \\
\hpl_\nrmu (\xi) \hx_\s^\nsnv (\xi) = \de_\m{}^\n
\de_{n(r)-n(s),0\,\text{mod }\r(\s)} \de_{u-v,0\,\text{mod }2} \\
\hO(\hx(\xi+2\pi))_\nrmu{}^\nsnv =e^{-\tp (\frac{n(r)-n(s)}{\r(\s)}
+\frac{u-v}{2})} \hO(\hx(\xi))_\nrmu{}^\nsnv \\
\he_\nrmu (\st,\hx(\xi+2\pi)) = e^{-\tp (\nrrsf +\frac{u}{2})} \tau_3
E(T,\s) \he_\nrmu (\st,\hx(\xi)) E^\ast (T,\s) \tau_3 \\
\he(\hx(\xi+2\pi))_\nrmu{}^\nsnv =e^{-\tp (\frac{n(r)-n(s)}{\r(\s)}
+\frac{u-v}{2})} \he(\hx(\xi))_\nrmu{}^\nsnv
\end{gather}
\begin{gather}
\hG_{\nrmu ;\nsnv} (\hx(\xi+2\pi)) =e^{-\tp (\frac{n(r)+n(s)}{\r(\s)}
+\frac{u+v}{2})} \hG_{\nrmu ;\nsnv}(\hx(\xi)) \\
\hB_{\nrmu ;\nsnv} (\hx(\xi+2\pi)) =e^{-\tp (\frac{n(r)+n(s)}{\r(\s)}
+\frac{u+v}{2})} \hB_{\nrmu ;\nsnv}(\hx(\xi)) \\
\hh_{\nrmu ;\nsnv ;\ntd w} (\hx(\xi+2\pi)) =e^{-\tp
(\frac{n(r)+n(s)+n(t)}{\r(\s)} +\frac{u+v+w}{2})} \hh_{\nrmu ;\nsnv ;\ntd
w}(\hx(\xi)) \\
\hat{j}^A (\xi+2\pi) = \hat{j}^A (\xi) \label{Eq 4.35l} \\
\hat{{\cL}}_\s^{WZW} (\hx (\xi+2\pi )) =\hat{{\cL}}_\s^{WZW} (\hx (\xi))
\,.
\end{gather}
\end{subequations}
The restriction \eqref{Eq 4.35a} to $h_\s^2 =1$ follows from the
discussion of Subsec.~$3.2$ (and in particular, from Eq.~\eqref{Eq2.29})
and our choice of coordinate system $x=\be$ in Eq.~\eqref{Eq4.2}.

For the reduced components $\hat{A}^{(w)}$ of the twisted fields, one finds
that
the monodromies
\begin{subequations}
\label{Eq4.36}
\begin{gather}
\hO^{(w)} (\hx(\xi+2\pi))_\nrm{}^\nsn =e^{-\tp (\frac{n(r)-n(s)}{\r(\s)}
+\frac{w}{2})} \hO^{(w)}(\hx(\xi))_\nrm{}^\nsn \\
\he^{(w)} (\hx(\xi+2\pi))_\nrm{}^\nsn =e^{-\tp (\frac{n(r)-n(s)}{\r(\s)}
+\frac{w}{2})} \he^{(w)}(\hx(\xi))_\nrm{}^\nsn \\
\hG_{\nrm;\nsn}^{(w)} (\hx(\xi+2\pi)) =e^{-\tp (\frac{n(r)+n(s)}{\r(\s)}
+\frac{w}{2})} \hG_{\nrm;\nsn}^{(w)} (\hx(\xi)) \\
\hB_{\nrm;\nsn}^{(w)} (\hx(\xi+2\pi)) =e^{-\tp (\frac{n(r)+n(s)}{\r(\s)}
+\frac{w}{2})} \hB_{\nrm;\nsn}^{(w)} (\hx(\xi)) \\
\hh_{\nrm;\nsn;\ntd}^{(w)} (\hx(\xi+2\pi)) =e^{-\tp
(\frac{n(r)+n(s)+n(t)}{\r(\s)} +\frac{w}{2})} \hh_{\nrm;\nsn;\ntd}^{(w)}
(\hx(\xi))
\end{gather}
\end{subequations}
follow from Eqs.~\eqref{Eq4.35} and \eqref{Eq4.30}.

Then following steps analogous to those given in Eq.~\eqref{Eq3.30}, the
monodromies \eqref{Eq 4.35c}, \eqref{Eq4.36} and the world-sheet parities
\eqref{Eq 4.28a},
\eqref{Eq4.31} give the {\it boundary conditions} for the twisted fields
on the strip:
\begin{subequations}
\label{Eq4.37}
\begin{gather}
\hx_\s^{\nrm 0}(0)=0 ,\quad \hx_\s^\nrmu (\pi) =0 \text{ unless } \nrrs
\in \Zint +\half \\
(\pl_t \hx_\s^{\nrm 0})(0)=0 ,\quad (\pl_t \hx_\s^\nrmu )(\pi) =0 \text{
unless } \nrrs \in \Zint +\half \\
(\pl_\xi \hx_\s^{\nrm 1})(0)=0 ,\quad (\pl_\xi \hx_\s^\nrmu )(\pi) =0
\text{ unless } \nrrs \in \Zint
\end{gather}\vspace{+0.02in}
\begin{gather}
[\ho^{(0)} (\hx) ,\ho^{(1)} (\hx)]=0 ,\quad \ho^{(0)}(\hx)^2
-\ho^{(1)}(\hx)^2 \!=\!\one \,\text{ at } \xi=0 \\
\left. \begin{array}{ccc}
\ho^{(0)}(\hx) E(\s) \ho^{(0)}(\hx) +\ho^{(1)}(\hx) E(\s) \ho^{(1)}(\hx)
=&E(\s) \\
\ho^{(0)}(\hx) E(\s) \ho^{(1)}(\hx) +\ho^{(1)}(\hx) E(\s) \ho^{(0)}(\hx)
=&0 \end{array} \right\} \text{ at } \xi=\pi \label{Eq 4.37e}
\end{gather}
\begin{gather}
\sum_{v=0}^1 \he^{(v)} (\hx(0))_\nrm{}^\ntd \left( \de_v^w \de_\ntd{}^\nsn
-(-1)^w \ho^{(w-v)} (\hx(0))_\ntd{}^\nsn \right)=0 \\
\sum_{v=0}^1 \he^{(v)} (\hx(\pi))_\nrm{}^\ntd \left( \de_v^w
\de_\ntd{}^\nsn e^{\tp (\frac{n(r)-n(t)}{\r(\s)})} -\ho^{(w-v)}
(\hx(\pi))_\ntd{}^\nsn \right)=0
\end{gather}\vspace{+0.02in}
\begin{gather}
\hG^{(1)}_{\nrm;\nsn}(\hx(0)) =0 ,\quad \hG^{(w)}_{\nrm;\nsn}(\hx(\pi))=0
\text{ unless } \srac{n(r)+n(s)}{\r(\s)} \in \Zint \\
(\pl_\xi \hG^{(0)}_{\nrm;\nsn} )(\hx(0)) =0 ,\quad (\pl_\xi
\hG^{(w)}_{\nrm;\nsn})(\hx(\pi))=0 \text{ unless }
\srac{n(r)+n(s)}{\r(\s)} \in \Zint +\half
\end{gather}
\begin{gather}
\hB^{(0)}_{\nrm;\nsn}(\hx(0)) =0 ,\quad \hB^{(w)}_{\nrm;\nsn}(\hx(\pi))=0
\text{ unless } \srac{n(r)+n(s)}{\r(\s)} \in \Zint +\half \\
(\pl_\xi \hB^{(1)}_{\nrm;\nsn})(\hx(0)) =0 ,\quad (\pl_\xi
\hB^{(w)}_{\nrm;\nsn})(\hx(\pi))=0 \text{ unless }
\srac{n(r)+n(s)}{\r(\s)} \in \Zint
\end{gather}
\begin{gather}
\hh^{(1)}_{\nrm;\nsn;\ntd}(\hx(0)) \!=\!0 ,\quad
\hh^{(w)}_{\nrm;\nsn;\ntd}(\hx(\pi))\!=\!0 \text{ unless }
\srac{n(r)+n(s)+n(t)}{\r(\s)} \!\in \!\Zint \\
(\pl_\xi \hh^{(0)}_{\nrm;\nsn;\ntd})(\hx(0)) =0 \\
(\pl_\xi \hh^{(w)}_{\nrm;\nsn;\ntd})(\hx(\pi)) =0 \text{ unless }
\srac{n(r)+n(s)+n(t)}{\r(\s)} \in \Zint +\half \,.
\end{gather}
\end{subequations}
The eigenvalue matrix $E(\s)$ which appears in Eq.~\eqref{Eq 4.37e} was
defined in Eq.~\eqref{Eq2.11}. Here we gave the $\pl_t$ boundary
conditions explicitly
for $\hx$ only, but the $\pl_t$ boundary conditions for $\hG, \hB$ and
$\hh$ are also the same as those for the fields themselves.

We also remark that the trivial monodromy \eqref{Eq 4.35l} and the
world-sheet parity \eqref{Eq 4.28h} of the twisted winding-number current
$\hat{j}$ show that the
$(3\!\leftrightarrow \!2)$-dimensional consistency relation \eqref{Eq
4.33b} is satisfied explicitly
\begin{gather}
\text{monodromy + world-sheet parity } \, \Rightarrow \,\, \hat{j}^\xi (0)
=\hat{j}^\xi (\pi) =0 \quad \label{Eq4.38}
\end{gather}
in parallel with the discussion of the untwisted case in Subsec.~$4.1$.

As noted above, all the boundary conditions of this subsection at $\xi =\pi$
are true only for $h_\s^2 =1$, while all the boundary conditions at $\xi = 0$
follow from the world-sheet parity alone -- and hence are valid for {\it
all}
$\hat{h}_\s$. Moreover, the boundary conditions of this subsection should
be considered as further substructure of the WZW orientation-orbifold
branes.

\section{Sigma-Model Orientation Orbifolds}

\subsection{Sigma Models with Orientation-Reversing Symmetry}

We consider here the nonlinear sigma model $A_M$ on a general target-space
manifold $M$ with local\footnote{A complete treatment of the nonlinear
sigma model should include discussion of global issues (in which the
discussion of the text pertains to a coordinate patch), but we will not do
so here.} Einstein coordinates $x^i ,\,i=1\ldots \text{dim }M$:
\begin{subequations}
\label{Eq5.1}
\begin{gather}
S= \frac{1}{8\pi} \!\int \!\!dt \!\int_0^{2\pi} \!\!\!d\xi \, \left(
G_{ij}(x) +B_{ij}(x) \right) \pl_+ x^i \pl_- x^j ,\quad \pl_\pm =\pl_t
   \pm \pl_\xi \label{Eq 5.1a} \\
H_{ijk}(x) =\pl_i B_{jk} (x) +\pl_j B_{ki}(x) +\pl_k B_{ij}(x) ,\quad x^i
(\xi +2\pi) =x^i (\xi)\,.
\end{gather}
\end{subequations}
We begin by requiring that the sigma model action is invariant under the
basic orientation-reversing automorphism:
\begin{subequations}
\label{Eq5.2}
\begin{gather}
\hat{h}_\s =\tau_1 \times \one \\
x^i (\xi)' = -x^i (-\xi) ,\quad G_{ij} (x)' = G_{ij}(x') ,\quad B_{ij}(x)'
=B_{ij}(x') ,\quad H_{ijk}(x)' = H_{ijk}(x') \,. \label{Eq 5.2b}
\end{gather}
\end{subequations}
Then one finds
\begin{subequations}
\label{Eq5.3}
\begin{align}
S' &= \frac{1}{8\pi} \int \!\!dt \int_0^{2\pi} \!\!d\xi \left( G_{ij}
(-x(-\xi)) +B_{ij}(-x(-\xi)) \right) \pl_+ x^i (-\xi) \pl_- x^j (-\xi) \\
&= \frac{1}{8\pi} \int \!\!dt \int_{-2\pi}^0 \!\!d\xi \left( G_{ij}
(-x(\xi)) +B_{ij} (-x(\xi)) \right) \pl_- x^i (\xi) \pl_+ x^j (\xi) \nn
\end{align}
\begin{align}
&= \frac{1}{8\pi} \int \!\!dt \int_0^{2\pi} \!\!d\xi \left( G_{ij}
(-x(\xi)) +B_{ij} (-x(\xi)) \right) \pl_- x^i (\xi) \pl_+ x^j (\xi) \nn \\
&= \frac{1}{8\pi} \int \!\!dt \int_0^{2\pi} \!\!d\xi \left( G_{ij}
(-x(\xi)) -B_{ij} (-x(\xi)) \right) \pl_+ x^i (\xi) \pl_- x^j (\xi) \equiv
S
\end{align}
\end{subequations}
where we needed the {\it space-time parities}
\begin{gather}
G_{ij}(-x) =G_{ij}(x) ,\quad B_{ij}(-x) =-B_{ij}(x) \,\longrightarrow \,
H_{ijk} (-x) =H_{ijk}(x) \label{Eq5.4}
\end{gather}
in the last step to obtain the invariance $S' =S$. It follows that the
space-time parity \eqref{Eq5.4} is a necessary and sufficient condition
for invariance of the sigma model under the basic orientation-reversing
automorphism. In other words, orientation reversal is in the automorphism
group of the isometry group of the manifold M
\begin{gather}
\tau_1 \times \one \in Aut (\text{Iso }M)  \label{Eq5.5}
\end{gather}
iff M satisfies the space-time parities in Eq.~\eqref{Eq5.4}.

We recall that the space-time parity \eqref{Eq5.4} was verified explicitly in
Subsec.~$5.1$ for the special case of the general WZW model, and hence
follows as well
for any principal chiral model. We have also checked (see
App.~\ref{CosetApp}) that the space-time parity \eqref{Eq5.4} holds for
all reductive $g/h$ coset conformal field
theories. In what follows, our discussion is restricted to such
orientation-symmetric sigma models, although the classification of these
models is beyond the scope of this paper.

As discussed above for WZW, the following two-component form  of the sigma
model on the strip
\begin{subequations}
\label{Eq5.6}
\begin{gather}
S=\!\int \!\!dt \int_0^{\pi} \!\!\!d\xi {\cL} \\
{\cL} = \frac{1}{8\pi} \left( G_{i\Id ;j\Jd}(x) +B_{i\Id ;j\Jd}(x) \right)
\pl_+ x^{i\Id} \pl_- x^{j\Jd} \label{Eq 5.6b} \\
x^{i\Id} (\xi) \equiv (-1)^\Id x^i ((-1)^\Id \xi) ,\quad x^{i\Id}(\xi
+2\pi) =x^{i\Id}(\xi) \\
G_{i\Id ;j\Jd}(x) \equiv \de_{\Id \Jd} G_{ij} (x^\Id ) ,\quad B_{i\Id
;j\Jd}(x) \equiv \de_{\Id \Jd} B_{ij} (x^\Id) \\
H_{i\Id ;j\Jd ;k\Kd}(x) \equiv \pl_{i\Id} B_{j\Jd ;k\Kd}(x) +\text{ cyclic
} =\de_{\Id \Jd} \de_{\Jd \Kd} H_{ijk} (x^{\Id}) ,\quad \Id ,\Jd ,\Kd \in
\{0,1\} \,
\end{gather}
\end{subequations}
is an equivalent form of the sigma model if
and only if the space-time parity
\eqref{Eq5.4} holds.

Using the space-time parities in Eq.~\eqref{Eq5.4} and the definitions
 in Eq.~\eqref{Eq5.6},
 we also find the world-sheet parities of all the two-component fields of
the sigma
model \eqref{Eq 5.6b}:
\begin{subequations}
\label{Eq5.7}
\begin{gather}
x^{i\Id} (-\xi) =-x^{i\Jd} (\xi) (\tau_1 )_\Jd{}^\Id ,\quad \pl_{i\Id}
(-\xi) = -(\tau_1 )_\Id{}^\Jd \pl_{i\Jd} (\xi) \\
G_{i\Id ;j\Jd} (x(-\xi)) =(\tau_1 )_\Id{}^{\Kd} (\tau_1 )_\Jd{}^{\dot{L}}
G_{i\Kd ;j\dot{L}} (x(\xi)) \\
H_{i\Id ;j\Jd ;k\Kd} (x(-\xi)) =(\tau_1 )_\Id{}^{\dot{L}} (\tau_1
)_\Jd{}^{\dot{M}} (\tau_1 )_\Kd{}^{\dot{N}}
H_{i\dot{L};j\dot{M};k\dot{N}}(x(\xi)) \\
B_{i\Id ;j\Jd} (x(-\xi)) =-(\tau_1 )_\Id{}^{\Kd} (\tau_1 )_\Jd{}^{\dot{L}}
B_{i\Kd;j\dot{L}} (x(\xi))\,.
\end{gather}
\end{subequations}
These forms are the same as those found above (see Eq.~\eqref{Eq4.4}) for the
corresponding two-component WZW fields.

In the two-component notation of Eq.~\eqref{Eq5.6}, the basic
orientation-reversing automorphism \eqref{Eq 5.2b} can be written as
\begin{subequations}
\label{Eq5.8}
\begin{gather}
x^{i\Id}(\xi)' =x^{i\Jd}(\xi) (\tau_1 )_\Jd{}^\Id \\
G_{i\Id ;j\Jd}(x)' =G_{i\Id ;j\Jd} (x') = (\tau_1 )_\Id{}^{\Kd} (\tau_1
)_\Jd{}^{\dot{L}} G_{i\Kd;j\dot{L}}(x) \\
B_{i\Id ;j\Jd}(x)' =B_{i\Id ;j\Jd} (x') = (\tau_1 )_\Id{}^{\Kd} (\tau_1
)_\Jd{}^{\dot{L}} B_{i\Kd;j\dot{L}}(x) \\
H_{i\Id ;j\Jd ;k\Kd}(x)' =H_{i\Id ;j\Jd ;k\Kd} (x') = (\tau_1
)_\Id{}^{\dot{L}} (\tau_1 )_\Jd{}^{\dot{M}} (\tau_1 )_\Kd{}^{\dot{N}}
   H_{i\dot{L};j\dot{M};k\dot{N}}(x)
\end{gather}
\end{subequations}
and the invariance of the strip action \eqref{Eq5.6} under \eqref{Eq5.8}
is transparent.

We may further require that the more general orientation reversal
\begin{subequations}
\label{Eq5.9}
\begin{gather}
x^{i\Id}(\xi)' =x^{j\Jd}(\xi) \om\hc (h_\s)_j{}^i (\tau_1 )_\Jd{}^\Id \\
G_{i\Id ;j\Jd}(x)' =G_{i\Id ;j\Jd}(x') =\ws_i{}^{k} \ws_j{}^{l} (\tau_1
)_\Id{}^{\Kd} (\tau_1 )_\Jd{}^{\dot{L}} G_{k\Kd ;l\dot{L}}(x) \\
B_{i\Id ;j\Jd}(x)' =B_{i\Id ;j\Jd}(x') =\ws_i{}^{k} \ws_j{}^{l} (\tau_1
)_\Id{}^{\Kd} (\tau_1 )_\Jd{}^{\dot{L}} B_{k\Kd ;l\dot{L}}(x) \\
H_{i\Id ;j\Jd ;k\Kd}(x)' \!=H_{i\Id ;j\Jd ;k\Kd}(x') \bigspc \bigspc
\bigspc \bigspc \bigspc \nn \\
\!=\!\ws_i{}^l \ws_j{}^m \ws_k{}^n (\tau_1 )_\Id{}^{\!\dot{L}} (\tau_1
)_\Jd{}^{\!\dot{M}} (\tau_1 )_\Kd{}^{\!\dot{N}}
   H_{l\dot{L};m\dot{M};n\dot{N}}(x)
\end{gather}
\end{subequations}
is a symmetry $S'\!=\!S$ of the strip form \eqref{Eq5.6} of the sigma
model. In particular, this requirement includes the following
symmetry conditions on the original fields
\vspace{-0.1in}
\begin{subequations}
\label{Eq5.10}
\begin{gather}
G_{ij} (x\om\hc) =\om_i{}^k \om_j{}^l G_{kl}(x) ,\quad B_{ij} (x\om\hc) =\om_i{}^k
\om_j{}^l B_{kl}(x) \\
H_{ijk} (x\om\hc) =\om_i{}^l \om_j{}^m \om_k{}^n H_{lmn}(x)
\end{gather}
\end{subequations}
which describe the class of nonlinear sigma models with a {\it linear}
symmetry \cite{Geom}. Linearity of the symmetry conditions can be maintained
only in certain preferred coordinate systems: In conventional terms, we
limit ourselves to the special case of a symmetry $h_\s$ with one fixed
point of $M$, and the preferred coordinate system is that in which the
fixed point is at the origin. In a somewhat more general language
\begin{gather}
h_\s \in Aut(\text{Iso }M) \label{Eq5.11}
\end{gather}
and $h_\s$ must also preserve the 2-form $B$ on $M$.

When $M$ is a group manifold and $h_\s \!\in \!Aut(g)$, we already know
(see Eq.~\eqref{Eq4.16}) that the linear symmetry conditions
\eqref{Eq5.10}
are satisfied for the WZW model on $g$, and hence for any principal chiral
model on $g$. In the case of the coset CFTs, the symmetry conditions
\eqref{Eq5.10} will also be satisfied so long as the $g/h$ coset
construction \cite{BH,2faces1,GKO} is $h_\s$-invariant: This means in
particular that the
subalgebra
$h\subset g$ is an $h_\s$-covariant subalgebra of $g$, as discussed for
the $H$-invariant coset constructions in Refs.~\cite{Coset,More,Fab}.

When the space-time parities \eqref{Eq5.4} and the linear symmetry
conditions \eqref{Eq5.10} are satisfied for all
\begin{gather}
\hat{h}_\s =\tau_1 \!\times \!h_\s \in H_- \subset Aut (\text{Iso }M)
\label{Eq5.12}
\end{gather}
we will say that the sigma model is $H_-$-symmetric and denote it by
$A_M(H_-)$. The discussion below is limited to the sigma models in this
class.

\subsection{The Fields of the Sigma-Model Orientation Orbifolds}

The sigma-model
eigenfields $\sG, \sB, \sH$  are constructed as shown in Eqs.~($4.17$h-j),
but now with
the WZW eigenvector matrix $U(\s)$ replaced by the eigenvector matrix of
the {\it Einstein-space $H$-eigenvalue problem} \cite{Geom}
\begin{gather}
\ws_i{}^j U\hc (\s)_j{}^\nrm =U\hc(\s)_i{}^\nrm E_{n(r)}(\s) ,\quad
U\hc(\s) U(\s) =\one ,\quad E_{n(r)}(\s) =e^{-\tp \nrrsf}  \label{Eq5.13}
\end{gather}
where $\ws$ appears in the general orientation-reversing automorphism
\eqref{Eq5.9}.

To go to the open-string sectors of the sigma-model orientation orbifold
$A_M (H_-)/H_-$,
we then apply the principle of local isomorphisms
\begin{gather}
\sx \dual \hx ,\quad \sG \dual \hG ,\quad \sB \dual \hB ,\quad \sH \dual
\hh  \label{Eq5.14}
\end{gather}
as above. For the twisted fields of these open-string sectors, we find the
 two-component structure
\begin{subequations}
\label{Eq5.15}
\begin{gather}
\hx_\s^\nrmu \equiv \hx^\nrmu_\s (\xi,t) ,\quad \hpl_\nrmu (\xi)
=\frac{\pl}{\pl \hx^\nrmu (\xi)} ,\quad \bar{u} = 0,1 \\
\hG_{\nrmu;\nsnv}(\hx) =\hG_{\nrm;\nsn}^{(u+v)} (\hx) ,\quad
\hB_{\nrmu;\nsnv}(\hx) =\hB_{\nrm;\nsn}^{(u+v)} (\hx) \\
\hh_{\nrmu;\nsnv ;\ntd w}(\hx) \bigspc \bigspc \bigspc \bigspc \bigspc
\bigspc  \quad \quad \quad \nn \\
   =\hpl_\nrmu \hB_{\nsnv;\ntd w}(\hx) +\hpl_\nsnv \hB_{\ntd w;\nrmu}(\hx)
+\hpl_{\ntd w}\hB_{\nrmu ;\nsnv}(\hx) \nn \\
   =\hh_{\nrm;\nsn;\ntd}^{(u+v+w)} (\hx) \bigspc \bigspc \bigspc \bigspc
\end{gather}
\end{subequations}
and the world-sheet parities
\begin{subequations}
\label{Eq5.16}
\begin{gather}
\hx_\s^\nrmu (-\xi) =-\hx_\s^{\nrm v} (\xi) (\tau_3 )_v{}^u = (-1)^{u+1}
\hx_\s^\nrmu (\xi) \\
\hG_{\nrm;\nsn}^{(w)} (\hx(-\xi)) =(-1)^w \hG_{\nrm;\nsn}^{(w)} (\hx(\xi))
\\
\hB_{\nrm;\nsn}^{(w)} (\hx(-\xi)) =(-1)^{w+1} \hB_{\nrm;\nsn}^{(w)}
(\hx(\xi)) \\
\hh_{\nrm;\nsn;\ntd}^{(w)} (\hx(-\xi)) =(-1)^w \hh_{\nrm;\nsn;\ntd}^{(w)}
(\hx(\xi))\,.
\end{gather}
\end{subequations}
Both of these results follow by local isomorphisms from the corresponding
properties
 of the eigenfields.

To discuss the twisted fields in further detail, we again recall that the
commuting diagrams \cite{Dual,Geom} of orbifold theory  hold as well for the
orientation
orbifolds. We mention in particular the commuting diagram shown in
Fig.~\ref{fig:coords} for the Einstein coordinates of any orbifold.

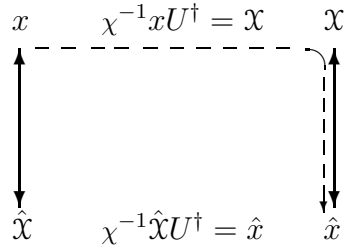
\begin{picture}(328,178)(0,0)
\put(143,165){$x$}
\put(152,158){\line(1,0){5}}
\put(173,158){\line(1,0){5}}
\put(193,158){\line(1,0){5}}
\put(213,158){\line(1,0){5}}
\put(233,158){\line(1,0){5}}
\put(162,158){\line(1,0){5}}
\put(182,158){\line(1,0){5}}
\put(203,158){\line(1,0){5}}
\put(223,158){\line(1,0){5}}
\put(243,158){\line(1,0){5}}
\put(254,150){\oval(15,15)[tr]}
\put(261,147){\line(0,-1){5}}
\put(261,139){\line(0,-1){5}}
\put(261,131){\line(0,-1){5}}
\put(261,123){\line(0,-1){5}}
\put(261,114){\line(0,-1){5}}
\put(261,106){\vector(0,-1){10}}
\put(177,165){$\schi^{-1}xU\hc = \sx$}
\thicklines
\put(146,158){\vector(0,-1){60}}
\put(146,98){\vector(0,1){60}}
\put(143,86){$\hat{\sx}$}
\put(261,165){$\sx$}
\put(177,86){$\schi^{-1}\hat{\sx}U\hc = \hx$}
\put(265,158){\vector(0,-1){60}}
\put(265,98){\vector(0,1){60}}
\put(261,86){$\hx$}
\put(104,70) {{\footnotesize Each vertical double arrow is a local
isomorphism }}
\put(93,59) {$\foot{x}$}
\put(104,59) {{\footnotesize = coordinates: mixed under automorphisms}}
\put(93,47) {$\sxfoot$}
\put(104,47) {{\footnotesize = eigencoordinates: diagonal under
automorphisms}}
\put(93,35) {$\foot{\hx}$}
\put(104,35) {{\footnotesize = twisted coordinates}}
\put(93,23) {${\hat{\sxfoot}}$}
\put(104,23) {{\footnotesize = coordinates with twisted boundary
conditions}}
\put(98,4) {Fig.\,\ref{fig:coords}: Coordinates and orbifold coordinates}
\end{picture}
\myfig{fig:coords}

\noindent The fields $\sxh$ in Fig.~\ref{fig:coords}, which are locally
isomorphic to the original untwisted Einstein coordinates
\begin{subequations}
\label{Eq5.17}
\begin{gather}
x^{i\Id} (\xi) \dual \sxh_\s^{i\Id} (\hx(\xi)) ,\quad \Id =0,1 \\
\sxh^{i\Id}_\s (\hx) \equiv \hx^\nrmu_\s \schisig_\nrm U(\s)_\nrm{}^i
(\sqrt{2} U_u{}^\Id ) \\
\hx_\s^\nrmu (\sxh) =\sxh_\s^{i\Id} \schisig_\nrm^{-1} U\hc(\s)_i{}^\nrm
(\srac{1}{\sqrt{2}} U\hc{}_\Id{}^u ) \\
\frac{\pl}{\pl \sxh^{i\Id}} =\schisig_\nrm^{-1} U\hc(\s)_i{}^\nrm
(\srac{1}{\sqrt{2}} U\hc{}_\Id{}^u ) \hpl_\nrmu \\
\hpl_\nrmu =\schisig_\nrm U(\s)_\nrm{}^i (\sqrt{2} U_u{}^\Id )
\frac{\pl}{\pl \sxh^{i\Id}}  \label{Eq 5.17e} \\
\sxh_\s^{i\Id} (\hx(-\xi)) =-\sxh_\s^{i\Jd} (\hx(\xi)) (\tau_1 )_\Jd{}^\Id
,\quad \sxh^{i\Id} (-\hx(\xi)) =-\sxh^{i\Id} (\hx(\xi))
\end{gather}
\end{subequations}
are generalizations of the Einstein coordinates with twisted boundary
conditions \cite{Geom} in space-time orbifold theory.

Using the diagram in Fig.~\ref{fig:coords} and following the construction
of Ref.~\cite{Geom}, we then obtain explicit formulas for the twisted
fields of the open-string orientation-orbifold sectors
\begin{subequations}
\label{Eq5.18}
\begin{gather}
\hG_{\nrmu;\nsnv} (\hx) =\!\schisig_\nrm \schisig_\nsn U(\s)_\nrm{}^i
U(\s)_\nsn{}^j \bigspc \bigspc \nn \\
\bigspc \bigspc \bigspc \times (\sqrt{2}U_u{}^\Id )(\sqrt{2}U_v{}^\Jd )
G_{i\Id ;j\Jd}(x \duals \sxh (\hx)) \\
\hB_{\nrmu ;\nsnv}(\hx) \!\equiv \!\schisig_\nrm \schisig_\nsn
U(\s)_\nrm{}^i U(\s)_\nsn{}^j \bigspc \bigspc \nn \\
\bigspc \bigspc \bigspc \times (\sqrt{2}U_u{}^\Id )(\sqrt{2}U_v{}^\Jd )
B_{i\Id ;j\Jd}(x \duals \sxh(\hx)) \\
\hh_{\nrmu ;\nsnv ;\ntd w}(\hx) \!\equiv \!\schisig_\nrm \schisig_\nsn
\schisig_\ntd U(\s)_\nrm{}^i U(\s)_\nsn{}^j U(\s)_\ntd{}^k \quad \nn \\
  \bigspc \bigspc \bigspc \times (\sqrt{2}U_u{}^\Id )(\sqrt{2}U_v{}^\Jd )
(\sqrt{2}U_w{}^\Kd ) H_{i\Id ;j\Jd ;k\Kd}(x \duals \sxh(\hx))
\end{gather}
\end{subequations}
in terms of the untwisted fields $G,B$, and $H$ of the symmetric sigma
model. If we insert the explicit untwisted forms of $G,B$ and $H$ given
in Eqs.~($4.7$c,d) for the special case of WZW, the general results in
Eq.~\eqref{Eq5.18} reduce to the explicit formulas ($4.29$c,d) given above
for the WZW orientation orbifolds.

In the general case, the explicit formulae \eqref{Eq5.18} can be further
evaluated as
\begin{subequations}
\label{Eq5.19}
\begin{gather}
\sxh_\s^{i\Id}(\hx) = (\hx_\s^{\nrm 0} +(-1)^{\Id} \hx_\s^{\nrm 1})
\schisig_\nrm U(\s)_\nrm{}^i  \label{Eq 5.19a}
\end{gather}
\begin{align}
& \bigspc \bigspc \bigspc \hG_{\nrmu;\nsnv}(\hx)
=\hG^{(u+v)}_{\nrm;\nsn}(\hx) \\
&\hG^{(w)}_{\nrm;\nsn}(\hx) \nn \\
&\bigspc = \schi_\nrm \schi_\nsn U_\nrm{}^i U_\nsn{}^j \left( G_{ij}(x^0
\duals \sxh_\s^0(\hx)) +(-1)^w G_{ij} (x^1 \duals \sxh_\s^1 (\hx))
  \right)
\end{align}
\begin{align}
&\bigspc \bigspc \bigspc \hB_{\nrmu;\nsnv}(\hx)
=\hB^{(u+v)}_{\nrm;\nsn}(\hx) \\
&\hB^{(w)}_{\nrm;\nsn}(\hx)  \nn \\
&\bigspc  = \schi_\nrm \schi_\nsn U_\nrm{}^i U_\nsn{}^j \left( B_{ij}(x^0
\duals \sxh_\s^0(\hx)) +(-1)^w B_{ij} (x^1 \duals \sxh_\s^1 (\hx))
  \right)
\end{align}
\begin{gather}
\hh_{\nrmu;\nsnv;\ntd w}(\hx) =\hh^{(u+v+w)}_{\nrm;\nsn}(\hx) \\
\hh^{(w)}_{\nrm;\nsn;\ntd}(\hx) = \schi_\nrm \schi_\nsn \schi_\ntd
U_\nrm{}^i U_\nsn{}^j U_\ntd{}^k \bigspc \bigspc \quad \quad  \nn \\
\bigspc \bigspc \times \left( H_{ijk}(x^0 \duals \sxh_\s^0(\hx)) +(-1)^w
H_{ijk} (x^1 \duals \sxh_\s^1 (\hx)) \right)
\end{gather}
\end{subequations}
where we have suppressed some $\s$-dependence. These results are easily
obtained from the forms in Eq.~\eqref{Eq5.18} by doing the sums on
the internal two-dimensional indices.

\subsection{The Sigma-Model Orientation-Orbifold Action}

The principle of local isomorphisms also gives us the {\it sigma-model
orientation-orbifold action}
\begin{subequations}
\label{Eq5.20}
\begin{gather}
S \dual \hat{S}_\s \\
\hat{S}_\s = \int \!\!dt \int_0^\pi \!\! d\xi \, \hat{{\cL}}_\s \\
\,\,\hat{{\cL}}_\s =\frac{1}{8\pi} \left( \hG_{\nrmu;\nsnv} (\hx)
+\hB_{\nrmu;\nsnv}(\hx) \right) \pl_+ \hx_\s^\nrmu \pl_- \hx_\s^\nsnv
\bigspc \nn \\
\bigspc =\frac{1}{8\pi} \sum_{w=0}^1 \left( \hG_{\nrm;\nsn}^{(w)}(\hx)
+\hB_{\nrm;\nsn}^{(w)}(\hx) \right) \sum_{u=0}^1 \pl_+ \hx_\s^\nrmu \pl_-
\hx_\s^{\nsn ,w-u} \, \\
\hat{{\cL}}_\s (-\xi) =\hat{{\cL}}_\s (\xi) \label{Eq 5.20d}
\end{gather}
\end{subequations}
by the standard derived isomorphism from the eigenfield form of $S$. This
action, which describes all open-string sectors
\begin{gather}
\hat{h}_\s = \tau_1 \!\times \!h_\s \in H_- ,\quad H_- \subset Aut
(\text{Iso }M) \label{Eq5.21}
\end{gather}
of the sigma-model orientation orbifold $A_M (H_-)/H_-$, is another central
result of
this paper.

For this action, the following {\it variational boundary conditions} which
describe the sigma-model orientation-orbifold branes
\begin{align}
& \de \hx^\nrmu (\xi) \left( \hG_{\nrmu ;\nsnv} (\hx(\xi)) \pl_\xi -
\hB_{\nrmu ;\nsnv} (\hx(\xi)) \pl_t \right) \hx^\nsnv (\xi) \bigspc \quad
\nn \\
& \quad \quad =\!\sum_{u,v=0}^1 \de \hx^\nrmu (\xi) \left(
\hG_{\nrm;\nsn}^{(u+v)} (\hx(\xi)) \pl_\xi \!- \!\hB_{\nrm ;\nsn}^{(u+v)}
(\hx(\xi)) \pl_t \right) \hx^\nsnv (\xi) \nn \\
& \quad \quad =0 \text{  at } \xi =0,\pi \label{Eq5.22}
\end{align}
are obtained in the standard manner by requiring that the bulk and
boundary terms cancel separately in a general variation of the fields. The
more explicit form of these boundary conditions
\begin{gather}
\Big{[} \de \hx^{\nrm 0} (\xi) \left( \hG_{\nrm;\nsn}^{(0)} (\hx(\xi))
\pl_\xi -\hB_{\nrm;\nsn}^{(0)}(\hx(\xi)) \pl_t \right) \hx^{\nsn 0}(\xi) +
\bigspc \bigspc \nn \\
\de \hx^{\nrm 0} (\xi) \left( \hG_{\nrm;\nsn}^{(1)} (\hx(\xi)) \pl_\xi
-\hB_{\nrm;\nsn}^{(1)} (\hx(\xi)) \pl_t \right) \hx^{\nsn 1}(\xi) +
\bigspc \bigspc  \nn \\
\de \hx^{\nrm 1} (\xi) \left( \hG_{\nrm;\nsn}^{(1)} (\hx(\xi)) \pl_\xi
-\hB_{\nrm;\nsn}^{(1)} (\hx(\xi)) \pl_t \right) \hx^{\nsn 0}(\xi) +
\bigspc \bigspc \nn \\
\de \hx^{\nrm 1} (\xi) \left( \hG_{\nrm;\nsn}^{(0)} (\hx(\xi)) \pl_\xi
-\hB_{\nrm;\nsn}^{(0)} (\hx(\xi)) \pl_t \right) \hx^{\nsn 1}(\xi) \Big{]}
=0 \text{  at }
  \xi =0,\pi \label{Eq5.23}
\end{gather}
will be useful in the following subsection.

The classical bulk equations of motion for the action \eqref{Eq5.20} are
given in Sec.~$6$, where we develop the necessary twisted Christoffel
symbols.

We finally mention the untwisted two-component interval $ds^2$ and the
orientation orbifold interval $d\hat{s}^2$
\begin{subequations}
\label{Eq5.24}
\begin{align}
ds^2 (\xi) &=G_{i\Id ;j\Jd}(x(\xi)) dx^{i\Id}(\xi) dx^{j\Jd} (\xi) \nn \\
&= G_{ij}(x(\xi)) dx^i (\xi) dx^j (\xi) +(\xi \rightarrow -\xi ) \label{Eq
5.24a} \\
d\hat{s}^2 (\xi) &= \hG_{\nrmu ;\nsnv} (\hx(\xi)) d\hx^\nrmu (\xi)
d\hx^\nsnv (\xi) \nn \\
&= \sum_{w=0}^1 \hG_{\nrm;\nsn}^{(w)} (\hx(\xi)) \sum_{u=0}^1 d\hx^\nrmu
(\xi) d\hx^{\nsn ,w-u}(\xi)
\end{align}
\end{subequations}
where the space-time parity \eqref{Eq5.4} was used to obtain \eqref{Eq
5.24a}. Both of these intervals are even under world-sheet parity
$\xi \leftrightarrow -\xi$.

\subsection{Monodromies and Extra Boundary Conditions}

In this subsection, we consider further structure of the sigma-model
orientation-orbifold branes, concentrating primarily but not exclusively
on the case
$h_\s^2 =1$.

For this discussion, we will need the $h_\s^2 =1$ monodromies
\begin{subequations}
\label{Eq5.25}
\begin{gather}
h_\s^2=1 \!: \quad \quad \text{automorphic responses } \dual \text{
monodromies} \quad \quad \quad \\
\hx_\s^\nrmu (\xi+2\pi)= \hx_\s^\nrmu (\xi) e^{\tp (\nrrsf +\frac{u}{2})}
,\quad \bar{u}=0,1 \\
\hG_{\nrm;\nsn}^{(w)} (\hx(\xi+2\pi)) =e^{-\tp (\frac{n(r)+n(s)}{\r(\s)}
+\frac{w}{2})} \hG_{\nrm;\nsn}^{(w)} (\hx(\xi)) \\
\hB_{\nrm;\nsn}^{(w)} (\hx(\xi+2\pi)) =e^{-\tp (\frac{n(r)+n(s)}{\r(\s)}
+\frac{w}{2})} \hB_{\nrm;\nsn}^{(w)} (\hx(\xi)) \\
\hh_{\nrm;\nsn;\ntd}^{(w)} (\hx(\xi+2\pi)) =e^{-\tp
(\frac{n(r)+n(s)+n(t)}{\r(\s)} +\frac{w}{2})} \hh_{\nrm;\nsn;\ntd}^{(w)}
(\hx(\xi)) \\
\hat{{\cL}}_\s (\xi+2\pi) =\hat{{\cL}}_\s (\xi)
\end{gather}
\end{subequations}
which follow in this case by local isomorphisms from the automorphic
responses of the eigenfields. These monodromies have the same forms as
those
found for WZW in Eqs.~\eqref{Eq4.35} and \eqref{Eq4.28}. We note in
particular that the
coordinates $\sxh$ with twisted boundary conditions
in Eq.~\eqref{Eq5.17} have mixed monodromy
\begin{gather}
\sxh_\s^{i\Id} (\hx(\xi+2\pi)) =\sxh_\s^{j\Jd}(\hx(\xi)) \om\hc(h_\s)_j{}^i
(\tau_1 )_\Jd{}^\Id \label{Eq5.26}
\end{gather}
and our coordinates $\hx$ with definite monodromy in Eq.~\eqref{Eq5.25}
are the monodromy decomposition of $\sxh$.

Using the world-sheet parities \eqref{Eq5.16} and the monodromies
\eqref{Eq5.25}, one finds the brane substructure
\begin{subequations}
\label{Eq5.27}
\begin{gather}
\hx_\s^{\nrm 0}(0)=0 ,\quad \hx_\s^\nrmu (\pi) =0 \text{ unless } \nrrs
\in \Zint +\half \\(\pl_t \hx_\s^{\nrm 0})(0)=0 ,\quad (\pl_t \hx_\s^\nrmu
)(\pi) =0 \text{
unless } \nrrs \in \Zint +\half \\
(\pl_\xi \hx_\s^{\nrm 1})(0)=0 ,\quad (\pl_\xi \hx_\s^\nrmu )(\pi) =0
\text{ unless } \nrrs \in \Zint
\end{gather}\vspace{+0.02in}
\begin{gather}
\hG^{(1)}_{\nrm;\nsn}(\hx(0)) =0 ,\quad \hG^{(w)}_{\nrm;\nsn}(\hx(\pi))=0
\text{ unless } \srac{n(r)+n(s)}{\r(\s)} \in \Zint \label{Eq 5.27d} \\
(\pl_t \hG^{(1)}_{\nrm;\nsn})(\hx(0)) =0 ,\quad (\pl_t
\hG^{(w)}_{\nrm;\nsn})(\hx(\pi))=0 \text{ unless }
\srac{n(r)+n(s)}{\r(\s)} \in \Zint \\
(\pl_\xi \hG^{(0)}_{\nrm;\nsn} )(\hx(0)) =0 ,\quad (\pl_\xi
\hG^{(w)}_{\nrm;\nsn})(\hx(\pi))=0 \text{ unless }
\srac{n(r)+n(s)}{\r(\s)} \in \Zint +\half
\end{gather}
\begin{gather}
\hB^{(0)}_{\nrm;\nsn}(\hx(0)) =0 ,\quad \hB^{(w)}_{\nrm;\nsn}(\hx(\pi))=0
\text{ unless } \srac{n(r)+n(s)}{\r(\s)} \in \Zint +\half \label{Eq 5.27g}
\\
(\pl_t \hB^{(0)}_{\nrm;\nsn})(\hx(0)) =0 ,\quad (\pl_t
\hB^{(w)}_{\nrm;\nsn})(\hx(\pi))=0 \text{ unless }
\srac{n(r)+n(s)}{\r(\s)} \in \Zint +\half \\
(\pl_\xi \hB^{(1)}_{\nrm;\nsn})(\hx(0)) =0 ,\quad (\pl_\xi
\hB^{(w)}_{\nrm;\nsn})(\hx(\pi))=0 \text{ unless }
\srac{n(r)+n(s)}{\r(\s)} \in \Zint
\end{gather}
\begin{gather}
\hh^{(1)}_{\nrm;\nsn;\ntd}(\hx(0)) \!=\!0 ,\quad
\hh^{(w)}_{\nrm;\nsn;\ntd}(\hx(\pi))\!=\!0 \text{ unless }
\srac{n(r)+n(s)+n(t)}{\r(\s)} \!\in \!\Zint \\
(\pl_t \hh^{(1)}_{\nrm;\nsn;\ntd})(\hx(0)) =0 \\
(\pl_t \hh^{(w)}_{\nrm;\nsn;\ntd})(\hx(\pi)) =0 \text{ unless }
\srac{n(r)+n(s)+n(t)}{\r(\s)} \in \Zint \\
(\pl_\xi \hh^{(0)}_{\nrm;\nsn;\ntd})(\hx(0)) =0 \\
(\pl_\xi \hh^{(w)}_{\nrm;\nsn;\ntd})(\hx(\pi)) =0 \text{ unless }
\srac{n(r)+n(s)+n(t)}{\r(\s)} \in \Zint +\half
\end{gather}
\end{subequations}
which has the same form as that found in Subsec.~$4.6$ for the special
case of the WZW orientation orbifolds. Note in particular that the
coordinates of the
basic sector with $h_\s =1$ are Dirichlet-Dirichlet and Neumann-Dirichlet
for $\hx^{0\m 0}_\s$ and $\hx^{0\m 1}_\s$ respectively, while the generic
sector with $h_\s^2 \!=\!1 ,\,h_\s \!\neq \!1$ contains all four
coordinate types, D-D, N-D, D-N and N-N\footnote{Half-integer moded scalar
fields \cite{2faces2}
and the corresponding twisted open strings with D-N or N-D boundary
conditions \cite{CF,WS} provided the first examples of twisted sectors of
orbifolds.}.

We turn next to explicitly verify the variational boundary conditions
\eqref{Eq5.22},\eqref{Eq5.23} given in the previous subsection, using the
$\hx ,\hG$ and
$\hB$ boundary
conditions above. In particular,  the
boundary conditions ($5.27$a-c) on $\hx$ allow us to reduce the variational
boundary condition \eqref{Eq5.23} at $\xi =0$ to the form
\begin{align}
&\sum_{u,v=0}^1 \de \hx^\nrmu (\xi) \left( \hG_{\nrm;\nsn}^{(u+v)}(\hx(\xi)) \pl_\xi
\!-\!\hB^{(u+v)}_{\nrm;\nsn}(\hx(\xi)) \pl_t \right) \hx^\nsnv (\xi) =\bigspc \label{Eq5.28} \\
\!&\!\!\de \hx^{\nrm 1} (\xi) \left( \hG_{\nrm;\nsn}^{(1)}(\hx(\xi)) \pl_\xi \hx^{\nsn 0}(\xi) 
\!-\!\hB_{\nrm;\nsn}^{(0)} (\hx(\xi)) \pl_t \hx^{\nsn
1}(\xi) \right) \text{  at } \xi =0 \,. \nn
\end{align}
and this quantity vanishes by the boundary conditions of $\hG$ and
$\hB$ given in Eqs.~\eqref{Eq 5.27d} and \eqref{Eq 5.27g}.

To study $\xi \!=\!\pi$, we consider first the generic $h_\s^2 \!=\!1$ sector
with $\r(\s)\!=\!2$ in Eq.~\eqref{Eq 2.30b}. In this case, we instead
expand the sum \eqref{Eq5.22}  in terms of the
spectral indices $\bar{n}(r),\bar{n}(s) \in \{0,1\}$:
\begin{align}
&\sum_{u,v=0}^1 \Big{[} \de \hx^{0\m u} (\xi) \left( \hG^{(u+v)}_{0\m
;0\n} (\hx(\xi)) \pl_\xi - \hB^{(u+v)}_{0\m ;0\n} (\hx(\xi)) \pl_t \right)
\hx^{0\n v}(\xi) + \bigspc \quad \quad \nn \\
& \quad \quad \quad \de \hx^{0\m u} (\xi) \left( \hG^{(u+v)}_{0\m ;1\n}
(\hx(\xi)) \pl_\xi - \hB^{(u+v)}_{0\m ;1\n} (\hx(\xi)) \pl_t \right)
\hx^{1\n v}(\xi) + \nn \\
& \quad \quad \quad \de \hx^{1\m u} (\xi) \left( \hG^{(u+v)}_{1\m ;0\n}
(\hx(\xi)) \pl_\xi - \hB^{(u+v)}_{1\m ;0\n} (\hx(\xi)) \pl_t \right)
\hx^{0\n v}(\xi) + \nn \\
& \quad \quad \quad \de \hx^{1\m u} (\xi) \left( \hG^{(u+v)}_{1\m ;1\n}
(\hx(\xi)) \pl_\xi - \hB^{(u+v)}_{1\m ;1\n} (\hx(\xi)) \pl_t \right)
\hx^{1\n v}(\xi) \Big{]} \,. \label{Eq5.29}
\end{align}
With this expansion and the explicit form of the boundary conditions
\begin{subequations}
\label{Eq5.30}
\begin{gather}
\hx^{0\m u}(\pi) =\de \hx^{0\m u}(\pi) =\pl_t \hx^{0\m u}(\pi) =0
\label{Eq 5.30a} \\
\pl_\xi \hx^{1\m u}(\pi) =\hG^{(u+v)}_{1\m ;0\n}(\hx(\pi))
=\hB^{(u+v)}_{1\m ;1\n}(\hx(\pi)) =0
\end{gather}
\end{subequations}
it is straightforward to check that this quantity vanishes at $\xi =\pi$.
For the basic open-string sector \eqref{Eq 2.30a} with $h_\s \!=\!\r(\s)
\!=\!1$, the same conclusion is reached
\begin{gather}
\hx^{0\m u}(\pi) =\de \hx^{0\m u}(\pi) =\pl_t \hx^{0\m u}(\pi) =0 \bigspc
\bigspc \bigspc \nn \\
\quad \Rightarrow \,\de \hx^{0\m u}(\pi) \left( \hG^{(u+v)}_{0\m
;0\n}(\hx(\pi)) \pl_\xi -\hB^{(u+v)}_{0\m ;0\n}(\hx(\pi)) \pl_t \right)
\hx^{0\n v}(\pi) =0 \label{Eq5.31}
\end{gather}
because only the $\bar{n}(r)=0$ terms survive in this case.

We finally emphasize that, as usual, all the boundary conditions at $\xi
=0$ in this subsection follow directly from world-sheet parity, and hence
hold for all $\hat{h}_\s$.

\subsection{Example: The Coset Orientation-Orbifold Action}

We give here a brief sketch of our results for coset orientation
orbifolds, following the discussion of ordinary coset orbifolds in
Refs.~\cite{Coset,More,Fab,Geom}.

To build the coset orientation orbifolds $A_{g/h}(H_-)/H_-$ at the
operator level, one begins with a left- and right-mover copy of any $g/h$
coset construction \cite{BH,2faces1,GKO, ICFT}
\begin{subequations}
\label{Eq5.32}
\begin{gather}
J_h \subset J_g ,\quad T_{g/h} =T_g -T_h \\
\bJ_h \subset \bJ_g ,\quad \bT_{g/h} =\bT_g -\bT_h \\
c_{g/h} =\bar{c}_{g/h} =c_g -c_h
\end{gather}
\end{subequations}
where $T_g$ and $T_h$ are the affine-Sugawara constructions [6,7,36-38,12] of $g\supset h$. For our
construction, we will assume that $G/H$ is a reductive coset space and that the coset construction is
$h_\s$-invariant:
\begin{gather}
\frac{g}{h} =\frac{g}{h} (h_\s) ,\quad h_\s \in Aut(g) ,\quad h_\s \in
Aut(h) \,. \label{Eq5.33}
\end{gather}
As studied in the orbifold program under the rubric of {\it $H$-invariant
coset constructions} \cite{Coset,More,Fab,Geom}, $h_\s$-invariance of the coset
construction means that $h_\s$ is an automorphism of $g$, and $h\subset g$
transforms covariantly under $h_\s$. Each such system is then
automatically invariant under the general orientation-reversing
automorphism $\hat{h}_\s =\tau_1 \!\times \!h_\s$, in agreement with the
results of App.~B. Then the method of
eigenfields and the principle of local isomorphisms gives the twisted
currents and stress tensors of sector $\hat{h}_\s$
\begin{gather}
\hj_{\hfrakh_O (\s)} \subset \hj_{\sgb_O (\s)} ,\quad
\hat{T}_{\sgb_O(\s)/\hfrakh_O(\s)} =\hat{T}_{\sgb_O (\s)}
-\hat{T}_{\hfrakh_O (\s)} ,\quad
  \hat{c}_{\sgb_O (\s)/\hfrakh_O (\s)} =2c_{g/h} \label{Eq5.34}
\end{gather}
as well as the half-integrally moded Virasoro generators of
Refs.~\cite{Chr,Orient1}. The individual stress tensors $\hat{T}_{\sgb_O(\s)}$
and
$\hat{T}_{\hfrakh_O(\s)}$ are the twisted affine-Sugawara constructions
 \cite{Orient1} associated to the twisted current algebras $\gfrakh_O (\s)
\supset \hfrakh_O (\s)$ of the corresponding WZW orientation orbifolds.

At the classical level, we begin with the general gauged WZW model
[39-43,23]
\begin{subequations}
\label{Eq5.35}
\begin{gather}
S_{g/h} [M,g, A_\pm ;\Gamma] =S_{WZW} [M,g;\Gamma] +\frac{1}{4\pi} \!\int
\!\!dt \int_0^{2\pi} \!\!d\xi Tr \Big{(} M\Big{(} g^{-1}\pl_+ g (iA_-) \nn
\\
  \bigspc \bigspc -iA_+ \pl_- g g^{-1} -g^{-1} A_+ gA_- +A_+ A_- \Big{)}
\Big{)} \\
A_+ =-ih_+^{-1} \pl_+ h_+ ,\quad A_- =-ih_- \pl_- h_-^{-1}
\end{gather}
\end{subequations}
for the $g/h$ coset construction, where $M$ is the data matrix in
Eq.~\eqref{Eq2.4}. In agreement with the operator argument above, this
action is invariant under the general orientation-reversing automorphism
$\hat{h}_\s =\tau_1 \!\times \!h_\s$
\begin{subequations}
\label{Eq5.36}
\begin{gather}
A_\pm (\xi)' =W(h_\s;T) A_\mp (-\xi) W\hc (h_\s;T) \\
h_\pm (\xi)' =W(h_\s;T) h_\mp^{-1} (-\xi) W\hc (h_\s;T) \\
S_{g/h} [M,g' ,A_{\pm}' ;\Gamma] =S_{g/h} [M,g,A_\pm ;\Gamma]
\end{gather}
\end{subequations}
when the $g/h$ coset construction is $h_\s$-invariant. Here $W$ is the
action of $h_\s$ in rep $T$ and $g'$ is given in Eq.~\eqref{Eq 2.8a}.

In this case, the appropriate two-component gauge fields are:
\begin{subequations}
\label{Eq5.37}
\begin{gather}
\tilde{A}_\pm (\xi) \equiv \left( \begin{array}{cc} A_\pm (\xi) & 0 \\ 0 &
A_\mp (-\xi) \end{array} \right) ,\quad \tilde{h}_\pm (\xi) \equiv \left(
\begin{array}{cc} h_\pm (\xi) & 0 \\ 0 & h_\mp^{-1} (-\xi) \end{array}
\right) \\
\tilde{A}_+ (\xi) =-i \tilde{h}_+^{-1} \pl_+ \tilde{h}_+ ,\quad
\tilde{A}_- (\xi) =-i \tilde{h}_- \pl_- \tilde{h}_-^{-1} \\
\tilde{A}_\pm (-\xi) =\tau_1 \tilde{A}_\mp (\xi) \tau_1 ,\quad
\tilde{h}_\pm (-\xi) =\tau_1 \tilde{h}_\mp (\xi) \tau_1 \\
\tilde{A}_\pm (\xi)' =\tau_1 W(h_\s;T) \tilde{A}_\pm (\xi) W\hc (h_\s;T)
\tau_1 \\
\tilde{h}_\pm (\xi)' =\tau_1 W(h_\s;T) \tilde{h}_\pm (\xi) W\hc (h_\s;T)
\tau_1 \,.
\end{gather}
\end{subequations}
A two-component gauge transformation matrix $\tilde{\psi}(\xi)$ is also
necessary
\begin{gather}
\tilde{\psi}(\xi) \equiv \left( \begin{array}{cc} \psi (\xi) &0 \\0& \psi
(-\xi) \end{array} \right) ,\quad
\tilde{A}_\pm \rightarrow \tilde{A}_\pm^{\tilde{\psi}} =\tilde{\psi}
\tilde{A}_\pm \tilde{\psi}^{-1} +i\pl_\pm \tilde{\psi} \psi^{-1}
\label{Eq5.38}
\end{gather}
to put the standard gauge transformation of $A_\pm$ into two-component
form. The two-component form of the gauged action on $\Gamma_{1/2}$ is
then easily worked out
\begin{subequations}
\label{Eq5.39}
\begin{align}
&S_{g/h} [M \!\otimes \!\one_2 ,\tg ,\tilde{A}_\pm ;\Gamma_\half] \!\equiv
\!S_{WZW} [M \!\otimes \!\one_2 ,\tg ;\Gamma_\half ] + \nn \\
&\,\,\, +\frac{1}{4\pi} \!\int \!\!dt \!\!\int_0^\pi \!\!\!d\xi
\,\widehat{Tr} \Big{(} (M \!\otimes \!\one_2) \Big{(} \tg^{-1} \pl_+ \tg
(i\tilde{A}_-) \!-\!i\tilde{A}_+ \pl_- \tg \tg^{-1} -\!\tg^{-1}
\tilde{A}_+ \tg \tilde{A}_- \!+\!\tilde{A}_+ \tilde{A}_- \Big{)} \Big{)} \\
&\quad \quad =S_{g/h} [M,g,A_\pm ;\Gamma ]
\end{align}
\end{subequations}
where the WZW action on $\Gamma_{1/2}$ is given in Eq.~\eqref{Eq3.4}.

The extra eigenfields $\sA_\pm, \hfrak_\pm$ and $\Psi$ associated with the
gauging
\begin{subequations}
\label{Eq5.40}
\begin{gather}
\Ord (\st,\xi,\s) \equiv UU(T,\s) \tilde{O} (T,\xi) U\hc (T,\s) U\hc \\
 \Ord = \sg,\sA_\pm ,\hfrak_\pm ,\Psi ,\quad \tilde{O} =\tg ,\tilde{A}_\pm
,\tilde{h}_\pm ,\tilde{\psi}
\end{gather}
\end{subequations}
are constructed in analogy to the eigengroup elements in Eq.~\eqref{Eq 2.16a}.
The usual structure of these eigenfields is easily worked out, but we omit
the details
for brevity. Then the principle of local isomorphisms
\begin{gather}
\sg \dual \hg ,\quad \sA_\pm \dual \hat{A}_\pm ,\quad \hfrak_\pm \dual
\hat{h}_\pm ,\quad \Psi \dual \hat{\psi} \label{Eq5.41}
\end{gather}
gives the following results in the open-string sectors of the general
coset orientation orbifold.

In the first place, we record the twisted Polyakov-Weigmann identity
\cite{Fab} on
$\Gamma_{1/2}$
\begin{gather}
\hat{S}_{\hg_O (\s)} [\sm \otimes \one_2 ,\hg \hat{h} ;\Gamma_{\half}]
=\hat{S}_{\hg_O(\s)} [\sm \otimes \one_2 ,\hg ;\Gamma_{\half}]
+\hat{S}_{\hg_O(\s)}
   [\sm \otimes \one_2 ,\hat{h};\Gamma_{\half}] \bigspc \nn \\
\bigspc \bigspc -\frac{1}{4\pi} \int \!\!dt \int_0^\pi \!\! d\xi \,
\widehat{Tr} \left( (\sm \otimes \one_2 ) \hg^{-1} \pl_+\hg \pl_-\hat{h}
\hat{h}^{-1} \right)  \label{Eq5.42}
\end{gather}
where the WZW orientation-orbifold action $\hat{S}_{\hg_O (\s)}$ on
$\Gamma_{1/2}$ is given in Eq.~\eqref{Eq3.9}. We also obtain the {\it general
 coset orientation-orbifold action}
\begin{subequations}
\label{Eq5.43}
\begin{align}
&\hat{S}_{\hg_O (\s)/\hat{h}_O(\s)} [\hg,\hat{A}_\pm ;\Gamma_{\half}]
\equiv \hat{S}_{\hg_O(\s)} [\sm \otimes \one_2 ,\hg ;\Gamma_{\half}]
\bigspc \bigspc \bigspc
  \bigspc \quad \quad \nn \\
&\quad +\! \frac{1}{4\pi} \!\int \!\!dt \!\int_0^\pi \!\!\! d\xi \,
\widehat{Tr} \left( (\sm \!\otimes \!\one_2 ) ( \hg^{-1} \pl_+ \hg
(i\hat{A}_- )
  \!-\!i\hat{A}_+ \pl_- \hg \hg^{-1} \!-\!\hg^{-1} \hat{A}_+ \hg \hat{A}_-
\!+\!\hat{A}_+ \hat{A}_- ) \right)
\end{align}
\begin{gather}
\hat{A}_+ \equiv -i \hat{h}_+^{-1} \pl_+ \hat{h}_+ ,\quad \hat{A}_- \equiv
-i \hat{h}_- \pl_- \hat{h}_-^{-1} \\
\hg \in \widehat{\mathfrak{G}}_O (\s) ,\quad \hat{h}_{\pm} \in
\widehat{\mathfrak{H}}_O (\s) \subset \widehat{\mathfrak{G}}_O (\s) ,\quad
\hat{A}_{\pm}
   \in \hat{h}_O (\s) \subset \hg_O (\s) \\
[\sm \otimes \one_2 ,\hg] =[\sm \otimes \one_2 ,\hat{h}_\pm ]=[ \sm
\otimes \one_2 ,\hat{A}_\pm ]=0
\end{gather}
\end{subequations}
which describes open-string sector $\hat{h}_\s$ of the general coset
orientation orbifold $A_{g/h}(H_-)/H_-$. Here
$\widehat{\mathfrak{G}}_O(\s)$ is the group formed by the set of all group
orbifold elements $\hg$ (exponentiated from $\hg_O(\s)$), the quantities
$\hat{h}_{\pm}$ are arbitrary subgroup orbifold elements in the subgroup
$\widehat{\mathfrak{H}}_O (\s)$ (exponentiated from $\hat{h}_O(\s)$) and
$\hat{A}_{\pm}$ are the {\it orientation-orbifold
matrix gauge fields}.

The coset orientation-orbifold action \eqref{Eq5.43} is a gauging of the
WZW orientation-orbifold action \eqref{Eq3.9} by a general
twisted vector gauge group
\begin{subequations}
\label{Eq5.44}
\begin{align}
\hg (\st,\xi,t,\s) \rightarrow \hg(\st,\xi,t,\s)^{\hat{\psi}} &\equiv
\hat{\psi} (\st,\xi,t,\s) \hg(\st,\xi,t,\s) \hat{\psi}^{-1}(\st,\xi,t,\s)
\\
\hat{h}_+(\st,\xi,t,\s) \rightarrow \hat{h}_+(\st,\xi,t,\s)^{\hat{\psi}}
&\equiv \hat{h}_+(\st,\xi,t,\s) \hat{\psi}^{-1}(\st,\xi,t,\s)  \\
\hat{h}_-(\st,\xi,t,\s) \rightarrow \hat{h}_-(\st,\xi,t,\s)^{\hat{\psi}}
&\equiv  \hat{\psi}(\st,\xi,t,\s) \hat{h}_-(\st,\xi,t,\s) \\
\hat{A}_\pm (\st,\xi,t,\s) \rightarrow \hat{A}_\pm
(\st,\xi,t,\s)^{\hat{\psi}} & \equiv \hat{\psi}(\st,\xi,t,\s) \hat{A}_\pm
(\st,\xi,t,\s)
  \hat{\psi}^{-1} (\st,\xi,t,\s) \nn \\
& \quad \quad \quad +i\pl_\pm \hat{\psi}(\st,\xi,t,\s) \hat{\psi}^{-1}
(\st,\xi,t,\s)
\end{align}
\begin{gather}
[\sm \otimes \one_2,\hat{\psi}(\st,\xi,t,\s)]  =0 \label{Eq 5.44e} \\
\hat{S}_{\hg_O(\s)/\hat{h}_O(\s)}[\sm \otimes \one_2 , \hg^{\hat{\psi}},
\hat{h}_{\pm}^{\hat{\psi}} ;\Gamma_{\half}] =
\hat{S}_{\hg_O(\s)/\hat{h}_O(\s)}
   [\sm \otimes \one_2, \hg, \hat{h}_{\pm} ;\Gamma_{\half}] \label{Eq
5.44f}
\end{gather}
\end{subequations}
where the twisted gauge transformation matrix $\hat{\psi} \in
\widehat{\mathfrak{H}}_O (\s)$ is any subgroup orbifold element. As seen
in Eq.~\eqref{Eq 5.44f}, the  general coset orientation-orbifold action is
invariant under the general twisted vector gauge transformation.

For the action of world-sheet parity on the twisted gauge fields and the
twisted gauge
transformation matrix, we find
\begin{subequations}
\label{Eq5.45}
\begin{gather}
\hat{A}_\pm (\st,-\xi) =\tau_3 \hat{A}_\mp (\st,\xi) \tau_3 ,\quad
\hat{h}_\pm (\st,-\xi) =\tau_3 \hat{h}_\mp^{-1} (\st,\xi) \tau_3 \\
\hat{\psi} (\st, -\xi) =\tau_3 \hat{\psi} (\st,\xi) \tau_3
\end{gather}
\end{subequations}
and each of the new twisted fields commutes with $\tau_1$, as usual. These
commutators tell us that the new fields also have only two independent
components in the two-dimensional space, for example:
\begin{subequations}
\label{Eq5.46}
\begin{gather}
\hat{A}_\pm (\st,\xi,t,\s) =\left( \begin{array}{cc} \hat{A}_\pm^{(0)}
(\st,\xi,t,\s) & \hat{A}_\pm^{(1)} (\st,\xi,t,\s) \\ \hat{A}_\pm^{(1)}
  (\st,\xi,t,\s) & \hat{A}_\pm^{(0)} (\st,\xi,t,\s) \end{array} \right) \\
\hat{A}_\pm^{(u)} (\st,-\xi,t,\s) =(-1)^u \hat{A}_\mp^{(u)} (\st,\xi,t,\s)
\,. \label{Eq 5.46b}
\end{gather}
\end{subequations}
In the special case of the class $h_\s^2 =1$, all the fields defined here
have the usual monodromy
\begin{gather}
h_\s^2 =1 ,\,E(T,\s)^2 =1 \!: \quad \hat{O} (\st,\xi+2\pi,\s) =\tau_3
E(T,\s) \hat{O}(\st,\xi,\s) E^\ast (T,\s) \tau_3 \nn \\
\hat{O} =\hg ,\hat{h}_\pm ,\hat{A}_\pm \text{ or } \hat{\psi}
\label{Eq5.47}
\end{gather}
as given for $\hg$ in Eq.~\eqref{Eq 2.25b}. It follows in particular that
each term of
 each
integrand of the coset orientation-orbifold action is $2\pi$-periodic.

To explicitly construct the twisted sectors of particular coset orientation
orbifolds, the standard procedure \cite{Fab}
is to incorporate the embedding
$\widehat{\mathfrak{H}}_O(\s)
\subset \widehat{\mathfrak{G}}_O(\s)$ at the tangent-space level
\begin{subequations}
\label{Eq5.48}
\begin{gather}
\hg =e^{i\hx^\nrmu \st_\nrmu } \,\longrightarrow \, \hat{h}=
e^{i\bh^{\hnrhmu} \st_{\hnrhmu}} \\
\hj_{\hnrhmu} (\xi,t,\s) \equiv R_r(\s)_{\hat{\m}}{}^\m \hj_\hnrmu
(\xi,t,\s) ,\quad u=0,1 ,\,\forall \hat{n}(r) ,\hat{\m} \in \hat{h}_O(\s)
   \subset \hg_O(\s) \\
\st_\hnrhmu (T,\s) \equiv R_r(\s)_{\hat{\m}}{}^\m \st_\hnrmu (T,\s) \\
\{ \hat{n} (r) \} \subset \{ n(r) \} ,\quad \text{dim} \{\hat{\m} \} \leq
\text{dim} \{\m \}
\end{gather}
\end{subequations}
where $R_r(\s)$ at fixed $\hat{n}(r)$ is the embedding matrix of both $\hfrakh_O(\s) \!\subset \!\gfrakh_O (\s)$ and $\hat{h}_O(\s) 
\!\subset \!\hg_O(\s)$. We emphasize with Ref.~\cite{Fab} that the orbifold affine algebras and their corresponding
orbifold Lie algebras
\begin{subequations}
\label{Eq5.49}
\begin{gather}
[\hj_{\sgb_O (\s)} (\cdot) , \hj_{\sgb_O (\s)} (\cdot) ] = i\scf_{\sgb_O
(\s)} \hj_{\sgb_O (\s)} (\cdot) +\sG_{\sgb_O (\s)} \longrightarrow
   [\st_{\hg_O (\s)} ,\st_{\hg_O (\s)} ]=i\scf_{\sgb_O (\s)} \st_{\hg_O
(\s)} \\
[\hj_{\hfrakh_O (\s)} (\cdot) ,\hj_{\hfrakh_O (\s)} (\cdot)
]=i\scf_{\hfrakh_O (\s)} \hj_{\hfrakh_O (\s)} (\cdot) +
   \sG_{\hfrakh_O (\s)} \longrightarrow [\st_{\hat{h}_O (\s)}
,\st_{\hat{h}_O (\s)}] = i\scf_{\hfrakh_O (\s)} \st_{\hat{h}_O (\s)}
\end{gather}
\end{subequations}
share the same twisted structure constants $\scf_{\gfrakh_O (\s)}$ and
$\scf_{\hfrakh_O (\s)}$.

We turn next to the {\it variational boundary conditions} implied by the
coset orientation-orbifold action. Assuming provisionally that the previous
boundary conditions from the WZW orientation-orbifold action continue to
hold, we find the following list of boundary conditions
\begin{subequations}
\label{Eq5.50}
\begin{align}
&\text{at } \xi=0,\pi \!: \nn \\
&\bigspc K(\xi) =R^\xi (\xi)=0 \label{Eq 5.50a} \\
& \bigspc K_1 (\xi) \equiv \widehat{Tr} \left( (\sm \!\otimes \!\one_2)
(\hg \hat{A}_+ +\hat{A}_- \hg^{-1} ) \de \hg \right) =0 \label{Eq 5.50b} \\
& \bigspc B_1 (\xi) \equiv \widehat{Tr} \left( (\sm \!\otimes \!\one_2 )
\hat{\psi}^{-1} \pl_t \hat{\psi} (\hg^{-1} \de \hg +\de \hg \hg^{-1})
\right) =0 \\
& \bigspc B_2 (\xi) \equiv \widehat{Tr} \left( (\sm \!\otimes \!\one_2 )
\hat{\psi}^{-1} \pl_\xi \hat{\psi} (\hg^{-1} \de \hg -\de \hg \hg^{-1})
\right) =0 \\
& \bigspc B_3 (\xi) \equiv \widehat{Tr} \left( (\sm \!\otimes \!\one_2 )
[\hg^{-1} ,\hat{\psi}^{-1} \pl_\xi \hat{\psi}] (\pl_\r \hg \hg^{-1} \de
\hg -\de \hg \hg^{-1} \pl_\r \hg) \right) =0
\end{align}
\end{subequations}
where we have neglected total time derivatives in the variation, as usual.
The first two boundary conditions in Eq.~\eqref{Eq 5.50a} were obtained
earlier in Eqs.~\eqref{Eq 3.12b} and \eqref{Eq 3.13d}. The boundary
condition on $K_1 (\xi)$ in Eq.~\eqref{Eq 5.50b} is obtained by varying
the gauge terms in the coset action \eqref{Eq5.43}, while the boundary
conditions on $B_i (\xi),\, i=1,2,3$ arise by requiring that the first
three boundary conditions are gauge-invariant.

Using the world-sheet parities in Eqs.~\eqref{Eq 2.21d} and
\eqref{Eq5.45}, it is straightforward to show that all these quantities
are odd under world-sheet parity
\begin{gather}
K(-\xi) =-K(\xi) ,\quad R^\xi (-\xi) =-R^\xi(\xi) ,\quad K_1 (-\xi)
=-K_1(\xi) ,\quad B_i (-\xi) =-B_i(\xi)  \label{Eq5.51}
\end{gather}
so that the variational boundary conditions in \eqref{Eq5.50} at $\xi =0$
are automatically satisfied. Moreover, for the class $h_\s^2 =1$, all these
quantities have trivial monodromy and therefore all the boundary
conditions \eqref{Eq5.50} are likewise satisfied explicitly at $\xi =\pi$.

Similarly, we obtain the boundary conditions on the twisted gauge fields
\begin{subequations}
\label{Eq5.52}
\begin{gather}
\hat{A}_-^{(u)} (\st,0) =(-1)^u \hat{A}_+^{(u)} (\st,0) \\
h_\s^2 =1 ,\,\, E(T,\s)^2 =1\!: \quad \hat{A}_-^{(u)} (\st,\pi) =E^\ast (T,\s) \hat{A}_+^{(u)} (\st,\pi) E(T,\s) \bigspc \quad \quad \quad
\end{gather}
\end{subequations}
from the world-sheet parity \eqref{Eq 5.46b} and the monodromy
\eqref{Eq5.47}.

Our final task is to integrate out the twisted matrix gauge fields
\begin{subequations}
\label{Eq5.53}
\begin{gather}
\ha_+ (T,\xi,t,\s) = \ha_+^\hnrhmu (\xi,t,\s) \st_\hnrhmu (T,\s) \in
\hat{h}_O(\s) \\
\ha_- (T,\xi,t,\s) = \ha_-^\hnrhmu (\xi,t,\s) \st_\hnrhmu (T,\s) (-1)^u \\
\ha_-^\hnrhmu (\xi,t,\s) = \ha_+^\hnrhmu (-\xi,t,\s)
\end{gather}
\end{subequations}
to find the equivalent orientation-orbifold sigma-model form of the coset
orientation-orbifold action. For this computation, we will need the
definitions
\begin{subequations}
\label{Eq5.54}
\begin{align}
&\!\!\sG_{\nrmu;\hnshnv} (\s) \!\equiv \!\widehat{Tr} \left((\sm \otimes
\one_2) \st_\nrmu (T,\s) \st_\hnshnv (T,\s) \right) \nn \\
&\bigspc = \!R_s(\s)_{\hat{\n}}{}^\n \sG_{\nrmu ;\hnsnv} (\s) \!=
\!R_s(\s)_{\hat{\n}}{}^\n \sG_{\hnsnv;\nrmu}(\s) \!\equiv \!\sG_{\hnshnv
;\nrmu} (\s)
\end{align}
\begin{align}
\sG_{\hnrhmu ;\hnshnv} (\s) & \equiv \widehat{Tr} \left( (\sm \otimes
\one_2 ) \st_\hnrhmu (T,\s) \st_\hnshnv (T,\s) \right) \nn \\
&= R_r(\s)_{\hat{\m}}{}^\m R_s(\s)_{\hat{\n}}{}^\n \sG_{\hnrmu ;\hnsnv}
(\s) = \sG_{\hnshnv ;\hnrhmu} (\s)
\end{align}
\begin{gather}
\sG_{\hnrhmu ;\nsnv} (\s) = \de_{u+v,0\,\text{mod }2} \de_{\hat{n}(r)
+n(s),0\, \text{mod }\r(\s)} \sG_{\hnrhmu ;-\hat{n}(r),\n ,-u}(\s) \\
\sG_{\hnrhmu ;\hnshnv} (\s) =\de_{u+v,0\,\text{mod }2} \de_{\hat{n}(r)
+\hat{n}(s),0\, \text{mod }\r(\s)} \sG_{\hnrhmu ;-\hat{n}(r) ,\hat{\n}
,-u} (\s)
\end{gather}
\begin{gather}
\ho (\hx)_\hnrhmu{}^{\!\!\!\nsnv} \!\equiv \!R_r(\s)_{\hat{\m}}{}^\m \ho
(\hx)_\hnrmu{}^{\!\!\!\nsnv} \\
\ho (\hx)_{\hnrhmu ;\hnshnv} \!\equiv \!\ho(\hx)_\hnrhmu{}^{\!\!\!\ntd w}
\sG_{\ntd w;\hnshnv} (\s)
\end{gather}
\end{subequations}
where $\sm$ is the twisted data matrix in \eqref{Eq 2.14f},
$\sG_{\hfrakh_O (\s)} \simeq \{\sG_{\hnrhmu ;\hnshnv}(\s)\}$ is the induced tangent-space
metric on the twisted affine subalgebra $\hfrakh_O (\s)$, and $\ho$ is the twisted adjoint action in Eq.~\eqref{Eq4.29}. The induced metric
$\sG_{\hfrakh_O (\s)}$ is invertible in the $\hat{h}_O(\s)$ subspace
\begin{equation}
\sG_{\hnrhmu ;\hat{n}(t)\hat{\de} w} (\s) \sG^{\hat{n}(t)\hat{\de}w ;\hnshnv} (\s) = \de_{\hat{\m}}{}^{\hat{\n}} \de_{u-v,0\,\text{mod }2}
   \de_{\hat{n}(r) -\hat{n}(s),0\, \text{mod }\r(\s)}  \label{Eq5.55}
\end{equation}
because the original $h_\s$-symmetric coset construction
$\frac{g}{h}(h_\s)$ was a reductive coset space.

Following the conventional procedure, we may then integrate out the
twisted matrix gauge fields by solving their equations
of motion:
\begin{subequations}
\label{Eq5.56}
\begin{gather}
\widehat{Tr} \left\{ (\sm \!\otimes \!\one_2 ) \left( -i\pl_- \hg \hg^{-1}
-\hg \ha_- \hg^{-1} +\ha_- \right) \st_\hnrhmu \right\} =0 \quad \quad
\quad \\
\widehat{Tr} \left\{ \!(\sm \!\otimes \!\one_2 ) \!\left( i\hg^{-1} \pl_+
\hg \!-\!\hg^{-1} \ha_+ \hg \!+\!\ha_+ \right) \!\st_\hnrhmu \right\}
\!=\!0  ,\,\,\,\, u=0,1,\,\,
\forall \hat{n}(r) ,\hat{\m} \!\in \!\hat{h}_O(\s) \,.
\end{gather}
\end{subequations}
After some algebra, we find that open-string sector $\hat{h}_\s$ of the
general coset orientation orbifold $A_{g/h} (H_-)/H_-$ is described by the
sigma model
orientation-orbifold action \eqref{Eq5.20} with the special values of the
twisted Einstein metric $\hG^{tot}$ and twisted $B$ field $\hB^{tot}$:
\begin{subequations}
\label{Eq5.57}
\begin{align}
&( \hG (\hx) \!+\!\hB(\hx) )_{\nrmu;\nsnv}^{tot} = ( \hG(\hx)
\!+\!\hB(\hx))_{\nrmu;\nsnv} \nn \\
&\bigspc \bigspc -2 \hat{Z}(\hx)_{\nrmu;\hat{n}(r')\hat{\m}'u'}
\hat{\bar{Z}}(\hx)_{\nsnv ;\hat{n}(s')\hat{\n} 'v'} \hat{X}^{-1}
    (\hx)^{\hat{n}(s')\hat{\n} 'v';\hat{n}(r') \hat{\m}'u'}
\end{align}
\begin{gather}
\hat{Z}(\hx)_{\nrmu;\nsnv} \!\equiv \! \he (\hx)_\nrmu {}^{\!\!\ntd w}
\sG_{\ntd w;\nsnv} (\s)\\
\hat{\bar{Z}}(\hx)_{\nrmu;\nsnv} \! \equiv \!\heb (\hx)_\nrmu{}^{\!\!\ntd
w} \sG_{\ntd w;\nsnv} (\s) \\
\hat{X}_{\hnrhmu ;\hnshnv} (\hx)  \!\equiv \!(\sG (\s) \!- \!\ho (\hx)
)_{\hnrhmu ;\hnshnv} \,. \label{Eq 5.57d}
\end{gather}
\end{subequations}
Here the quantities $\hG, \hB ,\he$ and $\ho$ are the WZW
orientation-orbifold fields defined in Subsec.~$5.3$, and the quantity
$\heb$ is given by:
\begin{subequations}
\label{Eq5.58}
\begin{gather}
-i\hg (\st,\hx) \hpl_\nrmu \hg^{-1} (\st,\hx) = \heb(\hx)_\nrmu{}^\nsnv
\st_\nsnv \\
\heb (\hx(\xi))_\nrmu{}^{\!\!\!\nsnv} \!\equiv \!-\Big{(} \he(\hx(\xi))
\hO (\hx(\xi)) \Big{)}_\nrmu{}^{\!\!\!\nsnv} \!=\! (-1)^{u+v+1} \he
(\hx(-\xi))_\nrmu{}^{\!\!\!\nsnv} \,.
\end{gather}
\end{subequations}
The matrix $\hat{X}(\hx)$ in Eq.~\eqref{Eq 5.57d} is invertible in the $\hat{h}_O (\s)$ subspace because the induced metric $\sG_{\hfrakh_O
(\s)} \simeq \{\sG_{\hnrhmu ;\hnshnv}(\s)\}$ is invertible. The same result \eqref{Eq5.57} for the sigma-model form of the coset orientation-orbifold
action can be obtained by eigenfields and local isomorphisms from the sigma-model form of the coset action (see App.~B).

The coset orientation-orbifold sigma models should be considered in fixed gauges such as:
\begin{gather}
0=\widehat{Tr} \left( (\sm \otimes \one_2 )\hx_\s^\nsnv (\xi,t) \st_\nsnv
\st_\hnrhmu \right) \bigspc \nn \\
\bigspc =\hx_\s^\nsnv \sG_{\nsnv ;\hnrhmu}(\s)
=2\hx_\s^{-\hat{n}(r),\hat{\n},-u} \sG_{-\hat{n}(r),\hat{\n};\hnrhm}(\s)
\,.  \label{Eq5.59}
\end{gather}
The contribution of the orientation-orbifold dilaton to the coset
orientation orbifolds is included in the discussion of Sec.6.

\subsection{Example: Twisted Open-String Free Bosons}

As another example in the sigma model, we consider the simple case of free
bosons with a
linear symmetry $\ws$
\begin{subequations}
\label{Eq5.60}
\begin{gather}
G_{ij}(x) = \text{ const.} ,\quad G_{ij} =\ws_i{}^k \ws_j{}^l G_{kl}
,\quad B_{ij} (x) =0 \\
S =\srac{1}{8\pi} \!\int \!\!dt \!\int_0^{2\pi} \!\!d\xi\, G_{ij} \pl_+ x^i
\pl_- x^j
\end{gather}
\end{subequations}
which may be considered as the abelian limit of the WZW models.

In open-string orientation-orbifold sector $\hat{h}_\s = \tau_1
\!\times \!h_\s$, the results of the preceding subsections
reduce to
\begin{subequations}
\label{Eq5.61}
\begin{align}
\hG_{\nrmu ;\nsnv} (\hx) &= \text{ const.} \\
& =\de_{n(r)+n(s) ,0\,\text{mod }\r(\s)} \de_{u+v,0\,\text{mod }2}
\hG^{(0)}_{\nrm ;\mnrn}
\end{align}
\begin{gather}
\hat{G}^{(0)}_{\nrm ;\nsn} = 2\sG_{\nrm; \nsn}(\s) \label{Eq 5.61c} \\
\hB_{\nrmu ;\nsnv} (\hx) =0 ,\quad \hx^\nrmu_\s (-\xi) = (-1)^{u+1}
\hx^\nrmu_\s (\xi) \\
\hat{S}_\s =\srac{1}{8\pi} \!\int \!\!dt \!\int_0^{\pi} \!\!d\xi\,
\hG^{(0)}_{\nrm;\mnrn} \sum_{u=0}^1 \pl_+ \hx^\nrmu \pl_- \hx^{\mnrn ,-u}
\label{Eq 5.61e} \\
\Rightarrow \pl_+ \pl_- \hx^\nrmu =0
\end{gather}
\end{subequations}
where $\hx$ are the twisted open-string free bosons and $\sG$ is the ordinary
twisted tangent space metric in Eq.~\eqref{Eq 2.19c} The variational
boundary condition \eqref{Eq5.22} takes the form
\begin{gather}
\hG_{\nrm;\mnrn}^{(0)} \sum_{u=0}^1 \pl_\xi \hx^\nrmu \de \hx^{\mnrn ,-u}
=0 \text{ at } \xi=0 ,\pi \label{Eq5.62}
\end{gather}
which also follows of course by variation of the action \eqref{Eq 5.61e}.
One way to consider these boundary conditions is to take $\de \hx$
infinitesimal and divide by a corresponding time increment $\de t$:
\begin{gather}
\hG_{\nrm;\mnrn}^{(0)} \sum_{u=0}^1 \pl_\xi \hx^\nrmu \pl_t \hx^{\mnrn
,-u} =0 \text{ at } \xi=0 ,\pi \label{Eq5.63} \,.
\end{gather}
Similarly, $\de \hx$ can be replaced by $\pl_t \hx$ in the general
boundary condition \eqref{Eq5.22}.

In our previous paper \cite{Orient1}, we gave the {\it correct} coordinate
boundary conditions or branes for these twisted bosons
\begin{subequations}
\label{Eq5.64}
\begin{gather}
\pl_t \hx^{\nrm 0} = \pl_\xi \hx^{\nrm 1} =0 \text{ at } \xi =0 \\
\pl_t \hx^\nrmu = i \tan (\nrrs \pi) \pl_\xi \hx^\nrmu \text{ at }
\xi=\pi  \label{Eq 5.64b}
\end{gather}
\end{subequations}
as well as the corresponding field expansions and quasi-canonical algebra.
The boundary conditions \eqref{Eq5.64} were derived from the equations of motion
\begin{align}
\!&\!\!\pl_+ \hx^\nrmu \!=\! 2\sG^{\nrmu ;\nsnv}(\s) \hj_\nsnv (\xi,t) ,\quad \pl_- \hx^\nrmu \!=\! 2\sG^{\nrmu ;\nsnv}(\s) \hj_\nsnv (-\xi,t) \label{Eq5.65}
\end{align}
using the monodromies of the currents -- which are known from the operator
theory. This situation is analogous to that which was encountered for WZW
orientation orbifolds in Eq.~\eqref{Eq3.21}.

At $\xi =0$, these coordinate boundary conditions are the same as those
which followed from world-sheet parity above, and they correspondingly
satisfy the variational boundary condition \eqref{Eq5.63} at $\xi =0$.

The coordinate boundary condition \eqref{Eq 5.64b} also solves the
variational boundary condition \eqref{Eq5.63} at $\xi =\pi$. To see this,
one needs only
the identity
\begin{gather}
\hG_{\nrm ;\mnrn}^{(0)} \tan (\nrrs \pi ) \sum_{u=0}^1 \pl_\xi \hx^\nrmu
\pl_\xi \hx^{\mnrn ,-u} =0 \label{Eq5.66}
\end{gather}
which follows because the tangent is odd under the relabelling $\nrmu
\leftrightarrow \mnrn ,-u$ of the dummy indices.

Although mixed boundary conditions such as Eq.~\eqref{Eq 5.64b} are
conventionally associated to non-zero $B$ field backgrounds, we emphasize
that there is {\it no twisted $\hB$ field} in the background of these twisted open-string
free bosons.

\section{The Orientation-Orbifold Einstein Equations}

In analogy to the Einstein equations [44-49,13] of
the untwisted sigma
model $A_M$ and the twisted Einstein equations \cite{Geom} of the ordinary
sigma-model orbifold $A_M (H_+)/H_+$, we will now construct the twisted
Einstein
equations of open-string sector $\hat{h}_\s$ of the sigma-model
orientation orbifold $A_M (H_-)/H_-$.

\subsection{The Two-Component Form of the Ordinary Einstein Equations}

We begin this construction\footnote{The Riemann and Ricci tensor conventions
 of Ref.~\cite{Jan} are followed here.} with the ordinary Einstein
equations [44-49,13]
\begin{subequations}
\label{Eq6.1}
\begin{gather}
R_{ij} (x) +\srac{1}{4} H_{ki}{}^l (x) H^k {}_{lj}(x) -2\nabla_{\!i} \phi_j (x)=0 \\
\left( \nabla_k -2 \phi_k(x) \right) H^k{}_{ij}(x) =0
\end{gather}
\end{subequations}
which describe one-loop conformal sigma models in the presence of a
general dilaton vector $\phi_i$ (see e.~g.~Ref.~\cite{Jan}). The special case of the ordinary dilaton solution is 
\begin{gather}
\phi_i =\nabla_{\!i} \phi =\pl_i \phi \label{Eq6.2}
\end{gather}
where $\phi$ is the dilaton.

We first ask which of these Einstein systems is invariant under the basic
orientation reversal
\begin{gather}
x^i (\xi)' =-x^i (-\xi) ,\quad O(x)' =O(x') \label{Eq6.3}
\end{gather}
where $O$ includes any tensor. To compare the original Einstein system
\eqref{Eq6.1}
 with the
image of the system under orientation reversal, we need the basic
space-time parities
\begin{gather}
G_{ij}(-x) =G_{ij}(x) ,\quad B_{ij}(-x) =-B_{ij}(x) ,\quad H_{ijk}(-x)
=H_{ijk}(x) \label{Eq6.4}
\end{gather}
(which we already know from the invariance of the sigma-model action) as
well as the derived space-time parities
\begin{subequations}
\label{Eq6.5}
\begin{gather}
G^{ij}(-x) =G^{ij}(x) ,\quad H_{ij}{}^k (-x) =H_{ij}{}^k(x) \\
\Gamma_{ij}{}^k(-x) =-\Gamma_{ij}{}^k (x) ,\quad \nabla_i (-x) =-\nabla_i
(x) \\
R_{ijkl}(-x) =R_{ijkl}(x) ,\quad R_{ij}(-x) =R_{ij}(x)
\end{gather}
\end{subequations}
which follow from the standard formulae of Riemannian geometry. Then we
find that the necessary and
sufficient condition for invariance of the Einstein system is the
following space-time parity of the dilaton vector:
\begin{gather}
\phi_i (-x) =-\phi_i (x) \,\longrightarrow \, \phi(-x) =\phi(x) \,.
\label{Eq6.6}
\end{gather}
We recall that the space-time parities \eqref{Eq6.4} were verified in
Subsec.~$4.1$ for WZW models and principal chiral models. Appendix B also
shows that the space-time parities in Eqs.~\eqref{Eq6.4}, \eqref{Eq6.5}
hold for all (reductive) coset CFTs.

As discussed for the sigma-model action in Subsec.~$5.1$, the
space-time parities above allow us to rewrite the Einstein system in the
two-component notation:
\begin{subequations}
\label{Eq6.7}
\begin{gather}
R_{i\Id ;j\Jd}(x) +\srac{1}{4} H_{k\Kd ;i\Id}{}^{l\dot{L}}(x)
H^{k\Kd}{}_{l\dot{L} ;j\Jd}(x) -2\nabla_{i\Id} \phi_{j\Jd}(x) =0 \\
\left( \nabla_{k\Kd} -2 \phi_{k\Kd}(x) \right) H^{k\Kd}{}_{i\Id ;j\Jd}(x)
=0 \\
x^{i\Id}(\xi)= (-1)^\Id x^i ((-1)^\Id \xi) ,\quad \pl_{i\Id}
=\frac{\pl}{\pl x^{i\Id}} ,\quad \Id =0,1 \\
G_{i\Id ;j\Jd}(x) =\de_{\Id \Jd} G_{ij}(x^\Id) ,\quad B_{i\Id ;j\Jd}(x)
=\de_{\Id \Jd}B_{ij} (x^\Id) \\
H_{i\Id ;j\Jd ;k\Kd}(x) =\pl_{i\Id} B_{j\Jd;k\Kd}(x) +\text{ cyclic }
=\de_{\Id \Jd} \de_{\Jd \Kd} H_{ijk}(x^\Id ) \\
\phi_{i\Id}(x) \equiv \phi_i (x^\Id) \, \rightarrow \, \frac{\pl}{\pl
x^{i\Id}} \phi (x^\Id) \,. \label{Eq 6.7f}
\end{gather}
\end{subequations}
In the two-component notation, we assume all the standard formulae of
Riemannian geometry with $i \rightarrow i\Id$ for every index -- for example:
\begin{subequations}
\label{Eq6.8}
\begin{gather}
G^{i\Id ;j\Jd} (x) =\de^{\Id \Jd} G^{ij}(x^\Id ) \\
\Gamma_{i\Id ;j\Jd}{}^{k\Kd}(x) \equiv \srac{1}{2} \left( \pl_{i\Id}
G_{j\Jd ;l\dot{L}}(x) +\pl_{j\Jd} G_{i\Id;l\dot{L}}(x) -\pl_{l\dot{L}}
   G_{i\Id;j\Jd}(x) \right) G^{l\dot{L};k\Kd}(x) \nn \\
   =\de_{\Id \Jd} \de_\Jd{}^\Kd \Gamma_{ij}{}^k (x^\Id) \\
\nabla_{i\Id} \phi_{j\Jd}(x) \equiv \pl_{i\Id} \phi_{j\Jd}(x)
-\Gamma_{i\Id;j\Jd}{}^{k\Kd} (x) \phi_{k\Kd}(x)
   =\de_{\Id \Jd} \left( \pl_{i\Id} \phi_j (x^\Id ) -\Gamma_{ij}{}^k
(x^\Id ) \phi_k (x^\Id ) \right)
\end{gather}
\begin{gather}
R_{i\Id ;j\Jd ;k\Kd ;l\dot{L}}(x) =\de_{\Id \Jd} \de_{\Jd \Kd} \de_{\Kd
\dot{L}} R_{ijkl} (x^\Id) ,\quad
   R_{i\Id ;j\Jd}(x) =\de_{\Id \Jd} R_{ij}(x^\Id ) \,.
\end{gather}
\end{subequations}
More generally, the two-component analogue of any Einstein field $O$ has
the form
\begin{subequations}
\label{Eq6.9}
\begin{gather}
O_{i_1 \ldots i_m}{}^{j_1 \ldots j_n} \,\rightarrow \, O_{i_1 \Id_1
;\ldots ;i_m \Id_m}^{\quad \quad j_1 \Jd_1 ;\ldots ;j_n \Jd_n}(x) \\
O_{i_1 \Id_1 ;\ldots ;i_m \Id_m}^{\quad \quad j_1 \Jd_1 ;\ldots ;j_n
\Jd_n}(x) =\de_{\Id_1 \Id_2} \ldots \de_{\Id_{m-1} \Id_m}
\de_{\Id_m}^{\,\,\,\Jd_1}
  \de^{\Jd_1 \Jd_2} \ldots \de^{\Jd_{n-1} \Jd_n} O_{i_1 \ldots i_m}^{\quad
j_1 \ldots j_n} (x^{\Id_1}) \\
\pl_{i\Id} O_{i_1 \Id_1 ;\ldots ;i_m \Id_m}^{\quad \quad j_1 \Jd_1 ;\ldots
;j_n \Jd_n}(x) =\de_{\Id \Id_1} \de_{\Id_1 \Id_2} \ldots \de_{\Id_{m-1}
\Id_m}
  \de_{\Id_m}^{\,\,\,\Jd_1} \de^{\Jd_1 \Jd_2} \ldots \de^{\Jd_{n-1} \Jd_n}
\pl_{i\Id} O_{i_1 \ldots i_m}^{\quad j_1 \ldots j_n} (x^{\Id})
\end{gather}
\end{subequations}
where the Kronecker deltas in these results set all $\Id ,\Jd$ indices
equal.

In this formulation, the action of the basic orientation-reversing
automorphism involves one $\tau_1$ for each $\Id$ index. This includes the
transformations in Eq.~\eqref{Eq5.8} and for example:
\begin{subequations}
\label{Eq6.10}
\begin{gather}
x^{i\Id}(\xi)' = -x^{i\Jd}(\xi) (\tau_1 )_\Jd{}^\Id ,\quad A(x)' =A(x') \\
\Gamma_{i\Id ;j\Jd}{}^{k\Kd}(x)' =\Gamma_{i\Id ;j\Jd}{}^{k\Kd} (x')
=(\tau_1 )_\Id{}^{\dot{L}} (\tau_1 )_\Jd{}^{\dot{M}} \Gamma_{i\dot{L}
;j\dot{M}}
   {}^{k\dot{N}}(x) (\tau_1 )_{\dot{N}}{}^\Kd \\
R_{i\Id ;j\Jd}(x)' =R_{i\Id ;j\Jd}(x') =(\tau_1 )_\Id{}^\Kd (\tau_1
)_\Jd{}^{\dot{L}} R_{i\Kd ;j\dot{L}}(x) \,.
\end{gather}
\end{subequations}
The invariance of the two-component Einstein system in \eqref{Eq6.7} under
the basic orientation-reversing automorphism is now immediately
transparent.

The space-time parities of the two-component fields
\begin{subequations}
\label{Eq6.11}
\begin{gather}
G_{i\Id ;j\Jd}(-x) =G_{i\Id ;j\Jd} (x) ,\quad B_{i\Id ;j\Jd}(-x) =-B_{i\Id
;j\Jd}(x) \\
H_{i\Id ;j\Jd ;k\Kd}(-x) =H_{i\Id ;j\Jd ;k\Kd}(x) \\
\Gamma_{i\Id ;j\Jd}{}^{k\Kd}(-x) =-\Gamma_{i\Id ;j\Jd}{}^{k\Kd}(x) ,\quad
\nabla_{i\Id} (-x) =-\nabla_{i\Id} (x) \\
R_{i\Id ;j\Jd}(-x) =R_{i\Id ;j\Jd}(x) ,\quad \phi_{i\Id}(-x)
=-\phi_{i\Id}(x)
\end{gather}
\end{subequations}
follow from Eqs.~\eqref{Eq6.4}-\eqref{Eq6.5}, and in fact form a symmetry
of the two-component Einstein system \eqref{Eq6.7}. We emphasize however
that the space-time parity is {\it not} a symmetry of the sigma-model
action \eqref{Eq 5.1a}, just as electromagnetic duality is not a symmetry
of the Maxwell action.

The world-sheet parities of the two-component fields include those in
Eq.~\eqref{Eq5.7} and
\begin{subequations}
\label{Eq6.12}
\begin{gather}
G^{i\Id ;j\Jd} (x(-\xi)) = G^{i\Kd ;j\dot{L}}(x(\xi)) (\tau_1 )_\Kd{}^\Id
(\tau_1 )_{\dot{L}}{}^\Jd \\
H_{i\Id ;j\Jd}{}^{k\Kd} (x(-\xi)) =(\tau_1 )_\Id{}^{\dot{L}} (\tau_1
)_\Jd{}^{\dot{M}} H_{i\dot{L} ;j\dot{M}}{}^{k\dot{N}}(x(\xi)) (\tau_1
)_{\dot{N}}{}^\Kd \\
\Gamma_{i\Id;j\Jd}{}^{k\Kd}(x(-\xi)) =-(\tau_1 )_\Id{}^{\dot{L}} (\tau_1
)_\Jd{}^{\dot{M}} \Gamma_{i\dot{L} ;j\dot{M}}{}^{k\dot{N}}(x(\xi)) (\tau_1
)_{\dot{N}}{}^\Kd  \\
\nabla_{i\Id} (-\xi) =-(\tau_1 )_\Id{}^{\Jd} \nabla_{i\Jd} (\xi) ,\quad
R_{i\Id;j\Jd}(x(-\xi)) =(\tau_1 )_\Id{}^\Kd (\tau_1 )_\Jd{}^{\dot{L}}
   R_{i\Kd ;j\dot{L}}(x(\xi)) \\
R_{i\Id ;j\Jd ;k\Kd ;l\dot{L}}(x(-\xi)) =(\tau_1 )_\Id{}^{\dot{M}} (\tau_1
)_\Jd{}^{\dot{N}} (\tau_1 )_\Kd{}^{\dot{P}} (\tau_1 )_{\dot{L}}{}^{\dot{Q}}
    R_{i\dot{M};j\dot{N};k\dot{P};l\dot{Q}}(x(\xi)) \\
\phi_{i\Id} (x(-\xi)) =-(\tau_1 )_\Id{}^\Jd \phi_{i\Jd} (x(\xi)) \,
\longrightarrow \, \phi (x^\Id (-\xi)) =\phi (x^\Jd (\xi) (\tau_1
)_\Jd{}^\Id )
\end{gather}
\end{subequations}
which follow from the definitions of the two-component objects. The
general rule for the world-sheet parities is one $\tau_1$ for each $\Id$
index and an extra minus sign when the {\it space-time} parity is odd.

We turn next to the action of the {\it general} orientation-reversing
automorphism $\hat{h}_\s$ on the two-component fields. For $x,G,B$ and
$H$, this action is given in Eq.~\eqref{Eq5.9}, and the actions on the
other fields
\begin{subequations}
\label{Eq6.13}
\begin{gather}
x^{i\Id}(\xi)' =x^{j\Jd}(\xi) \om\hc (h_\s)_j{}^i (\tau_1 )_\Jd{}^\Id ,\quad
A(x)' =A(x') \\
\!\!\!\!\Gamma_{i\Id;j\Jd}{}^{\!k\Kd}(x)' \!=\!\Gamma_{i\Id
;j\Jd}{}^{\!k\Kd} (x') \!=\!\ws_i{}^l \ws_j{}^{\!m} (\tau_1
)_\Id{}^{\!\dot{L}}
  (\tau_1 )_\Jd{}^{\!\dot{M}} \Gamma_{l\dot{L}
;m\dot{M}}{}^{\!\!n\dot{N}}(x) \om\hc (h_\s)_n{}^k (\tau_1
)_{\dot{N}}{}^{\!\Kd} \,\,\, \\
R_{i\Id ;j\Jd ;k\Kd ;l\dot{L}}(x)' =R_{i\Id ;j\Jd ;k\Kd ;l\dot{L}}(x')
\bigspc \bigspc \bigspc \bigspc \bigspc \quad \quad \quad \nn \\
\quad =\ws_i{}^m \ws_j{}^n \ws_k{}^p \ws_l{}^q (\tau_1 )_\Id{}^{\dot{M}}
(\tau_1 )_\Jd{}^{\dot{N}} (\tau_1 )_\Kd{}^{\dot{P}}
  (\tau_1 )_{\dot{L}}{}^{\dot{Q}}
R_{m\dot{M};n\dot{N};p\dot{P};q\dot{Q}}(x) \\
R_{i\Id ;j\Jd}(x)' =R_{i\Id ;j\Jd}(x') =\ws_i{}^k \ws_j{}^l (\tau_1
)_\Id{}^\Kd (\tau_1 )_\Jd{}^{\dot{L}} R_{k\Kd ;l\dot{L}}(x)
\end{gather}
\end{subequations}
follow by the standard formulae of Riemannian geometry. With the
corresponding transformation of the dilaton vector
\begin{gather}
\phi_{i\Id}(x)' =\phi_{i\Id}(x') =\ws_i{}^j (\tau_1 )_\Id{}^\Jd
\phi_{j\Jd}(x)  \label{Eq6.14}
\end{gather}
we find that the two-component Einstein system \eqref{Eq6.7} is covariant
under this automorphism. The general rule for the transformation of the
Einstein fields is a factor $\om\tau_1$ for each down index and a factor
$\om\hc \tau_1$ for each up index.

As above, the space-time parities \eqref{Eq6.11} can be used to see
that the linear symmetry conditions \eqref{Eq5.10} for $G,B$ and $H$, as
well as the
corresponding linear symmetry conditions for the other fields
\begin{subequations}
\label{Eq6.15}
\begin{gather}
G^{ij} (xw\hc ) =G^{kl}(x) (\om\hc )_k{}^i (\om\hc )_l{}^j \\
\Gamma_{ij}{}^k (x\om\hc) =\om_i{}^l \om_j{}^m \Gamma_{lm}{}^n(x) (\om\hc )_n{}^k
\\
R_{ijkl}(x\om\hc )=\om_i{}^m \om_j{}^n \om_k{}^p \om_l{}^q R_{mnpq}(x) ,\quad
R_{ij}(x\om\hc) =\om_i{}^k \om_j{}^l R_{kl}(x) \\
\phi_i (x\om\hc )=\om_i{}^j \phi_j (x) \, \longrightarrow \, \phi (x\om\hc)
=\phi(x)
\end{gather}
\end{subequations}
are incorporated in the transformations \eqref{Eq6.13}and \eqref{Eq6.14}.
The general rule for the linear symmetry conditions involves one $\om$ for
each down index and one $\om\hc$ for each up index.

\subsection{Riemannian Eigenfields}

Using the eigenvector
matrix of the Einstein-space
$H$-eigenvalue problem \eqref{Eq5.13}, we next define the corresponding
Riemannian eigenfields
\begin{subequations}
\label{Eq6.16}
\begin{align}
&\bigspc \quad \sx_\s^\nrmu (\xi) \!\equiv \!x^{i\Id}(\xi)
\schisig_\nrm^{-1} U\hc (\s)_i{}^\nrm (\srac{1}{\sqrt{2}} U\hc {}_\Id{}^u
) ,\quad \bar{u}=0,1 \\
\!\!&\!\!\sG_{\nrmu ;\nsnv}(\sx) \!\equiv \!\schisig_\nrm \schisig_\nsn
U(\s)_\nrm{}^{\!i} U(\s)_\nsn{}^{\!j} (\sqrt{2}U_u{}^{\!\!\Id}
)(\sqrt{2}U_v{}^{\!\!\Jd} ) G_{i\Id ;j\Jd}(x(\sx))  \\
\!\!&\!\!\sB_{\nrmu ;\nsnv}(\sx) \!\equiv \!\schisig_\nrm \schisig_\nsn
U(\s)_\nrm{}^{\!i} U(\s)_\nsn{}^{\!j} (\sqrt{2}U_u{}^{\!\!\Id}
)(\sqrt{2}U_v{}^{\!\!\Jd} ) B_{i\Id ;j\Jd}(x(\sx))  \\
&\sH_{\nrmu ;\nsnv ;\ntd w}(\sx) \equiv \schisig_\nrm \schisig_\nsn
\schisig_\ntd U(\s)_\nrm{}^i U(\s)_\nsn{}^j U(\s)_\ntd{}^k \nn \\
&  \bigspc \bigspc \bigspc \times (\sqrt{2}U_u{}^\Id )(\sqrt{2}U_v{}^\Jd )
(\sqrt{2}U_w{}^\Kd ) H_{i\Id ;j\Jd ;k\Kd}(x(\sx))
\end{align}
\begin{gather}
\gamma_{\nrmu ;\nsnv}{}^{\ntd w}(\sx) \equiv \schi_\nrm \schi_\nsn
U_\nrm{}^i U_\nsn{}^j (\sqrt{2}U_u{}^\Id )(\sqrt{2}U_v{}^\Jd ) \times
\bigspc \nn \\
  \bigspc \bigspc \bigspc \times \Gamma_{i\Id;j\Jd}{}^{k\Kd}(x(\sx))
\schi^{-1}_\ntd (U\hc )_k{}^\ntd (\srac{1}{\sqrt{2}} U\hc {}_\Kd{}^w ) \\
{{\cal R}}_{\nrmu;\nsnv}(\sx) \equiv \schi_\nrm \schi_\nsn U_\nrm{}^i
U_\nsn{}^j (\sqrt{2}U_u{}^\Id )(\sqrt{2}U_v{}^\Jd ) R_{i\Id ;j\Jd}(x(\sx))
\\
\Phi_\nrmu (\sx) \equiv \schi_\nrm U_\nrm{}^i (\sqrt{2} U_u{}^\Id)
\phi_{i\Id}(x(\sx)) ,\quad \bar{u},\bar{v},\bar{w} \in \{0,1\}
\end{gather}
\end{subequations}
where the $\schi$'s are arbitrary normalization constants. The general
rule for the eigenfields is a factor $\schi U(\s) \sqrt{2}U$ for each down
index and a factor $\schi^{-1} U\hc(\s) \srac{1}{\sqrt{2}} U\hc$ for each
up index. This prescription diagonalizes the automorphic response of all
the eigenfields, for example:
\begin{subequations}
\label{Eq6.17}
\begin{gather}
\gamma_{\nrmu ;\nsnv}{}^{\ntd w}(\sx)' =e^{-\tp (
\frac{n(r)+n(s)-n(t)}{\r(\s)} +\frac{u+v-w}{2})} \gamma_{\nrmu
;\nsnv}{}^{\ntd w}(\sx) \\
{{\cal R}}_{\nrmu ;\nsnv ;\ntd w;n(q)\ep y}(\sx)' = \bigspc \bigspc
\bigspc \bigspc \nn \\
\bigspc =e^{-\tp (\frac{n(r)+n(s)+n(t)+n(q)}{\r(\s)} +\frac{u+v+w+y}{2})}
{{\cal R}}_{\nrmu ;\nsnv ;\ntd w;n(q)\ep y}(\sx) \\
{{\cal R}}_{\nrmu;\nsnv}(\sx)' =e^{-\tp (\frac{n(r)+n(s)}{\r(\s)}
+\frac{u+v}{2})} {{\cal R}}_{nrmu;\nsnv}(\sx) \\
\Phi_\nrmu (\sx)' =e^{-\tp (\frac{n(r)}{\r(\s)} +\frac{u}{2})} \Phi_\nrmu
(\sx) \,.
\end{gather}
\end{subequations}
The general rule for the automorphic responses of the eigenfields is a
negative (positive) phase for each down (up) index.

As above, these eigenfields are two-component fields in the
two-dimensional space. For example one has the forms of $\sG,\sB$ and
$\sH$ in \eqref{Eq4.21},
as well as:
\begin{subequations}
\label{Eq6.18}
\begin{gather}
\sG^{\nrmu ;\nsnv}(\sx) =\sG_{(u+v)}^{\nrm;\nsn}(\sx) \\
\gamma_{\nrmu;\nsnv}{}^{\ntd w}(\sx) = \gamma_{\nrm;\nsn}^{(u+v-w)}{}^\ntd
(\sx) \\
{{\cal R}}_{\nrmu ;\nsnv ;\ntd w ;n(q)\ep y}(\sx) = {{\cal R}}_{\nrm ;\nsn
;\ntd ;n(q)\ep}^{(u+v+w+y)} (\sx) \\
{{\cal R}}_{\nrmu;\nsnv}(\sx) ={{\cal R}}_{\nrm;\nsn}^{(u+v)}(\sx) \,.
\end{gather}
\end{subequations}
These relations are special cases of the general eigenfield identities
\begin{subequations}
\label{Eq6.19}
\begin{gather}
\Ord_{n(r_1)\m_1 u_1 ;\ldots ;n(r_m)\m_m u_m}^{\quad \quad \quad n(s_1
)\n_1 v_1 ;\ldots ;n(s_n )\n_n v_n} =\Ord_{n(r_1)\m_1 ;\ldots ;n(r_m)\m_m
}^{(\Sigma u_i -\Sigma v_j) \quad \quad n(s_1 )\n_1 ;\ldots ;n(s_n )\n_n}
\\
\frac{\pl}{\pl \sx^\nrmu} \Ord_{n(s_1 )\n_1 v_1 ;\ldots ;n(s_m )\n_m
v_m}^{\quad \quad \quad n(t_1 )\de_1 w_1 ;\ldots ;n(t_n )\de_n w_n}
   ={{\cal C}}_{\nrm ;n(s_1 )\n_1 ;\ldots n(s_m) \n_m}^{(u +\Sigma v_i
-\Sigma w_j) \quad \quad n(t_1 )\de_1 ;\ldots ;n(t_n )\de_n}
\end{gather}
\end{subequations}
which follow from Eq.~\eqref{Eq6.9} and show the two-component structure
for any eigenfield $\Ord$.

\subsection{Open-String Twisted Riemannian Geometry}

We then use the principle of local isomorphisms
\begin{subequations}
\label{Eq6.20}
\begin{gather}
\sx \dual \hx ,\quad \sG \dual \hG ,\quad \sB \dual \hB ,\quad \sh \dual
\hh \\
\gamma \dual \hat{\Gamma} ,\quad {{\cal R}} \dual \hat{R} ,\quad \Phi
\dual \hat{\phi}
\end{gather}
\end{subequations}
to obtain the corresponding twisted fields in the open-string sectors.
Relations among
the twisted fields include
\begin{subequations}
\label{Eq6.21}
\begin{gather}
\hat{\Gamma}_{\nrmu ;\nsnv}{}^{\ntd w}(\hx) =\srac{1}{2} \Big{(}
\hpl_\nrmu \hG_{\nsnv ;n(q)\ep x}(\hx) +\hpl_\nsnv \hG_{\nrmu ;n(q)\ep
x}(\hx) \bigspc \bigspc
  \nn \\
\bigspc \bigspc -\hpl_{n(q)\ep x} \hG_{\nrmu;\nsnv} (\hx) \Big{)}
\hG^{n(q)\ep x;\ntd w}(\hx) \\
\hat{\nabla}_\nrmu \hat{\phi}_\nsnv (\hx) =\hpl_\nrmu \hat{\phi}_\nsnv
(\hx) -\hat{\Gamma}_{\nrmu ;\nsnv}{}^{\ntd w}(\hx) \hat{\phi}_{\ntd
w}(\hx)
\end{gather}
\end{subequations}
where $\hat{\Gamma}$ is the twisted Christoffel symbol and $\hat{\phi}$ is
the orientation-orbifold dilaton vector. The twisted Riemann tensor is
constructed as usual from the twisted Christoffel symbol.

We note that the twisted Christoffel symbols appear in the classical bulk
equations of
motion
\begin{subequations}
\label{Eq6.22}
\begin{gather}
\left( \pl_\pm \de_\nsnv{}^{\!\!\nrmu} -\pl_\pm \hx^{\ntd w}
(\hat{\Gamma}^{(\mp)} )_{\ntd w;\nsnv}{}^{\!\!\nrmu} \right) \pl_\mp
\hx^\nsnv =0 \\
\hat{\Gamma}^{(\pm )} \equiv \hat{\Gamma} \pm \half \hh
\end{gather}
\end{subequations}
which follow, as promised in Subsec.~$5.3$, by variation of the
sigma-model orientation-orbifold action.

Further applying the principle of local isomorphisms, we now discuss other
properties of the twisted fields, beginning with the space-time parities
\begin{gather}
\hat{O}_\pm (-\hx) =\pm \hat{O}_\pm (\hx) \nn \\
\hat{O}_+ = \hG, \hh \text{ and the } \hat{R}'s ,\quad \quad \hat{O}_-=
\hx, \hB ,\hat{\Gamma} \text{ and } \hat{\phi}_\nrmu  \label{Eq6.23}
\end{gather}
which are the same as those of the corresponding untwisted fields.

Similarly, each twisted field has only two independent  reduced components
\begin{subequations}
\label{Eq6.24}
\begin{gather}
\hpl_\nrmu (\xi) =\frac{\pl}{\pl \hx^\nrmu (\xi)} ,\quad \hpl_\nrmu (\xi)
\hx_\s^\nsnv (\xi)=\de_{n(r)-n(s),0\,\text{mod }\r(\s)} \de_\m{}^\n
   \de_{u-v,0\,\text{mod }2} \\
\hG_{\nrmu;\nsnv}(\hx) =\hG_{\nrm;\nsn}^{(u+v)}(\hx) \\
\hB_{\nrmu;\nsnv}(\hx) =\hB_{\nrm;\nsn}^{(u+v)}(\hx) \\
\hh_{\nrmu;\nsnv;\ntd w}(\hx) =\hh_{\nrm;\nsn;\ntd}^{(u+v+w)}(\hx) \\
\hat{\Gamma}_{\nrmu ;\nsnv}{}^{\ntd w}(\hx)
=\hat{\Gamma}_{\nrm;\nsn}^{(u+v-w)}{}^\ntd (\hx) \\
\hat{R}_{\nrmu;\nsnv;\ntd w;n(q)\ep y}(\hx) =\hat{R}_{\nrm;\nsn;\ntd
;n(q)\ep}^{(u+v+w+y)}(\hx) \\
\hat{R}_{\nrmu;\nsnv}(\hx) =\hat{R}_{\nrm;\nsn}^{(u+v)}(\hx) \\
\hat{O}_{n(r_1 )\m_1 u_1 ;\ldots ;n(r_m )\m_m u_m}^{\quad \quad n(s_1
)\n_1 v_1 ;\ldots ;n(s_n )\n_n v_n} =\hat{C}_{n(r_1 )\m_1 ;\ldots n(r_m)
   \m_m}^{(\Sigma u_i -\Sigma v_j ) \quad \quad n(s_1 )\n_1 ;\ldots ;n(s_n
)\n_n}
\end{gather}
\end{subequations}
and we have checked in detail that the reduced two-component structures are
consistent with
the formulae of Riemannian geometry. In this demonstration, we found the
following formulae to be useful
\begin{subequations}
\label{Eq6.25}
\begin{gather}
\hpl_\nrmu \hat{O}_{n(s_1 )\n_1 v_1 ;\ldots ;n(s_m )\n_m v_m}^{\quad \quad
n(t_1 )\de_1 w_1 ;\ldots ;n(t_n )\de_n w_n} =\hat{C}_{\nrm ;n(s_1 )\n_1
;\ldots n(s_m)
   \n_m}^{(u +\Sigma v_i -\Sigma w_j ) \quad \quad n(t_1 )\de_1 ;\ldots
;n(t_n )\de_n}  \label{Eq 6.25a} \\
\hat{\Gamma}^{(w)}_{\nrm ;\nsn}{}^\ntd (\hx) =\srac{1}{2} \sum_{u=0}^1
\Big{(} \hpl_{\nrmu} \hG_{\nsn ;n(q)\ep}^{(w)}(\hx) +\hpl_{\nsn u}
   \hG_{\nrm ;n(q)\ep}^{(w)} (\hx) \bigspc \bigspc \quad \quad \nn \\
\quad \quad \bigspc \bigspc -\hpl_{n(q)\ep u} \hG_{\nrm;\nsn}^{(w)} (\hx)
\Big{)} \hG^{n(q)\ep ;\ntd}_{(u)} (\hx) \\
\hat{\nabla}_\nrmu \hat{O}_{n(s_1 )\n_1 v_1 ;\ldots ;n(s_m )\n_m
v_m}^{\quad n(t_1 )\de_1 w_1 ;\ldots ;n(t_n )\de_n w_n} =\hat{D}_{\nrm
;n(s_1 )\n_1 ;\ldots n(s_m)
   \n_m}^{(u +\Sigma v_i -\Sigma w_j) \quad \quad n(t_1 )\de_1 ;\ldots
;n(t_n )\de_n} \label{Eq 6.25c} \\
\hat{\nabla}_\nrmu \hat{\phi}_\nsnv (\hx) =\hpl_{\nrm 0} \hat{\phi}_{\nsn
,u+v}(\hx) +\sum_{w=0}^1 \hat{\Gamma}^{(u+v-w)}_{\nrm;\nsn}{}^\ntd (\hx)
   \hat{\phi}_{\ntd w}(\hx)
\end{gather}
\end{subequations}
where the general relation in \eqref{Eq 6.25a} follows by local
isomorphisms from Eq.~\eqref{Eq6.19}.

In terms of the reduced quantities, we find the world-sheet parities
\begin{subequations}
\label{Eq6.26}
\begin{gather}
\hat{\nabla}_\nrmu (-\xi) =(-1)^{u+1} \hat{\nabla}_\nrmu (\xi) \\
\hG_{(w)}^{\nrm;\nsn} (\hx(-\xi)) =(-1)^w \hG_{(w)}^{\nrm;\nsn} (\hx(\xi))
\\
\hat{\Gamma}^{(w)}_{\nrm;\nsn}{}^\ntd (\hx(-\xi)) =(-1)^{w+1}
\hat{\Gamma}^{(w)}_{\nrm;\nsn}{}^\ntd (\hx(\xi)) \\
\hat{R}^{(w)}_{\nrm;\nsn;\ntd ;n(q)\ep} (\hx(-\xi)) =(-1)^w
\hat{R}^{(w)}_{\nrm;\nsn;\ntd ;n(q)\ep} (\hx(\xi)) \\
\hat{R}^{(w)}_{\nrm;\nsn} (\hx(-\xi)) = (-1)^w \hat{R}^{(w)}_{\nrm;\nsn}
(\hx(\xi)) \\
\hat{\phi}_\nrmu (\hx(-\xi)) = (-1)^{u+1} \phi_\nrmu (\hx(\xi)) ,\quad
\bar{u},\bar{w} \in \{ 0,1\}
\end{gather}
\end{subequations}
which should be considered together with those given earlier in
Eq.~\eqref{Eq5.16}.

Following the procedure shown in Fig.~\ref{fig:coords}, we have already
given the explicit forms of $\hG, \hB$ and $\hh$ in Eqs.~\eqref{Eq5.18},
\eqref{Eq5.19}, and we similarly give the  explicit forms of the other
twisted fields:
\begin{subequations}
\label{Eq6.27}
\begin{gather}
\hG^{\nrmu;\nsnv}(\hx) =G^{i\Id ;j\Jd}(x\duals \sxh(\hx)) \schi^{-1}_\nrm
\schi^{-1}_\nsn U\hc {}_i{}^\nrm U\hc {}_j{}^\nsn (\srac{1}{\sqrt{2}}
   U\hc{}_\Id{}^u ) (\srac{1}{\sqrt{2}} U\hc {}_\Jd{}^v ) \\
\hG_{(w)}^{\nrm;\nsn}(\hx) \!=\!\srac{1}{4} \!\left( \!G^{ij} (x^0 \duals
\sxh^0 (\hx)) +(-1)^w G^{ij} (x^1 \duals \sxh^1 (\hx)) \!\right) \times
\bigspc
   \bigspc \quad \quad \nn \\
   \bigspc \bigspc \bigspc \quad \quad \times \schi^{-1}_\nrm
\schi^{-1}_\nsn U\hc {}_{\!i}^{\,\nrm} U\hc {}_{\!j}^{\,\nsn}
\end{gather}
\begin{align}
&\hat{\Gamma}_{\nrmu;\nsnv}{}^{\ntd w}(\hx) =\schi_\nrm \schi_\nsn
\schi_\ntd^{-1} U_\nrm{}^i U_\nsn{}^j U\hc {}_k{}^\ntd \times \nn \\
&\bigspc \bigspc \bigspc \times (\sqrt{2} U_u{}^\Id )(\sqrt{2} U_v{}^\Jd
)(\srac{1}{\sqrt{2}} U\hc{}_\Kd{}^w ) \Gamma_{i\Id ;j\Jd}{}^{k\Kd} (x
\,\duals \sxh(\hx)) \\
&\hat{\Gamma}_{\nrm;\nsn}^{(w)}{}^\ntd (\hx) = \srac{1}{2} \schi_\nrm
\schi_\nsn \schi_\ntd^{-1} U_\nrm{}^i U_\nsn{}^j U\hc {}_k{}^\ntd \times
\bigspc \nn \\
&\bigspc \bigspc \bigspc \times \left( \Gamma_{ij}{}^k (x^0 \duals
\sxh^0(\hx)) +(-1)^w \Gamma_{ij}{}^k (x^1 \duals \sxh^1(\hx)) \right) \\
& \hat{R}_{\nrmu ;\nsnv}(\hx) =\!\schi_\nrm \schi_\nsn U_\nrm{}^i
U_\nsn{}^j (\sqrt{2}U_u{}^\Id )(\sqrt{2}U_v{}^\Jd )
   R_{i\Id ;j\Jd}(x \,\duals \sxh (\hx)) \\
&\hat{R}^{(w)}_{\nrm;\nsn}(\hx) =\!\schi_\nrm \schi_\nsn U_\nrm{}^i
U_\nsn{}^j \left( R_{ij}(x^0 \duals \sxh^0(\hx)) \!+\!(-1)^w R_{ij}(x^1
\duals \sxh^1(\hx)) \right)
\end{align}
\begin{align}
\hat{\phi}_\nrmu (\hx) &=\schi_\nrm U_\nrm{}^i (\sqrt{2} U_u{}^\Id )
\phi_{i\Id} (x \duals \sxh(\hx)) \nn \\
&= \schi_\nrm U_\nrm{}^i (\sqrt{2} U_u{}^\Id ) \phi_i (x^\Id \duals
\sxh^\Id (\hx)) \,. \label{Eq 6.27g}
\end{align}
\end{subequations}
Here $G^{ij}, \Gamma_{ij}{}^k, R_{ij}$ and $\phi_i$ are the original
untwisted Einstein fields and $\sxh$ is the Einstein coordinate with
twisted boundary conditions (see
Fig.~\ref{fig:coords} and Eq.~\eqref{Eq5.17}). Including the results in
Eq.~\eqref{Eq5.18}, the general rule for the unreduced explicit forms
(with the full set of $\nrmu$ indices) is $x \dual \sxh (\hx)$ and a
factor $\schi U(\s) \sqrt{2}U$
or $\schi^{-1} U\hc(\s) \srac{1}{\sqrt{2}} U\hc$ respectively for each set
of down or up indices.

For the special case of the dilaton solution, we obtain the particular
form of the orientation-orbifold dilaton vector
\begin{subequations}
\label{Eq6.28}
\begin{gather}
\hat{\phi}_\nrmu (\hx) = \schi_\nrm U_\nrm{}^i (\sqrt{2} U_u{}^\Id )
\frac{\pl}{\pl \sxh^{i\Id}} \phi(\sxh^\Id (\hx)) \bigspc \bigspc \bigspc
\label{Eq 6.28a} \\
=\schi_\nrm U_\nrm{}^i (\sqrt{2} U_u{}^\Id ) \frac{\pl}{\pl \sxh^{i\Id}}
(\phi (\sxh^0 (\hx)) +\phi (\sxh^1 (\hx))) \\
= \hpl_\nrmu \hat{\phi} (\hx) \bigspc \bigspc \bigspc \bigspc \label{Eq
6.28c} \\
\hat{\phi} (\hx) \equiv \phi (\sxh^0 (\hx)) +\phi (\sxh^1 (\hx))
\end{gather}
\end{subequations}
where we used Eqs.~\eqref{Eq 6.7f}, \eqref{Eq 6.27g} to obtain
Eq.~\eqref{Eq 6.28a}, and Eq.~\eqref{Eq 5.17e} to obtain the final form in
Eq.~\eqref{Eq 6.28c}.
Here $\phi(x)$ is the original untwisted dilaton and $\hat{\phi}(\hx)$ is
the {\it orientation-orbifold dilaton}. The world-sheet and space-time
parities of
the orientation-orbifold dilaton
\begin{gather}
\hat{\phi} (\hx(-\xi)) =\hat{\phi} (\hx(\xi)) ,\quad \hat{\phi}
(-\hx(\xi)) =\hat{\phi} (\hx(\xi)) \label{Eq6.29}
\end{gather}
follow from the corresponding properties of the orientation-orbifold
dilaton vector and the dilaton therefore falls in the class $\hat{O}_+$
(see Eq.~\eqref{Eq6.23}) of orientation-orbifold operators.

\subsection{The Twisted Einstein Equations of Open-String Sector
$\hat{h}_\s$}

The two-component Einstein equations in Eq.~\eqref{Eq6.7} are easily
reexpressed in terms of the corresponding Riemannian eigenfields, and then
the principle
of local isomorphisms gives the {\it orientation-orbifold Einstein
equations}
\begin{subequations}
\label{Eq6.30}
\begin{align}
& \hat{R}_{\nrmu;\nsnv}(\hx)+ \bigspc \bigspc \bigspc \bigspc \bigspc \nn
\\
&\quad \quad \quad +\srac{1}{4} \hh_{\ntd w;\nrmu}{}^{\!\!n(q)\ep x} (\hx)
\hh^{\ntd w}{}_{\!\!n(q)\ep x;\nsnv}(\hx) -2 \hat{\nabla}_{\!\nrmu}
   \hat{\phi}_\nsnv (\hx)=0 \\
&\bigspc \quad \quad \left( \hat{\nabla}_{\ntd w} -2\hat{\phi}_{\ntd
w}(\hx) \right) \hh^{\ntd w}{}_{\nrmu;\nsnv}(\hx) =0
\end{align}
\end{subequations}
for each open-string sector $\hat{h}_\s$ of $A_M (H_-)/H_-$. When the
corresponding
 untwisted Einstein fields solve the untwisted Einstein equations, the
principle of
local isomorphisms guarantees that the explicit forms of the twisted fields in
Eqs.~\eqref{Eq5.18} and \eqref{Eq6.27} solve the orientation-orbifold
Einstein equations.

As always, the orientation-orbifold Einstein equations \eqref{Eq6.30} have
only two independent components in the two-dimensional space, and the
reduced two-component
form of this system is
\begin{subequations}
\label{Eq6.31}
\begin{gather}
\hat{R}^{(w)}_{\nrm ;\nsn}(\hx) \!+\!\srac{1}{2} \sum_{u=0}^1 \!\hh_{\ntd
;\nrm}^{(u)}{}^{\!\!n(q)\ep}(\hx) \hh^\ntd{}_{\!n(q)\ep ;\nsn}^{\quad
(w-u)} (\hx)
   \!-\! 2\hat{\nabla}_{\nrm 0} \hat{\phi}_{\nsn w}(\hx) \!=\!0 \\
\hat{\nabla}_{\ntd 0} \hh^\ntd{}_{\nrm;\nsn}^{\,\,\,\,(w)} (\hx)
-\sum_{u=0}^1 \hat{\phi}_{\ntd u}(\hx)
\hh^{\ntd}{}_{\nrm;\nsn}^{\,\,\,\,(w-u)}(\hx)=0 ,\quad
\bar{w}=0,1
\end{gather}
\end{subequations}
where we have used the identity in Eq.~\eqref{Eq 6.25c}. The different
numerical factors here are the result of performing internal sums on $u,v$
indices.

We know that the orientation-orbifold Einstein equations \eqref{Eq6.30},
\eqref{Eq6.31} hold in each open-string sector $\hat{h}_\s$ when the
original untwisted
closed string theory was conformal through one loop. Moreover,
Ref.~\cite{Orient1} gives a general argument (starting from {\it any}
conformal
closed string theory)
that each resulting open-string orientation-orbifold sector $\hat{h}_\s$
always contains the {\it full} orbifold Virasoro algebra
\begin{gather}
[\hat{L}_u (m\!+\!\srac{u}{2}) ,\hat{L}_v (n\!+\!\srac{v}{2}) ] =(m\!-\!n
\!+\!\srac{u-v}{2}) \hat{L}_{u+v} (m\!+\!n \!+\! \srac{u+v}{2}) \bigspc
\quad \quad \nn \\
\bigspc \quad \quad +\de_{m+n+\srac{u+v}{2},0} \frac{2c}{12}
(m\!+\!\srac{u}{2}) ((m\!+\!\srac{u}{2} )^2 -1) \label{Eq6.32}
\end{gather}
with the characteristically-doubled central charge $\hat{c}=2c$. Therefore
one expects that the orientation-orbifold Einstein system is the necessary
and sufficient condition that the open-string orientation-orbifold sector
satisfies the orbifold Virasoro algebra through one loop. It would be
interesting to check this conclusion explicitly at one loop using the
Feynman diagrams associated to the sigma-model orientation-orbifold action
\eqref{Eq5.20}.

\subsection{Monodromies and Boundary Conditions}

When $h_\s^2 =1$, the principle of local isomorphisms also gives us the
monodromies of all the twisted fields in the orientation-orbifold Einstein
equations:
\begin{subequations}
\label{Eq6.33}
\begin{gather}
h_\s^2 =1\!: \quad \text{automorphic responses } \dual \text{ monodromies}
\bigspc \\
\hx_\s^\nrmu (\xi +2\pi) =\hx_\s^\nrmu (\xi) e^{\tp (\nrrsf +\frac{u}{2})}
,\quad \hpl_\nrmu (\xi+2\pi) =e^{-\tp (\nrrsf +\frac{u}{2})} \hpl_\nrmu
(\xi) \\
\hG_{\nrm;\nsn}^{(w)} (\hx(\xi+2\pi)) =e^{-\tp (\frac{n(r)+n(s)}{\r(\s)}
+\frac{w}{2})} \hG_{\nrm;\nsn}^{(w)} (\hx(\xi)) \\
\hB_{\nrm;\nsn}^{(w)} (\hx(\xi+2\pi)) =e^{-\tp (\frac{n(r)+n(s)}{\r(\s)}
+\frac{w}{2})} \hB_{\nrm;\nsn}^{(w)} (\hx(\xi)) \\
\hh_{\nrm;\nsn;\ntd}^{(w)} (\hx(\xi+2\pi)) =e^{-\tp
(\frac{n(r)+n(s)+n(t)}{\r(\s)} +\frac{w}{2})} \hh_{\nrm;\nsn;\ntd}^{(w)}
(\hx(\xi))
\end{gather}
\begin{gather}
\hat{\Gamma}_{\nrm;\nsn}^{(w)}{}^\ntd (\hx(\xi+2\pi)) =e^{-\tp
(\frac{n(r)+n(s)-n(t)}{\r(\s)} +\frac{w}{2})}
\hat{\Gamma}_{\nrm;\nsn}^{(w)}{}^\ntd (\hx(\xi)) \\
\hat{R}_{\nrm;\nsn;\ntd ;n(q)\ep}^{(w)}(\hx(\xi +2\pi)) =e^{-\tp
(\frac{n(r)+n(s)+n(t)+n(q)}{\r(\s)} +\frac{w}{2})}
\hat{R}_{\nrm;\nsn}^{(w)} (\hx(\xi)) \\
\hat{R}_{\nrm;\nsn}^{(w)}(\hx(\xi +2\pi)) =e^{-\tp
(\frac{n(r)+n(s)}{\r(\s)} +\frac{w}{2})} \hat{R}_{\nrm;\nsn}^{(w)}
(\hx(\xi)) \\
\hat{\phi}_\nrmu (\hx(\xi+2\pi)) =e^{-\tp (\frac{n(r)}{\r(\s)}
+\frac{u}{2})} \hat{\phi}_\nrmu (\hx(\xi)) \,\rightarrow \,\hat{\phi}
(\hx(\xi+2\pi)) = \hat{\phi} (\hx(\xi)) \,.
\end{gather}
\end{subequations}
The general rule here, as in space-time orbifold theory, is a negative
phase for each down index and a positive phase for each up index. We note
in particular
that the orientation-orbifold dilaton, like the space-time orbifold
dilaton \cite{Geom}, has trivial monodromy.

Using the monodromies \eqref{Eq6.33} along with the world-sheet parities
\eqref{Eq5.16}, \eqref{Eq6.26} in the form
\begin{subequations}
\label{Eq6.34}
\begin{gather}
\hat{O}_\pm^{(w)} (\hx(-\xi)) =\pm (-1)^w \hat{O}_\pm^{(w)} (\hx(\xi)) \\
\hat{O}_+ = \hG ,\hh, \text{ the } \hat{R}'s, \,\hat{\phi}, \, \pl_t
\hat{O}_+ \text{ and } \pl_\xi \hat{O}_- \\
\hat{O}_- =\hx ,\hB ,\hat{\Gamma}, \hat{\phi}_\nrmu ,\,\pl_t \hat{O}_-
\text{ and } \pl_\xi \hat{O}_+
\end{gather}
\end{subequations}
we may derive, as above, the following boundary conditions on the twisted
fields:
\begin{subequations}
\label{Eq6.35}
\begin{gather}
\text{For } \hat{O}_{n(r_1)\m_1 ;\ldots ;n(r_m )\m_m}^{(w) \quad \quad
n(s_1 )\n_1 ;\ldots ;n(s_n )\n_n} (\hx) , \text{ define } N \equiv
  \sum_{i=1}^m n(r_i ) -\sum_{j=1}^n n(s_j ) \nn \\
\text{then} \bigspc \hat{O}_+^{(1)} (\hx(0)) ,\, \hat{O}_-^{(0)} (\hx(0))
=0 \bigspc \quad \label{Eq 6.35a} \\
\hat{O}_+^{(w)} (\hx(\pi)) =0 \,\text{ unless } \srac{N}{\r(\s)} \in \Zint
\\
\hat{O}_-^{(w)} (\hx(\pi)) =0 \,\text{ unless } \srac{N}{\r(\s)} \in \Zint
+\srac{1}{2} \,.
\end{gather}
\end{subequations}
We mention in particular that the orientation-orbifold dilaton
$\hat{\phi}$ falls in the class $\hat{O}_+$ with the specific values
$N=w=0$. We finally note that
as usual, the boundary conditions \eqref{Eq 6.35a} at $\xi =0$ follow
directly from world-sheet parity alone, and hence hold for all
$\hat{h}_\s$.

\section{Discussion}

In this paper we have provided a description at the action level of the
WZW orientation orbifolds, which were constructed at the operator level in
Ref.~\cite{Orient1}. We have
moreover extended this classical description to include the action
formulation of a large class of sigma-model orientation orbifolds, and
their associated
orientation-orbifold Einstein equations.

Directions for future work include the following:

\noindent $1)$ For the WZW orientation orbifolds it may be possible to
find more restrictive boundary conditions for the basic fields $\hg$ at
$\xi =\pi$ when
$h_\s^2 \!\neq \!1$, consistent with the variational and twisted current
boundary conditions discussed in the text. Such boundary conditions on
$\hg$ may follow
by a detailed analysis of the classical equations of motion
\eqref{Eq3.20}, using the known monodromies of the currents, as seen for the
free bosonic strings in
Ref.~\cite{Orient1} and Subsec.~$5.6$.

\noindent $2)$ It may similarly be possible to find more restrictive boundary
conditions for the basic fields of the coset orientation orbifolds.
Additionally, by
relaxing our provisional assumption that the WZW orientation-orbifold
boundary conditions survive the gauging, more general variational boundary
conditions can
also be studied.

\noindent $3)$ More restrictive boundary conditions on the basic fields of
the general sigma-model orientation orbifolds are particularly desirable
-- because for $h_\s^2 \!\neq \!1$ at
$\xi=\pi$ we have so far only found the variational boundary conditions.
As seen for WZW and the free boson models, possibilities here include
study of various sigma-model analogues of the currents.

\noindent $4)$ As seen for ordinary orbifolds in Ref.~\cite{Geom},
 the linear
 symmetry conditions
\eqref{Eq5.10} imply {\it selection rules} for the coefficients in the moment
expansions of the
twisted Einstein tensors.
In ordinary orbifolds such selection rules are known to connect the
monodromies of
the twisted Einstein tensors with the monodromy of the twisted Einstein
coordinates $\hx$.
 Although we will not pursue this topic here,
the corresponding selection rules for the orientation orbifolds may similarly
 connect the boundary
conditions of the twisted Einstein tensors with those of $\hx$.

\vspace{-.02in}
\bigskip

\noindent
{\bf Acknowledgements}

For helpful discussions, we thank J.~de Boer, O.~Ganor, P.~Ho\v{r}ava and
E.~Rabinovici.

This work was supported in part by the Director, Office of Energy
Research, Office of High Energy and Nuclear Physics, Division of High
Energy Physics of the U.S. Department of Energy under Contract
DE-AC03-76SF00098 and in part by the National Science Foundation under
grant PHY00-98840.

\appendix

\section{Useful Identities}

In the computations of the text, we found the following identities useful
\begin{equation}
\r_0 = \left( \begin{array}{cc} 1&0 \\ 0&0 \end{array} \right) ,\quad \r_1
=\left(
    \begin{array}{cc} 0&0 \\ 0&1 \end{array} \right) ,\quad \tau_1 \r_\Id
\tau_1 =\one -\r_\Id = (\tau_1 )_{\Id \Jd} \r_\Jd \label{EqA.1}
\end{equation}
\begin{gather}
(\tau_1 )_\Id{}^\Kd (\tau_1 )_\Jd{}^\Kd =\de_{\Id \Jd} (\tau_1 )_\Id
{}^\Kd ,\quad \Id ,\Jd ,\Kd \in \{ 0,1\} \label{EqA.2}
\end{gather}
\begin{subequations}
\label{EqA.3}
\begin{gather}
\sqrt{2} \sum_{\Id =0}^1 U_u{}^\Id U\r_\Id U\hc = \tau_u ,\quad \sqrt{2}
\sum_{\Id =0}^1 U_u{}^\Id U(\one -\r_\Id ) U\hc =(-1)^u \tau_u
  ,\quad u=0,1 \label{Eq A.3a} \\
U\tau_1 =\tau_3 U ,\quad \sqrt{2} \sum_{\Id =0}^1 U_u{}^\Id U_v{}^\Id
(U\hc)_\Id{}^w =\de_{u+v+w,0\, \text{mod }2} \label{Eq A.3b}\\
2^{n/2} \sum_{\Id =0}^1 U_{u_1}{}^\Id \cdots U_{u_n}{}^\Id f(x^\Id )
=f(x^0 ) +(-1)^{\sum_{i=1}^n u_i} f(x^1) \label{Eq A.3c}
\end{gather}
\end{subequations}
where $\vec{\tau}$ are the Pauli matrices and $\tau_0 \!=\!\thickone$ is
the 2x2 unit matrix.

\section{Space-Time Parity in Coset CFTs \label{CosetApp}}

In this appendix, we will study the space-time parities of the Einstein
fields $G,B$ and the dilaton $\phi$ for any $g/h$ coset CFT
\cite{BH,2faces1,GKO,ICFT}
with $G/H$ a reductive coset space.

We begin with the well-known form [39-43] of the total $G$
and $B$ fields in the coset CFT:
\begin{subequations}
\label{EqB.1}
\begin{gather}
(G(x) +B(x) )_{ij}^{tot} = (G(x) +B(x))_{ij} +\Delta (G(x)+B(x))_{ij} \\
\Delta (G(x)+B(x))_{ij} \equiv -2(e(x)_i{}^{\hat{a}} G_{\hat{a}\hat{c}})
(\bar{e}(x)_j{}^{\hat{b}} G_{\hat{b}\hat{d}}) X^{-1}(x)^{\hat{c}\hat{d}}
\label{Eq B.1b} \\
X(x)_{\hat{a}\hat{b}} \equiv G_{\hat{a}\hat{b}}
-\Omega(x)_{\hat{a}\hat{b}} ,\quad \Omega (x)_{ab} =\Omega(x)_a{}^c G_{cb}
\\
i,j =1\ldots \text{dim }g ,\quad a,b=1\ldots \text{dim }g ,\quad
\hat{a},\hat{b}=1\ldots \text{dim }h \,.
\end{gather}
\end{subequations}
Here the quantities $G(x),B(x),e(x),\bar{e}(x)$ and $\Omega(x)$ are the
standard WZW quantities in Eq.~\eqref{Eq4.1}. The tangent-space metric
$G_{ab}$ of $g$ is block diagonal in $h$ and $g/h$ and
$G_{\hat{a}\hat{b}}$ is the invertible induced metric on $h \subset g$. We
already know (see Eq.~\eqref{Eq 4.6b}) that
the WZW metric and $B$ field exhibit the space-time parities
\begin{gather}
G_{ij}(-x) =G_{ij}(x) ,\quad B_{ij}(-x) =-B_{ij}(x) \label{EqB.2}
\end{gather}
so it is only necessary to study the extra contributions $\Delta (G+B)$ in
\eqref{Eq B.1b}.

In this discussion the following identities are useful
\begin{subequations}
\label{EqB.3}
\begin{gather}
\Omega^T (x)_{ac} =\Omega (x)_{ca} =\Omega (x)_c{}^d G_{da} = \Omega^{-1}
(x)_a{}^d G_{dc} = \Omega^{-1} (x)_{ac} \\
\Omega (-x)_{ab} =\Omega^{-1} (x)_{ab} =\Omega^T (x)_{ab} \\
e(-x)_i{}^a = -\bar{e}(x)_i{}^a ,\quad \bar{e}(-x)_i{}^a =-e(x)_i{}^a
\end{gather}
\end{subequations}
where superscript $T$ is transpose. Because $G/H$ is reductive, all these
relations can be restricted to $h$, e.~g.
\begin{gather}
\Omega (x)_{\hat{c}}{}^{\hat{d}} G_{\hat{d}\hat{a}} =
\Omega^{-1}(x)_{\hat{a}}{}^{\hat{d}} G_{\hat{d}\hat{c}} \label{EqB.4}
\end{gather}
and it follows that:
\begin{gather}
X(-x)_{\hat{a}\hat{b}} =X^T (x)_{\hat{a}\hat{b}} ,\quad
X^{-1}(-x)^{\hat{a}\hat{b}} = X^{-1T}(x)^{\hat{a}\hat{b}} \,. \label{EqB.5}
\end{gather}
Then we may compute
\begin{subequations}
\label{EqB.6}
\begin{align}
\Delta (G(-x) +B(-x))_{ij} &= -2(e(-x)_i{}^{\hat{a}}
G_{\hat{a}\hat{c}})(\bar{e}(-x)_j{}^{\hat{b}} G_{\hat{b}\hat{d}})
X^{-1}(-x)^{\hat{c}\hat{d}} \\
&=-2(\bar{e}(x)_i{}^{\hat{a}} G_{\hat{a}\hat{c}}) (e(x)_j{}^{\hat{b}}
G_{\hat{b}\hat{d}}) X^{-1T}(x)^{\hat{c}\hat{d}} \nn \\
&=-2(e(x)_j{}^{\hat{a}} G_{\hat{a}\hat{c}})(\bar{e}(x)_i{}^{\hat{b}}
G_{\hat{b}\hat{d}}) X^{-1}(x)^{\hat{c}\hat{d}} \nn \\
&= \Delta (G(x)+B(x))_{ji}
\end{align}
\end{subequations}
and this result implies the space-time parities of $\Delta G$ and $\Delta
B$ separately
\begin{gather}
\Delta G_{ij} (-x) =\Delta G_{ij}(x) ,\quad \Delta B_{ij} (-x) =-\Delta
B_{ij} (x) \label{EqB.7}
\end{gather}
because they are respectively the symmetric and antisymmetric parts of
$\Delta (G+B)$.

The final result for the space-time parities of the total coset $G$ and $B$
\begin{gather}
G_{ij}^{tot}(-x) = G_{ij}^{tot}(x) ,\quad B_{ij}^{tot}(-x) =
-B_{ij}^{tot}(x)  \label{EqB.8}
\end{gather}
then follows immediately from Eqs.~\eqref{EqB.2} and \eqref{EqB.7}.

The formula for the coset dilaton is also well-known [39-43,50]
\begin{gather}
\phi (x) =-\srac{1}{2} \ln \det (X(x)) ,\quad \phi_i (x) \equiv
\nabla_{\!i} \phi(x) =\pl_i \phi (x)  \label{EqB.9}
\end{gather}
where this $\phi_i$ is the special dilaton vector associated to the
dilaton. The
space-time parities of the dilaton and dilaton vector
\begin{gather}
\phi(-x) =-\srac{1}{2} \ln \det (X^T(x)) =\phi(x) ,\quad \phi_i (-x)
=-\phi_i (x)  \label{EqB.10}
\end{gather}
follow immediately from Eq.~\eqref{EqB.5}.

These space-time parities and the discussion of the Subsecs.~$5.1$ and
$6.1$ show explicitly that all (reductive) $g/h$ coset CFTs admit the
basic orientation-reversing automorphism, at
least through one loop. We know of course from the operator form of the
coset orientation-orbifold stress tensors in Eq.~\eqref{Eq5.34} that the
basic
orientation-reversal is an automorphism of the coset CFT to all orders.

\vskip .5cm
\addcontentsline{toc}{section}{References}

\renewcommand{\baselinestretch}{.4}\rm
{\footnotesize

\providecommand{\href}[2]{#2}\begingroup\raggedright\endgroup

\pagebreak

\end{document}